\documentclass[12pt]{article}
\usepackage{amsmath}
\usepackage{graphicx,psfrag,epsf}
\usepackage{enumerate}
\usepackage{natbib}
\usepackage{hyperref}
\usepackage{hypernat}
\usepackage{url} 
\usepackage{xcolor}
\graphicspath{{figures/}}
\usepackage{bbm}
\usepackage{amsfonts}
\usepackage{amssymb}
\usepackage{amsthm}
\usepackage{amsmath}
\newcommand{\mb}[1]{\boldsymbol{#1}}
\newcommand{\Cov}{\textrm{Cov}}
\DeclareMathOperator*{\argmax}{arg\,max}
\DeclareMathOperator*{\argmin}{arg\,min}

\usepackage{setspace}

\hypersetup{breaklinks=true,
            pdfauthor={Drew Yarger (Sandia National Laboratories), J. Derek Tucker (Sandia National Laboratories)},
            pdfkeywords = {changepoint analysis, functional data analysis},
            pdftitle={Detecting changepoints in globally-indexed functional time series},
            colorlinks=true,
            citecolor=blue,
            urlcolor=blue,
            linkcolor=magenta,
            pdfborder={0 0 0}}
            
\usepackage{xcolor}
\definecolor{Blue}{RGB}{0,119,255}

\newcommand{\checklater}[1]{#1}
\usepackage{etoolbox}

\newif\ifabbreviation
\pretocmd{\thebibliography}{\abbreviationfalse}{}{}
\AtBeginDocument{\abbreviationtrue}

\newcommand{\blind}{0}

\def\spacingset#1{\renewcommand{\baselinestretch}%
{#1}\small\normalsize} \spacingset{1}

\begin{document}



\if0\blind
{
  \title{\bf Detecting changepoints\\ in globally-indexed functional time series}
  \author{Drew Yarger and J. Derek Tucker\hspace{.2cm}\\
    Statistical Sciences, Sandia National Laboratories}
  \maketitle
} \fi

\if1\blind
{
  \bigskip
  \bigskip
  \bigskip
  \begin{center}
    {\LARGE\bf Detecting changepoints\\ in globally-indexed functional time series}
\end{center}
  \medskip
} \fi

\bigskip
\begin{abstract}
In environmental and climate data, there is often an interest in determining if and when changes occur in a system. 
Such changes may result from localized sources in space and time like a volcanic eruption or climate geoengineering events. 
Detecting such events and their subsequent influence on climate has important policy implications. 
However, the climate system is complex, and such changes can be challenging to detect. 
One statistical perspective for changepoint detection is functional time series, where one observes an entire function at each time point. 
We will consider the context where each time point is a year, and we observe a function of temperature indexed by day of the year. 
Furthermore, such data is measured at many spatial locations on Earth, which motivates accommodating sets of functional time series that are spatially-indexed on a sphere. 
Simultaneously inferring changes that can occur at different times for different locations is challenging. 
We propose test statistics for detecting these changepoints, and we evaluate performance using varying levels of data complexity, including a simulation study, simplified climate model simulations, and climate reanalysis data. 
We evaluate changes in stratospheric temperature globally over 1984-1998.
Such changes may be associated with the eruption of Mt.\ Pinatubo in 1991. 
\end{abstract}

\noindent%
{\it Keywords:} atmospheric science, changepoint analysis, spatially-dependent functional data
\vfill

\newpage
\spacingset{1.8} 
\section{Introduction}\label{sec:intro}

In environmental and climate data, there is often an interest in determining if and when a change occurs in a system. 
One type of such changes in climate may be due to localized sources in space and time like a volcanic eruption. 
The sulfur dioxide aerosols injected into the stratosphere by volcanic eruptions can absorb incoming solar radiation, increasing temperatures in the stratosphere and decreasing surface temperatures \citep{marshall_2022}. 
In climate simulation experiments, volcanic eruptions lead to a variety of changes in climate, including fewer temperature extremes \citep{paik_2018} and changes to the El Ni\~no cycle \citep{khodri2017tropical}. 
Climate geoengineering events, a broad class of potential human interventions on the Earth's climate, may also lead to changes in the climate \citep{national2021reflecting, lenton2009radiative}. 
Of these events, an injection of aerosols into the stratosphere is the most similar to a volcanic eruption. 
Other interventions include marine cloud brightening (increasing the reflectance of low-level clouds) and removal of carbon dioxide from the atmosphere \citep{lenton2009radiative}.
The effects of such interventions are often not well understood, and approaches for evaluating their downstream effects in the climate system are needed when planning such interventions.
Furthermore, methodology for detecting such an intervention and evaluating its effect on the climate can inform policy and security interests, since such interventions may disproportionately affect different countries and could be done unilaterally \citep{national2021reflecting}. 

In this work, we propose and evaluate statistical methodology in the context of the June 1991 eruption of Mt.\ Pinatubo in the Philippines. 
The eruption injected approximately 20 million tons of sulfur dioxide in the stratosphere, which spread over the tropics in the following weeks \citep{bluth_1992}. 
The eruption has been linked using observations to increases in stratospheric temperature \citep{labitzke1992stratospheric} and decreases in surface temperatures \citep{parker1996impact}. 
The eruption size and the availability of atmospheric observations at the time make it an attractive test case for statistical methodology evaluating climate impacts from localized sources. 
We are interested in a few questions. 
Can we detect changes in the climate associated with Mt.\ Pinatubo in the presence of complex climate variability? 
What is that heterogeneous effect on the climate? 
What does this test case tell us about our ability to detect and evaluate climate geoengineering?

Various approaches have been proposed for detecting changes in climate data. 
The work \cite{reeves_2007} compares approaches for basic univariate time series. 
More recently, \cite{lund_2022} discusses important aspects and challenges of changepoint detection in climate time series data, including controlling for seasonality. 
Both reviews focus on time series and ignore the potential spatial nature of the data. 
\cite{majumdar2005spatio} considers the spatial problem yet estimates a change time that is shared across locations.
One work that addresses the spatial problem and heterogeneity across locations is \cite{moradi_2023}, though adjusting for seasonality is not addressed.

\begin{figure}
\centering
\includegraphics[width =.7\textwidth]{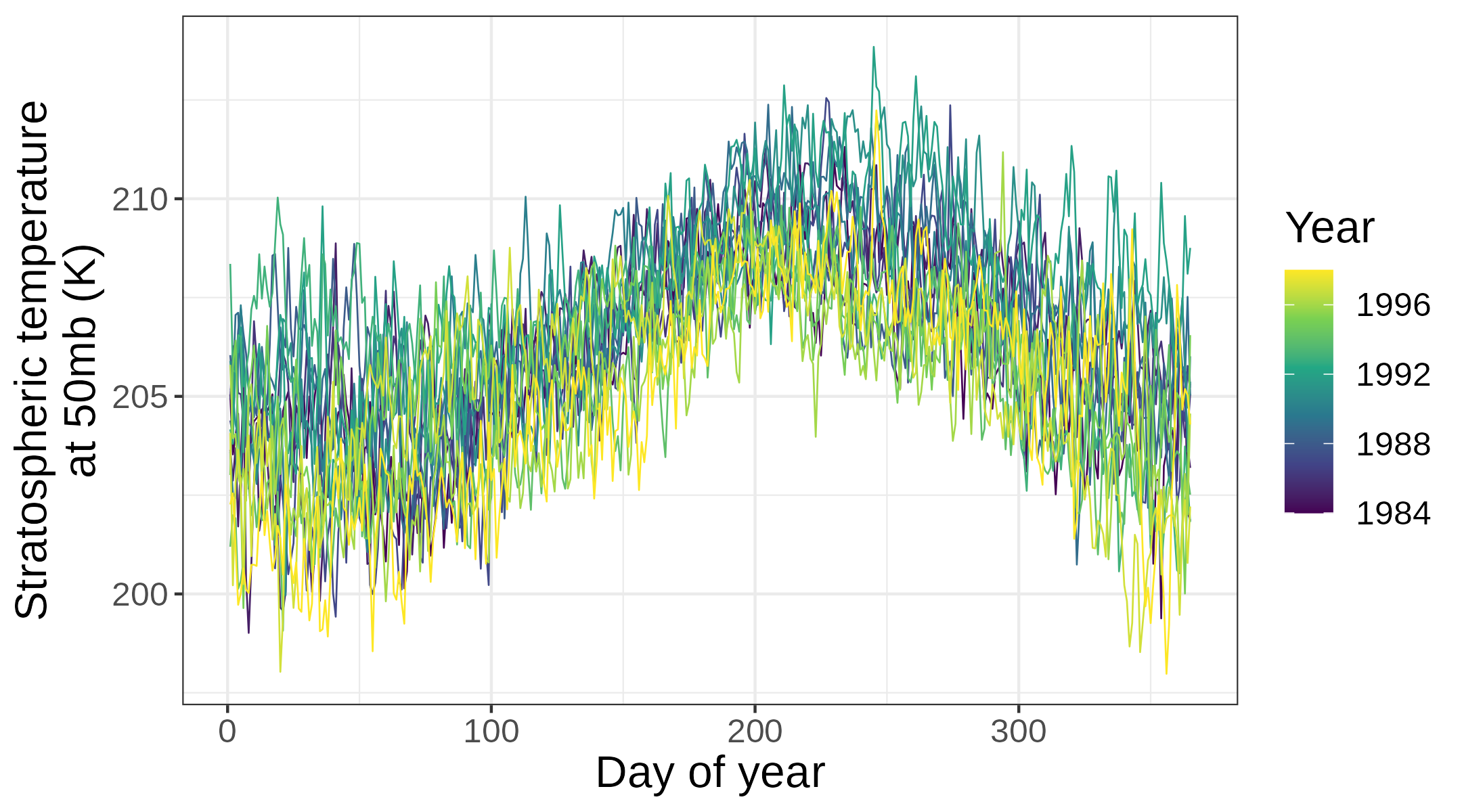}
\caption{MERRA-2 climate reanalysis data presented as a functional time series \citep{merra2} at $115^\circ$E and $10^\circ$N.}\label{fig:merra2_intro}
\end{figure}

In the statistics literature, one perspective for changepoint estimation is functional time series \citep[for example, ][]{hormann2012functional}, where we wish to detect a distributional change in a sequence of functions. 
In Figure \ref{fig:merra2_intro}, we present stratospheric temperature data from the MERRA-2 climate reanalysis at one location as a functional time series \citep{merra2}. 
In this setting, we let $k=1, \dots, N$ denote the year and observe a function of temperature $Y_k(u)$ of $u \in [0,1]$ where $u \cdot 365.25$ is the day of the year. 
The functional time series at-most-one-change (AMOC) model is \begin{align*}
Y_k(u) &= \mu(u) + \mathbbm{1}(k > \tau)\delta(u) + \epsilon_k(u).
\end{align*}
Here $\mu(u)$ is the mean function of the data before the changepoint $\tau$, after which the mean is $\mu(u) 
+ \delta(u)$.
The error $\epsilon_k(u)$ is assumed to have zero mean and may be dependent in $u$ and $k$. 
Approaches for detecting whether a change exists and for estimating $\tau$ have been proposed in various works including \cite{berkes:2009, aue2009estimation, Aue_2018, aston_2012}, and we review these in more detail in Section \ref{sec:method_independent}. 

In this work, we consider the problem where we observe such functional time series at many spatial locations, and we aim to evaluate changes at each location. 
That is, instead of looking at data at only one location as in Figure \ref{fig:merra2_intro}, we are interested in heterogenous effects of Mt.\ Pinatubo at different locations across the globe. 
There are some opportunities and challenges with this additional structure to the problem.
Spatial coherence in the changepoints should allow one to borrow strength from different locations to improve their estimates. 
There may also be spatial dependence in nuisance variability which challenges our ability to borrow this strength.
The full heterogenous model will allow for a spatially-varying time and size of a detected change. 
Inferring changepoints for this model is a challenging multiple testing problem with null hypotheses $H_{0i}: \{\textrm{no change at location }\mb{s}_i\}$ for $i=1$, $2$, $\dots$, $n$, where we observe data at $n$ different locations indexed by $\mb{s}_i$. 

Changepoint detection for spatially-dependent functional time series began in \cite{gromenko:2017} which was later extended in \cite{zhao2019composite} and \cite{dette_2022}. 
While these approaches address spatial dependence in the functional data, one changepoint time for all locations is assumed \citep[as in][]{majumdar2005spatio}, resulting in no spatial heterogeneity and a single determination if a changepoint occurred. 
This may be appropriate for data defined on small spatial regions, but it is reductive for globally-defined data. 
We are particularly inspired by the model recently proposed by \cite{li_changepoint_2022}. 
It has many attractive components mentioned earlier: a spatially varying changepoint time $\tau(\mb{s})$, spatially-varying mean function $\mu(\mb{s},u)$ and change function $\delta(\mb{s}, u)$, and spatially-dependent error $\epsilon_k(\mb{s},u)$.
For the data $Y_k(\mb{s}_i, u)$ now indexed by $\mb{s}$, the model is \begin{align}
Y_k(\mb{s}_i, u) &= \mu(\mb{s}_i, u) +\mathbb{I}(k > \tau(\mb{s}_i))\delta(\mb{s}_i, u)+ \epsilon_k(\mb{s}_i, u). \label{eq:orig_model}
\end{align}
We also assume that we observe the function $Y_k(\mb{s}_i, u)$ at points $u_{j}$, $j=1, \dots, m$, with $u_j$ evenly spaced on $[0,1]$. 
\cite{li_changepoint_2022} propose a Bayesian hierarchical model that vastly improves estimates of $\tau(\mb{s})$ compared to approaches that ignore space.

We will propose an approach for estimation of the model in \eqref{eq:orig_model} that is more broad, applicable, and direct compared to \cite{li_changepoint_2022}.   
For example, \cite{li_changepoint_2022} consider a single family of test statistics called ``fully-functional,'' and we will develop approaches for both ``fully-functional'' and ``score-based'' statistics. 
While \cite{li_changepoint_2022} focus on the spatial estimation of $\tau(\mb{s})$, we also show that one can improve changepoint detection using spatial information. 
One major contrast is that, while \cite{li_changepoint_2022} model the changepoint test statistics, we directly model the original data. 
There are benefits and drawbacks for each approach. 
For example, to facilitate modeling the original data, we assume its normality, a much more restrictive condition than considered in \cite{li_changepoint_2022}. 
Conversely, modeling the test statistics poses its own challenges. 
\cite{li_changepoint_2022} use a piecewise-linear and normal approximation to nonnegative test statistics; while they show that this works in practice, such a formulation is somewhat unnatural. 
This also results in a relatively complicated and computationally-intensive Bayesian hierarchical model. 
We will show that by modeling the dependence in the original data, we can improve both changepoint detection and estimates of the changepoint time using relatively simple tools for estimating the model \eqref{eq:orig_model} without proposing a new model for the test statistics. 

We evaluate our approach in the context of a statistical simulation study, a simplified climate simulation, and climate reanalysis, which represent successively more challenging applications. 
The simulation study shows that, by sharing information across space, one can dramatically increase power for changepoint detection, use multiple-testing corrections to control the false discovery rate, and estimate spatially-varying changepoints with more accuracy. 
We next apply changepoint techniques to simplified climate model output based on the Held-Suarez-Williamson model \citep[HSW, ][]{hsw_1994, williamson_1998} in \cite{hollowed_2022} (referred to as HSW++). 
This simulates a Mt.\ Pinatubo-like injection of aerosols in the stratosphere with simplified climate processes. 
With multiple-testing corrections on spatially-predicted test statistics, we detect spatially-varying changes in stratospheric temperature in line with expectations. 
The analysis on climate reanalysis data \citep[MERRA-2,][]{merra2} is the most challenging setting in this work. 
When using spatial information, we detect stratospheric temperature increases and evaluate them in the context of Mt.\ Pinatubo by comparing estimates of $\tau(\mb{s})$ and $\delta(\mb{s}, u)$. 
Overall, this work establishes the spatially-correlated functional time series changepoint detection as a useful tool to detect changes in the climate due to localized sources.

We discuss the organization of the rest of the paper.
In Section \ref{sec:changepoint_procedures}, we introduce the changepoint detection procedures used, including the proposed spatially-predicted test statistics.
In Section \ref{sec:sphere}, we propose a basic estimation procedure for a spatially-dependent functional data model indexed on the sphere (written as $\mathbb{S}^2$), which to our knowledge has not yet been addressed.
Section \ref{sec:simulation} details the simulation study, and Section \ref{sec:data_analysis} presents the two data analyses. 
We discuss areas of improvement for future analyses in Section \ref{sec:discussion}. 
Accompanying the paper is a supplement with additional details. 

\section{Changepoint detection for functional time series} \label{sec:changepoint_procedures}


\subsection{Application at each location separately}\label{sec:method_independent}

Our baseline detection procedure computes test statistics using data only at one location at a time, ignoring spatial coherency and dependence in the data. 
We review this literature, focusing on ``score-based" and ``fully-functional" statistics. 
Other approaches for detecting changes in functional data include the fused-lasso approach of \cite{harris_2022}.

\subsubsection{Score-based approaches}

The score-based group of detection procedures was proposed in \cite{berkes:2009} and \cite{aue2009estimation}. 
This approach estimates a model under the null hypothesis and uses functional principal component scores to evaluate a change.  
That is, one estimates the model at location $\mb{s}_i$ of $
Y_{k}(\mb{s}_i, u_{j}) = \mu^{\mb{s}_i}(u_{j}) + \sum_{q=1}^Q Z^{\mb{s}_i}_{qk} \phi_q^{\mb{s}_i}(u_{j})$ for some fixed $Q$. 
We refer to $\{\phi_{q}^{\mb{s}_i}(\cdot)\}_{q}$ and $\{Z_{qk}^{\mb{s}_i}\}_{q}$ as the functional principal components and scores of the data at $\mb{s}_i$, respectively.  
Here, the superscript $\mb{s}_i$ indicates that the estimates only use the data at $\mb{s}_i$.
To test the null hypothesis of $\delta(\mb{s}_i, u) = 0$ for all $u \in [0,1]$, one may use the test statistic \begin{align}
  T_{score}^{\mb{s}_i}= \frac{1}{N^2} \sum_{q = 1}^Q \frac{1}{\lambda_q^{\mb{s}_i}}  \sum_{t=1}^N \left( \sum_{k=1}^t Z_{qk}^{\mb{s}_i} - \frac{t}{N}\sum_{k=1}^NZ_{qk}^{\mb{s}_i}\right)^2,\label{eq:score_individ}
\end{align}where $\lambda_q^{\mb{s}_i} = \textrm{Var}(Z^{\mb{s}_i}_{qk})$ is the variance of the $q$-th principal component at location $\mb{s}_i$. 
The test statistic is a normalized sum of CUSUM (cumulative sum) statistics of the first $Q$ scores. 
Under conditions on the error process $\epsilon_{k}(\mb{s}_i, u)$, \cite{berkes:2009} show that under the null hypothesis $
T_{score}^{\mb{s}_i} \overset{d}{\to} \sum_{q = 1}^Q \int_0^1 B_{q}^2(x) dx$ as $N\to \infty$ where $\{B_q(\cdot)\}_{q=1}^Q$ are independent, standard Brownian bridges and $ \overset{d}{\to}$ represents convergence in distribution. 
Furthermore, they establish that if $\delta(\mb{s}_i, \cdot)$ is not orthogonal to the subspace spanned by $\{ \phi_q^{\mb{s}_i}\}_{q=1}^Q$, then $T_{score}^{\mb{s}_i}$ converges in probability to $\infty$ as $N \to \infty$. 
To estimate a changepoint time, \cite{aue2009estimation} propose and provide asymptotic analysis of the estimator \begin{align*}
\hat{\tau}^{\mb{s}_i}_{score}& =\argmax_{t=1, \dots, N} \frac{1}{N} \sum_{q=1}^Q  \frac{1}{\lambda_q^{\mb{s}_i}} \left( \sum_{k=1}^t Z_{qk}^{\mb{s}_i} - \frac{t}{N}\sum_{k=1}^NZ_{qk}^{\mb{s}_i}\right)^2.
\end{align*}

Various work has expanded upon this score-based approach. 
For example, \cite{jiao_2022} aim to find a favorable set of basis functions for projection that may not necessarily be eigenfunctions. 
\cite{stoehr_2021} and \cite{jiao_2022b} propose detection procedures for changes in the functional covariance. 
\cite{chiou_2019} use score-based test statistics to find multiple changepoints.
\cite{aston_2012} analyze a ``epidemic''-type changepoint, where the mean later returns to its original level after a second time point. 

\subsubsection{Fully-functional approaches}
A newer approach for changepoint detection for functional data was introduced in \cite{Aue_2018}. 
This ``fully-functional'' test statistic relies upon the raw data instead of estimated principal component scores, but the test statistic takes a similar CUSUM-like form. 
\cite{Aue_2018} introduce the test statistic of $T_{ff}^{\mb{s}_i} = \max_t\lVert S_{t, ff}^{\mb{s}_i}(\cdot) \rVert^2$, where \begin{align}
    S_{t, ff}^{\mb{s}_i}(u) = \frac{1}{\sqrt{N}}\left\{\sum_{k=1}^t Y_k^{\mb{s}_i}(u) -\frac{t}{N}\sum_{k=1}^N Y_k^{\mb{s}_i}(u) \right\},\  t=0, \dots, N,\label{eq:ff}
\end{align}$Y_k^{\mb{s}_i}(u) = Y_k(\mb{s}_i, u)$ is viewed as data at only that location, and $\lVert f \rVert^2 = \int_0^1 f^2(u) du$. 
To estimate the distribution of $T_{ff}^{\mb{s}_i}$ under the null hypothesis, one must estimate the long-run variance of the leading principal components. 
These principal component variances $\{\lambda_{q,ff}^{\mb{s}_i}\}$ are estimated assuming the {\it alternative} hypothesis is true. 
In particular, it is shown that $T_{ff}^{\mb{s}_i} \overset{d}{\to}\displaystyle \sup_{0 \leq x\leq 1} \sum_{q=1}^\infty \lambda_{q,ff}^{\mb{s}_i} B_q^2(x)$ where $\{B_q(x)\}$ again are independent standard Brownian bridges. 
In practice, the infinite sum must be truncated and approximated with the first $Q$ terms. 
To estimate $\tau(\mb{s}_i)$, a natural estimator is $
\hat{\tau}_{ff}^{\mb{s}_i} = \argmax_t\lVert S_{t, ff}^{\mb{s}_i}(\cdot) \rVert^2,$ and \cite{Aue_2018} also justify this estimator in theoretical results.  

The fully-functional approach has also spawned further research. 
\cite{horvath_2022}, \cite{aue_2020}, and \cite{dette_2021} detect changes in the covariance function using fully-functional or similar test statistics.
\cite{rice_2022} estimate multiple changepoints, and \cite{wegner_2022} propose a robust fully-functional test statistic. 
\cite{tucker_elastic_2022} develop test statistics for functional data with phase variability.

\subsection{Assumed model for error}\label{sec:assumed_model}

To share information across locations, some structure in the error process $\epsilon_k(\mb{s}, u)$ is assumed, which influences how one analyzes the data. 
Most critically, we will incorporate spatial dependence in $\epsilon_{k}(\mb{s}, u)$ through its principal component scores.
We assume that \begin{align*}
\epsilon_{k}(\mb{s}_{i}, u_{j}) &= \sum_{q=1}^Q Z_{qk}(\mb{s}_{i}) \phi_q(u_{j}) + W_{ijk},
\end{align*}where $W_{ijk} \sim \mathcal{N}(0, \xi^2)$ is normally-distributed measurement error with variance $ \xi^2 > 0$, $\{\phi_q(\cdot)\}$ are a set of principal component functions, and $\{Z_{qk}(\mb{s})\}$ are spatial fields of principal component scores. 
The scores are further decomposed $Z_{qk}(\mb{s}_{i}) = \tilde{Z}_{qk}(\mb{s}_i) + U_{iqk}$, where, for each $q$ and $k$, $\tilde{Z}_{qk}(\mb{s}_i) $ is a spatially-dependent random field with a Mat\'ern covariance function, and $\{U_{iqk}\}$ are independent across all $i$, $q$, and $k$ and distributed $\mathcal{N}(0, \gamma_q^2(\mb{s}))$. 
As a simplified example, if the smoothness of the Mat\'ern covariance is set to $1/2$, we obtain
 \begin{align}
 \begin{split}
\Cov(Z_{q_1k_1}(\mb{s}_{1}), Z_{q_2k_2}(\mb{s}_{2})) &= \mathbbm{1}(q_1 = q_2, k_1 = k_2) \sigma_{q_1}(\mb{s}_1)\sigma_{q_1}(\mb{s}_2) \textrm{exp}\left(- \frac{\left\lVert \mb{s}_1 - \mb{s}_2\right\rVert}{\alpha_{q_1}}\right)  \\
&~~~~+ \mathbbm{1}(q_1 = q_2, k_1 = k_2, \mb{s}_1 = \mb{s}_2) \gamma_{q_1}^2(\mb{s}_1). \end{split}\label{eq:cov_fun_simple}
\end{align}
Here $\sigma_{q_1}^2(\mb{s}_1)$ and $\gamma_{q_1}^2(\mb{s}_1)$ are spatially-varying variance terms and $\alpha_q$ is a range or scale parameter describing the decay of dependence. 
The spatially-varying variance term is based on \cite{stein2005nonstationary} generalizing the approach of \cite{paciorek2006spatial} that introduces nonstationary forms for the parameter $\alpha_q$. 
We assume that $\{W_{ijk}\}$, $\{U_{iqk}\}$, and $\{\tilde{Z}_{qk}(\mb{s}_i)\}$ are jointly normal and independent from each other. 
A discussion and comparison of these assumptions is included in Section \checklater{S1} of the supplement. We also evaluate the normality assumption in Section \checklater{S5}. 

\subsection{An approach for spatially-dependent functional time series}

We next introduce our approach for improving upon test statistics using spatial information, consisting of three steps detailed in the following three subsections. 
The motivation for spatial prediction is to reduce the independent error in the principal component scores $U_{iqk}$ or measurement error $W_{ijk}$. 
We will focus on data observed on a fine grid, so spatial interpolation is less of a motivation. 
Also, it is unclear if one should estimate the model under the null or alternative hypothesis. 
We will evaluate both in our simulation study. 

\subsubsection{Estimate spatio-functional model}\label{sec:estimate_model}

We outline our approach for estimating the entire model \eqref{eq:orig_model} under the null or an alternative hypothesis, resulting in the estimate $\hat{\mu}(\mb{s}, u)$, possibly $\hat{\delta}(\mb{s}, u)$, and an estimated covariance structure of $\epsilon_k(\mb{s}, u)$. 
To estimate a model under an alternative hypothesis, one should first specify a ``pilot'' estimate of $\hat{\tau}(\mb{s}_i)$. 
For data at one location, one can estimate each of the $N$ models taking $\hat{\tau}(\mb{s}_1) \in \{1, 2, \dots, N\}$. 
With spatial dependence, there are $N^n$ models, and one cannot estimate all of them. 
Sensible pilot estimates of $[\hat{\tau}(\mb{s}_i)]_{i=1}^n$ are those in Section \ref{sec:method_independent}.
Assuming the null hypothesis requires no pilot estimate as $\hat{\tau}(\mb{s}_i) = N$ for all $i$. 

First, the fixed effects $\mu(\mb{s}, u)$ and $\delta(\mb{s}, u)$ are estimated. 
We use penalized least squares estimators for these terms.
Exact formulation of estimators may depend on the domain of $\mb{s}$ and $u$. 
In Section \ref{sec:sphere_mean}, we specifically outline an approach where $\mb{s} \in \mathbb{S}^2$ and $u \in [0,1]$. 

We use a similar approach to estimate the marginal covariance structure in $u$. 
If we define $\mb{Y}_k(\mb{s}_i) =\left[Y_{k}(\mb{s}_i, u_{j})\right]_{j=1}^m$ and $\hat{\mb{\mu}}_k(\mb{s}_i) =  \left[\hat{\mu}(\mb{s}_i, u_{j}) + \mathbbm{1}(k > \hat{\tau}(\mb{s}_i)) \hat{\delta}(\mb{s}_i, u_{j})\right]_{j=1}^m$, an empirical covariance estimate in $u$ collapsed over all years and spatial locations is
\begin{align*}
\hat{\mb{C}}_u &= \frac{1}{n} \sum_{i=1}^n  \frac{1}{N}\sum_{k=1}^N \left(\mb{Y}_k(\mb{s}_i) - \hat{\mb{\mu}}_k(\mb{s}_i) \right)\left(\mb{Y}_k(\mb{s}_i) - \hat{\mb{\mu}}_k(\mb{s}_i) \right)^\top .
\end{align*}
In practice we include a smoothing penalty on the resulting covariance estimate, and we implement generalized cross-validation \citep[c.f.][]{wahba1990spline} to choose smoothing parameters.  

Let $\hat{\mb{\Phi}} \in \mathbb{R}^{m \times Q}$ be the matrix of the first $Q$ functional principal components based on $\hat{\mb{C}}_u$ with $Q < m$ evaluated at $[u_{j}]_{j=1}^m$. 
The principal component scores are estimated with $[Z_{qk}(\mb{s}_i)]_{q=1}^Q =  \hat{\mb{\Phi}}^\top  \left(\mb{Y}_k(\mb{s}_i) - \hat{\mb{\mu}}_k(\mb{s}_i) \right)$ omitting the ``hat'' on $Z_{qk}(\mb{s})$ for notational flexibility later. 
We then form estimators $ \hat{\gamma}_q^2(\mb{s})$ and $\hat{\sigma}_q^2(\mb{s})$ based on Remark \checklater{S2} of the supplement. 

We next turn to the estimation of the dependence structure of $\{Z_{qk}(\mb{s})\}$, for example, the parameters $\{\alpha_q\}_{q=1}^Q$ and smoothness parameters $\{\nu_q\}_{q=1}^Q$ assumed constant over $k$. 
Let $\mb{\theta}_q = (\alpha_q, \nu_q, \hat{\gamma}_q^2(\mb{s}), \hat{\sigma}_q^2(\mb{s}))$ denote the covariance's parameters, yet we treat $\hat{\gamma}_q^2(\mb{s})$ and $\hat{\sigma}_q^2(\mb{s})$ as fixed since their flexible structure is estimated previously. 
Let $\mb{Z}_{qk} = [Z_{qk}(\mb{s}_i)]_{i=1}^n$ and $\mb{C}_s(\mb{\theta}_q) = \textrm{Cov}(\mb{Z}_{qk}, \mb{Z}_{qk} | \mb{\theta}_q)$ be its covariance matrix using $\mb{\theta}_q$ and specified by a form like \eqref{eq:cov_fun_simple}. 
Then, the Gaussian likelihood for the $q$-th principal component is
\begin{align*}
\ell_q(\mb{\theta}_q) &= -\frac{1}{2}\sum_{k=1}^N \left( \log(\textrm{det}(\mb{C}_s(\mb{\theta}_q))) + \mb{Z}_{qk}^\top \mb{C}_s(\mb{\theta}_q)^{-1}\mb{Z}_{qk} + n \log(2\pi)\right).
\end{align*} 
In our simulation study, we directly evaluate the likelihood, yet for the data application some approximation is necessary as detailed in Section \ref{sec:sphere}. 
Maximum likelihood estimation through numerical optimization can then proceed separately for each $q=1, \dots, Q$.

\subsubsection{Use model for prediction}

Referring to the notation in Section \ref{sec:assumed_model}, our goal is to form an estimate of $\tilde{\mb{Z}}_{qk} = \left[\tilde{Z}_{qk}(\mb{s}_i)\right]_{i=1}^n$, a spatially-smoothed version of the scores $\mb{Z}_{qk}$. 
Based on the multivariate normal distribution, the conditional expectation using estimated parameters $\hat{\mb{\theta}}_q$ for each $q$ and $k$ is
\begin{align*}
\hat{\mb{Z}}_{qk} =\left[\hat{Z}_{qk}(\mb{s}_i)\right]_{i=1}^n= \mathbb{E}\left[\tilde{\mb{Z}}_{qk}\middle| \mb{Z}_{qk}, \hat{\mb{\theta}}_q\right] &=\left[\tilde{\mb{C}}_s\left(\hat{\mb{\theta}}_q\right)\right]\left[ \mb{C}_s\left(\hat{\mb{\theta}}_q\right)\right]^{-1} \mb{Z}_{qk}.
\end{align*}
Here, the $n\times n$ matrix $
\tilde{\mb{C}}_s\left(\hat{\mb{\theta}}_q\right) = \textrm{Cov}\left( \tilde{\mb{Z}}_{qk},  \mb{Z}_{qk} \middle |\hat{\mb{\theta}}_q\right)$ uses the estimated covariance function. 
Then, one can form predictions of the original data with reduced variance \begin{align*}
\hat{Y}_k(\mb{s}_i, u_{j}) &= \hat{\mu}(\mb{s}_i, u_{j}) + \mathbbm{1}(k > \hat{\tau}(\mb{s}_i)) \hat{\delta}(\mb{s}_i, u_{j}) + \sum_{q=1}^Q  \hat{Z}_{qk}(\mb{s}_i) \hat{\phi}_q(u_{j}).
\end{align*}

\subsubsection{Recompute test statistics using predictions}\label{sec:recompute_test}

Using the predicted scores $\left\{\hat{Z}_{qk}(\mb{s})\right\}$ and predicted data $\left\{\hat{Y}_k(\mb{s}_i, u_{j})\right\}$, we now recompute test statistics.
Our primary score-based predicted test statistic at location $\mb{s}_i$ is \begin{align}
 \hat{T}_{score}(\mb{s}_i)= \frac{1}{N^2} \sum_{q = 1}^Q \frac{1}{\hat{\lambda}^*_q(\mb{s}_i)}  \sum_{t=1}^N \left( \sum_{k=1}^t \hat{Z}_{qk}(\mb{s}_i) - \frac{t}{N}\sum_{k=1}^N\hat{Z}_{qk}(\mb{s}_i)\right)^2. \label{eq:score_pred_ts}
\end{align}where $\hat{\lambda}^*_q(\mb{s}_i) = \textrm{Var}\left(\hat{Z}_{qk}(\mb{s}_i)\right)$ is a local-regression estimate of the marginal variance of the predicted scores.
We consider two additional flavors of a predicted score-based test statistic. 
The first replaces $\hat{\lambda}^*_q(\mb{s}_i)$ with the estimate $\hat{\lambda}_q(\mb{s}_i) = \textrm{Var}(Z_{qk}(\mb{s}_i))$ defined in Remark \checklater{S2}; we say this has ``unadjusted variances'' since it uses the variance of the original scores rather than the predicted scores. 
In addition, one can recompute score-based test statistics at each location using the predicted data $\left\{\hat{Y}_k(\mb{s}_i, u_{j})\right\}$; we refer to this approach as ``recomputing individual test statistics.''
For score-based test statistics, we use predictions of scores and data assuming the null hypothesis is true; the predicted scores can still be appropriately used directly in \eqref{eq:score_pred_ts} to detect a change. 
Otherwise, the change may be partitioned in $\hat{\mu}(\mb{s}_i, u_{j}) + \mathbbm{1}(k > \hat{\tau}(\mb{s}_i)) \hat{\delta}(\mb{s}_i, u_{j})$. 
For fully functional test statistics, we use the predictions $\left\{\hat{Y}_k(\mb{s}_i, u_{j})\right\}$ to form and evaluate test statistics. 

\subsection{Multiple testing}

Inferring changes at multiple locations is fundamentally a multiple testing problem. 
Throughout our simulation and data analysis, we will evaluate both unadjusted for multiple testing and unadjusted for multiple testing p-values. 
We adjust p-values using the Bonferroni and Benjamini-Hochberg procedures to control family-wise error rate and false discovery rate, respectively \citep{benjamini1995controlling}. 
These procedures do not inherently consider the spatial nature of the data, yet using them on predicted data may improve results compared to using them without prediction. 
There is also substantial literature on spatial multiple testing procedures, in the setting of predetermined spatial clusters \citep{benjamini2007false}, a Bayesian framework \citep{risser2019spatially}, and an EM framework \citep{yun2022detection}.
Since multiple testing is not the main focus of this work, we evaluate the more simple Bonferroni and Benjamini-Hochberg approaches. 

\section{Spherically-indexed functional data framework} \label{sec:sphere}

Here, we introduce our framework to estimate the functional data model \eqref{eq:orig_model} when the spatial information $\mb{s}$ is indexed on the unit sphere $\mathbb{S}^2$, which is used to approximate the surface of the Earth in our analysis. 
In the spatial statistics community, there has been substantial interest in modeling random fields on $\mathbb{S}^2$. 
See \cite{jeong_2017} for a review, and more recent contributions include \cite{porcu2019axially} and \cite{alegria2020turning}. 
Modeling random fields on $\mathbb{S}^2$ can be more challenging in some aspects compared to modeling those on $\mathbb{R}^d$. 
For example, the choice of distance (i.e.\ chordal or geodesic distance) on $\mathbb{S}^2$ is thoroughly debated \citep[e.g.][]{Krivoruchko_2020}.
Axial symmetric models \citep{jeong_2017} are an example of models specific to $\mathbb{S}^2$. 

We are interested in modeling random fields of functional data $Y_k(\mb{s}, u)$ indexed on the sphere with $\mb{s} \in \mathbb{S}^2$ and $u \in [0,1]$. 
Though we are unaware of previous literature on this, many have extended random fields on a sphere to more complicated settings.
For example, modeling sphere-time processes has received attention, though not from the functional-data perspective; see \cite{porcu_2018b}. 
Other work treats the sphere $\mathbb{S}^2$ as the functional {\it domain} of the data \citep[see][and references within]{caponera_2022} or treats {\it trajectories} on $\mathbb{S}^2$ as functional data \cite[for example,][]{dai_2018, su2014statistical}. 

As we build our approach, we have a few guiding principles. 
For modeling the principal component scores, a nonstationary model is necessary, especially with respect to their variances. 
We also aim for an implementation that is possible and straightforward on the scale of global data with potentially tens of thousands of locations. 
We base the model for each principal component score on \cite{cao_2022} to reach these goals.
We first discuss the estimation of the mean and change functions. 

\subsection{Estimation of fixed effects}\label{sec:sphere_mean}
To estimate $\mu(\mb{s},u)$ and $\delta(\mb{s}, u)$, we propose a standard penalized optimization-based estimator. For illustration, we focus on $\mu(\mb{s},u)$, and estimate
\begin{align*}
\hat{\mu}(\mb{s}, u) =\argmin_{\beta} \sum_{i=1}^n \sum_{j=1}^m \left(\overline{Y}^\mu(\mb{s}_i, u_{j}) - \beta(\mb{s}_i, u_{j})\right)^2 + \textrm{Pen}(\zeta).
\end{align*}where $\overline{Y}^\mu(\mb{s}_i, u) = \textrm{mean}\left\{Y_k(\mb{s}_i, u) \textrm{ such that } k \leq \hat{\tau}(\mb{s})\right\} $ is the empirical average estimate based on an estimate $\hat{\tau}(\mb{s})$, and $\textrm{Pen}(\zeta)$ is a smoothness penalty on the solution $\beta$ with smoothing parameter $\zeta >0$. 
A similar estimator is used to estimate $\delta(\mb{s}, u)$.
We represent $\beta(\mb{s}, u)$ using a product basis of spherical harmonic functions \citep[of $\mb{s}$, \texttt{sphunif},][]{sphunif_r, garcia_2018} and cubic B-splines (of $u$).
For the penalty $\textrm{Pen}(\zeta)$, we apply smoothing with respect to $u$ only.
Since spherical harmonic functions are not local, using a large basis can be computationally infeasible, so we use instead a regression approach similar to \cite{jeong_2017}.
We provide details in Section \checklater{S2}.
		        						
\subsection{Estimation of spatio-functional dependence structure}\label{sec:sphere_cov}

We focus on modeling the spatial dependence in principal component scores $Z_{qk}(\mb{s}_i)$; mostly following the notation of \cite{cao_2022}, we take \begin{align*}
\textrm{Cov}(Z_{qk}(\mb{s}_1), Z_{qk}(\mb{s}_2) | \mb{\theta}_q) &= \sigma_q(\mb{s}_1) \sigma_q(\mb{s}_2) c_q(\mb{s}_1, \mb{s}_2) \rho_q(d_q(\mb{s}_1, \mb{s}_2)) + \mathbbm{1}(\mb{s}_1 = \mb{s}_2) \gamma_q^2(\mb{s}_1),\end{align*}
where $c_q(\mb{s}_1, \mb{s}_2) $ is a normalizing constant, $\rho_q(\cdot)$ is an isotropic covariance function, and $d_q(\mb{s}_1, \mb{s}_2)$ is a distance function. 
Here, we take \begin{align*}
d_q(\mb{s}_1, \mb{s}_2)^2 &= 2(\mb{s}_1 - \mb{s}_2)^\top (\mb{\Sigma}_q(\mb{s}_1) + \mb{\Sigma}_q(\mb{s}_2))^{-1} (\mb{s}_1 - \mb{s}_2) \\
c_q(\mb{s}_1, \mb{s}_2)^2 &= | \mb{\Sigma}_q(\mb{s}_1)|^{1/2} |\mb{\Sigma}_q(\mb{s}_2)|^{1/2} |(\mb{\Sigma}_q(\mb{s}_1) + \mb{\Sigma}_q(\mb{s}_2))/2|^{-1}.
\end{align*}
The matrix $\mb{\Sigma}_q(\mb{s})$ specifies the spatially-varying anisotropy and depends on the angle parameter $\kappa_q(\mb{s})$ and length scale parameters $\chi_{q1}(\mb{s})$ and $\chi_{q2}(\mb{s})$. 
For the exact construction of $\mb{\Sigma}_q(\mb{s})$ from these parameters, we refer the reader to \cite{cao_2022}. 
Following their implementation, we take $\kappa_q(\mb{s})  = \kappa_q$. 
For the length scale parameters, we take 
$\chi_{qr}(\mb{s}) =  \textrm{exp}\left(\beta_{qr0} + \beta_{qr1}\sin(l) +\beta_{qr2}L+ \beta_{qr3} L^2\right)$,
for $r=1, 2$ and $l$ is the longitude and $L$ is the latitude of $\mb{s}$.
This differs slightly from \cite{cao_2022} as we introduce a quadratic term of latitude, allowing for behavior in parameters symmetric about the equator. 
For each $\rho_q(\cdot)$, a Mat\'ern covariance function is used, each with a smoothness parameter $\nu_q$ also estimated. 
Therefore, $\mb{\theta}_q = \{\nu_q$, $\kappa_q$, $\{\beta_{qr\ell}\}_{r=1, \ell=0}^{2, 3}$,  $\hat{\sigma}_q^2(\cdot)$, $\hat{\gamma}_q^2(\cdot) \}$.

This covariance is attractive because it allows for a spatially-varying variance and decay parameters as well as control over the smoothness of the process. 
This covariance is also implemented with a Vecchia approximation in \cite{cao_2022}, which makes implementation straightforward for analyses on the scope considered in this work. 
Instead of working with and inverting the entire dense covariance matrix $\mb{C}_s(\mb{\theta}_q)\in \mathbb{R}^{n\times n}$ for each likelihood evaluation or prediction, this instead builds a sparse representation of the precision matrix $(\mb{C}_s(\mb{\theta}_q))^{-1}$ based on a neighborhood structure \citep{gpvecchia}. 
We also avoid recomputations over $k$ when working with $ \{Z_{qk}(\mb{s})\}$ for all $k=1, \dots, N$.
The parameters $\mb{\theta}_q$ (except $ \hat{\sigma}_q^2(\cdot)$ and $\hat{\gamma}_q^2(\cdot)$) are estimated by maximum likelihood as detailed in Section \ref{sec:estimate_model}. 

\section{Simulation} \label{sec:simulation}


In this section, we consider a simulation study which finds that spatially-predicted versions of test statistics improve both the detection of changepoints and the estimation of $\tau(\mb{s})$. 
For illustrative use, we consider a spatial domain on a grid in $\mathbb{R}^2$. 

\subsection{Simulation design}

\begin{figure}[h]
    \centering
    \includegraphics[width = .4875\textwidth]{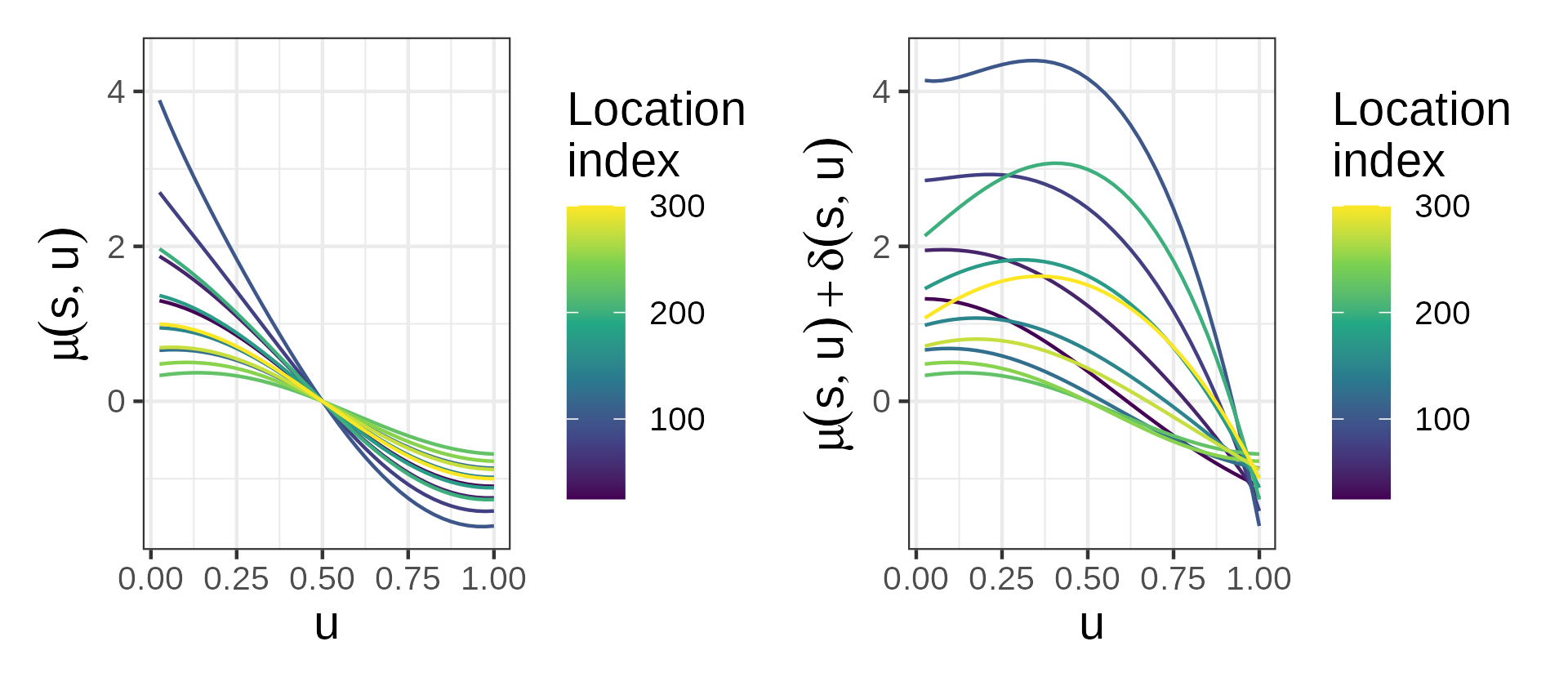}
    \includegraphics[width = .2475\textwidth]{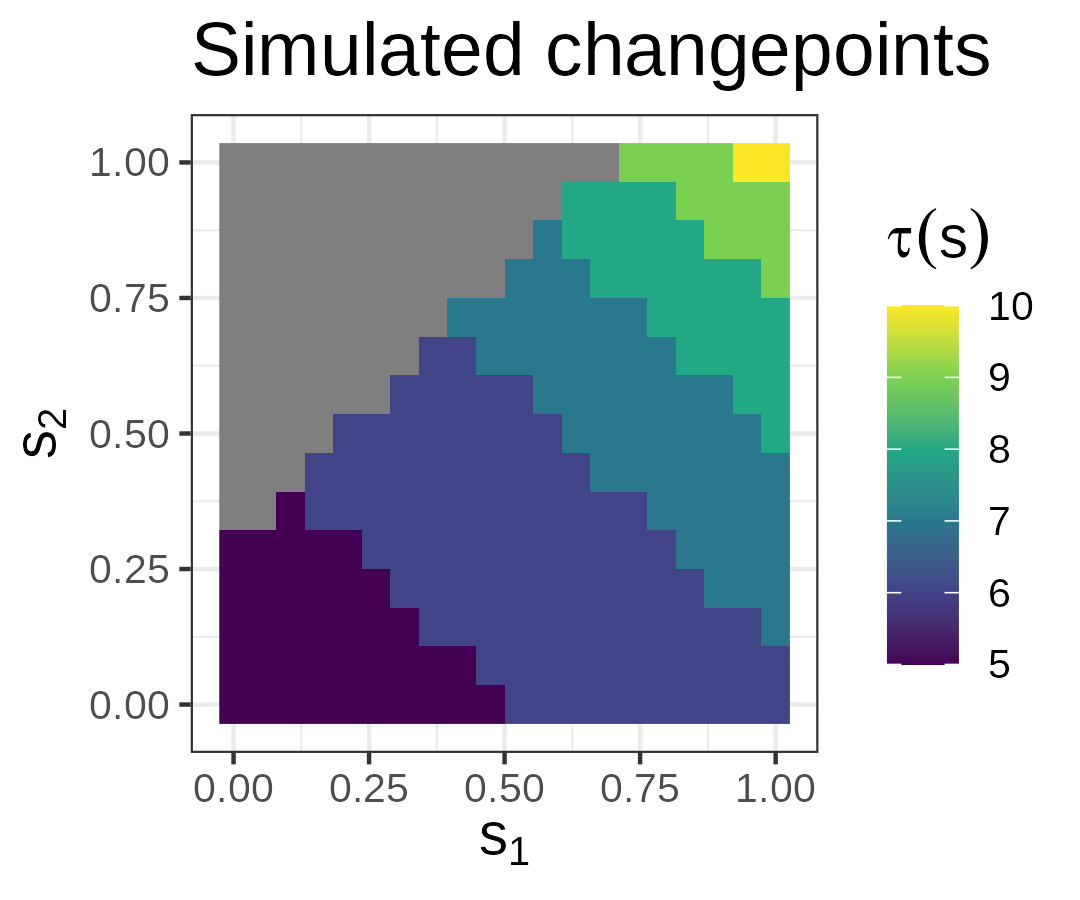}
    \includegraphics[width = .25\textwidth]{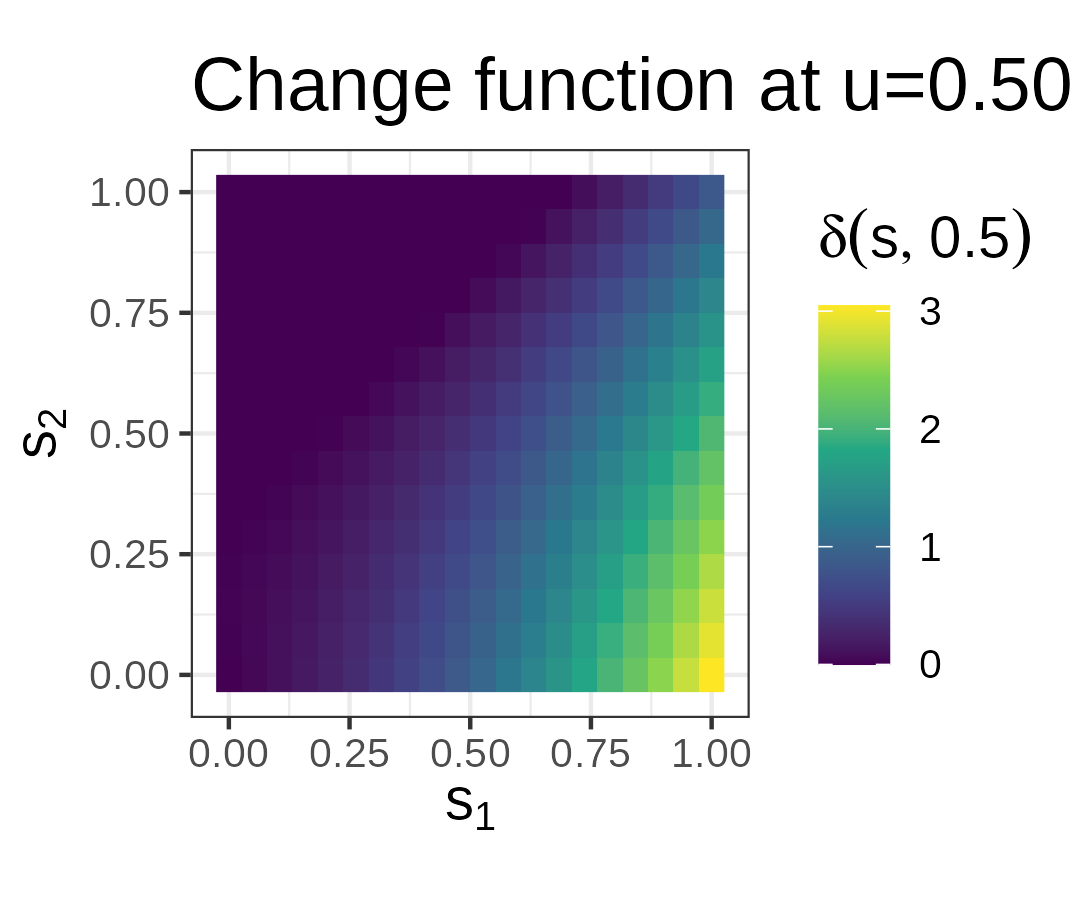}

    \caption{(Far Left) Original means and (Left) means after the changepoint for selected locations, with $\eta = 10$. (Right) Changepoint surface $\tau(\mb{s})$, with gray tiles having no change. (Far Right) Change function for a fixed $u = 0.5$.}
    \label{fig:sim_mean}
\end{figure}

We again take the model defined in \eqref{eq:orig_model}.
We consider $N=15$, $n = 300$ locations on a $15\times 20$ grid on $[0,1] \times [0,1]$, and $m =40$ measurements in $u$ per location evenly-spaced on $[0,1]$. 
To specify the fixed components, we take, with $\mb{s} = (s_1, s_2)^\top$, \begin{align*}
    \mu(\mb{s}, u) &= \cos(\pi u) \textrm{exp}\left(\frac{2(s_1 - s_2)}{2u + 1}\right), \\
    \delta(\mb{s}, u) &= \underbrace{\eta}_{\textrm{signal strength} }\underbrace{\left(u s_1 + u^2  s_2 - u^3(s_1 +  s_2)\right)}_{\textrm{varying change function} }\underbrace{\mathbbm{1}(s_1 - s_2> -0.3) (s_1 - s_2 + 0.3)}_{\textrm{taper near boundary}}, \\
            \tau(\mb{s})&=\frac{1}{4} \left\lceil 15 + 3\textrm{exp}(s_1 + s_2)\right\rceil   \textrm{ if } s_1 - s_2 > -.3, \textrm{ and } N \textrm{ (no changepoint) otherwise} ,
\end{align*}where $\lceil x\rceil$ is the ceiling function of $x$. 
We plot these functions in Figure \ref{fig:sim_mean}. 
The design of $\tau(\mb{s})$ and $\delta(\mb{s}, u)$ avoids sudden changes in space between locations with or without a change, which is realistic in our application, since circulation prevents hard spatial boundaries in stratospheric temperature. 
Thus, the locations with a change near the boundary with ``no change'' are hardest to detect.
We vary the signal strength with $\eta \in\{0.5, 2, 6, 10, 14, 18, 22 \}$.

Turning to the specification of $\epsilon_k(\mb{s}, u)$,  we take $Q=3$ principal components based on \begin{align*}
    \epsilon_k(\mb{s}_i, u_{j}) &= Z_{1k}(\mb{s}_i) + Z_{2k}(\mb{s}_i) 2^{-\frac{1}{2}}\sin(2\pi u_{j}) + Z_{3k}(\mb{s}_i)2^{-\frac{1}{2}} \cos(2\pi u_{j}) + W_{ijk}.
\end{align*}
We decompose $Z_{qk}(\mb{s}_i) = \tilde{Z}_{qk}(\mb{s}_i) + U_{iqk}$, and the variances of $\tilde{Z}_{qk}(\mb{s}_i)$ are spatially-varying: $\sigma_{1}(\mb{s}) = |s_1| + |s_2| +0.01$, $\sigma_{2}(\mb{s}) = (25)^{-1} (3|s_1| + 27|s_2|) +0.01$, and $ \sigma_{3}(\mb{s}) =(25)^{-1}(9|s_1| + |s_2|)  + 0.01$, 
while $\textrm{Var}(U_{iqk}) = \gamma_q^2(\mb{s}_i) = 0.3$ for all $q$ and $\mb{s}_i$. We take $\textrm{Var}(W_{ijk}) = 0.16$. 

\begin{figure}[h]
    \centering
    \includegraphics[width = .48\textwidth]{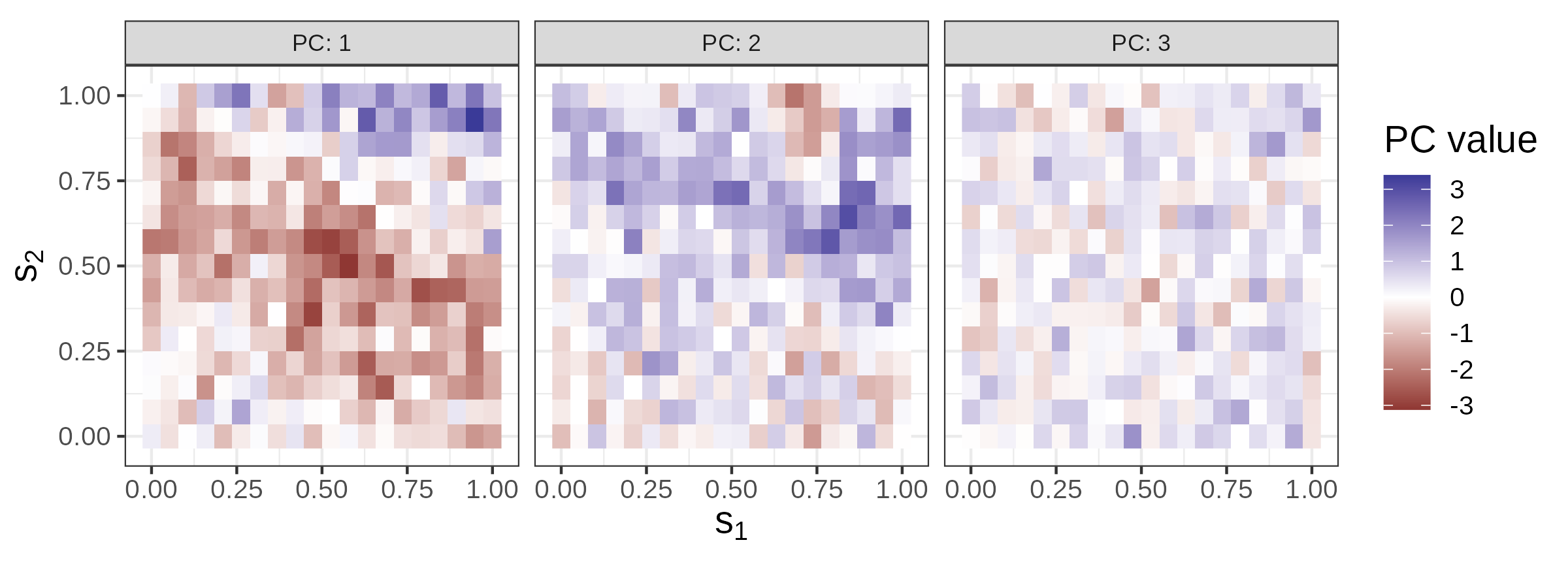}
        \includegraphics[width = .48\textwidth]{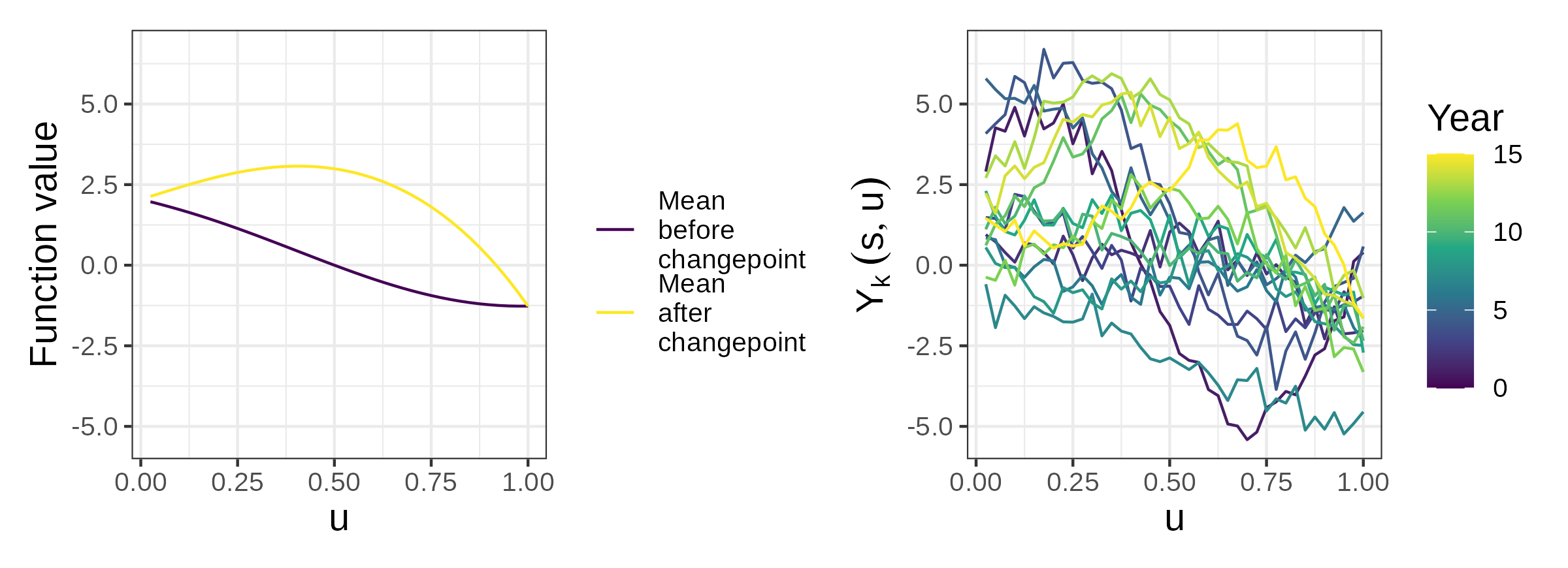}
    \caption{(Left) Simulated principal component scores $Z_{qk}(\mb{s})$ for $q= 1$, $2$, and $3$ and $k = 1$. (Middle Right) Plots of the mean function before and after the changepoint at one location $\mb{s} = (1, 0.643)$ and (Far Right) simulated data for that location with $\eta = 10$ and $\tau(\mb{s}) = 8$.}
    \label{fig:sim_pc_values}
\end{figure}

We consider two settings of the absence or presence of spatial dependence in $\epsilon_k(\mb{s},u)$:
\begin{itemize}
\item Independent model: We take $
    \textrm{Cov}(Z_{q_1k_1}(\mb{s}_1), Z_{q_2k_2}(\mb{s}_2)) =  \left(\sigma_{q_1}^2(\mb{s}_1) + \gamma_{q_1}^2(\mb{s}_1)\right) $ if $q_1 =  q_2$, $k_1 =  k_2$, and $\mb{s}_1 = \mb{s}_2$, and $0$ otherwise. 
    Here, $\sigma_q^2(\mb{s})$ and $\gamma_q^2(\mb{s})$ are not individually identifiable. 
    Using Remark \checklater{S2}, $\hat{\sigma}_{q}^2(\mb{s})$ estimates $0$ while $\hat{\gamma}_{q}^2(\mb{s})$ estimates $\sigma_{q}^2(\mb{s}) + \gamma_{q}^2(\mb{s})$.  
    
\item Dependent model: We take the model defined in \eqref{eq:cov_fun_simple} with $(\alpha_1, \alpha_2, \alpha_3) = (0.4,0.3,0.6)$. 
\end{itemize}
An example of simulated $Z_{qk}(\mb{s}_i)$ under the dependent model are shown in the left panel of Figure \ref{fig:sim_pc_values}. 
We plot an example of simulated data at one location in Figure \ref{fig:sim_pc_values}. 
We complete 100 replicates for each signal strength and for the independent and dependent model.

\subsection{Approaches compared}

We first consider a variety of score-based test statistics. 
Our baseline approach for comparison is the score-based test statistic computed at each location while ignoring data at all other locations as in \eqref{eq:score_individ}. 
For our predicted test statistics, we use predictions assuming the null hypothesis is true. 
We also vary the number of estimated principal components, taking $Q = 3$ (the ``true'' number), $Q = 4$, and $Q=5$; we also present results for $Q=4$ for the ``unadjusted variances'' and ``recomputed'' variations of the predicted test statistics described in Section \ref{sec:recompute_test}. 
When estimating the model under the null hypothesis, the change function may be effectively absorbed into the covariance structure, which motivates using $Q > 3$ to ensure we capture both the change and residual variability. 

For fully-functional test statistics, our baseline is again the test statistic computed at each location. 
We also consider predictions under the global null and four alternative models using results from the individual fully-functional test statistic at each location: (``Alt'') a p-value $<0.05$, (``BH'') a Benjamini-Hochberg-corrected p-value $<0.05$, (``Bonf'') a Bonferroni-corrected p-value $<0.05$, (``All'') for all locations regardless of p-value.
These models lie on a spectrum between ``the null is true everywhere'' and ``the alternative is true everywhere.''
We also compute the test statistics of \cite{gromenko:2017} for each simulation. 
We also tried applying \cite{li_changepoint_2022}, but we were not able to obtain convergent MCMC results that improved the estimate of $\tau(\mb{s})$ in this setting. 

\subsection{Changepoint testing performance}

In our simulation, we evaluate detection performance, with full results presented in Section \checklater{S3}. 
We find that spatially predicting score-based test statistics dramatically increases the detection power with slight degradations in Type I error. 
Also, applying multiple testing criteria leads to control of false discovery rate.
After correction, only spatially-predicted test statistics have substantial power. 
Finally, we compare results with \cite{gromenko:2017}, which, by testing for a single changepoint, is an oversimplification in this setting.

\subsection{Changepoint estimation performance}

\begin{figure}
\includegraphics[width = .48\textwidth]{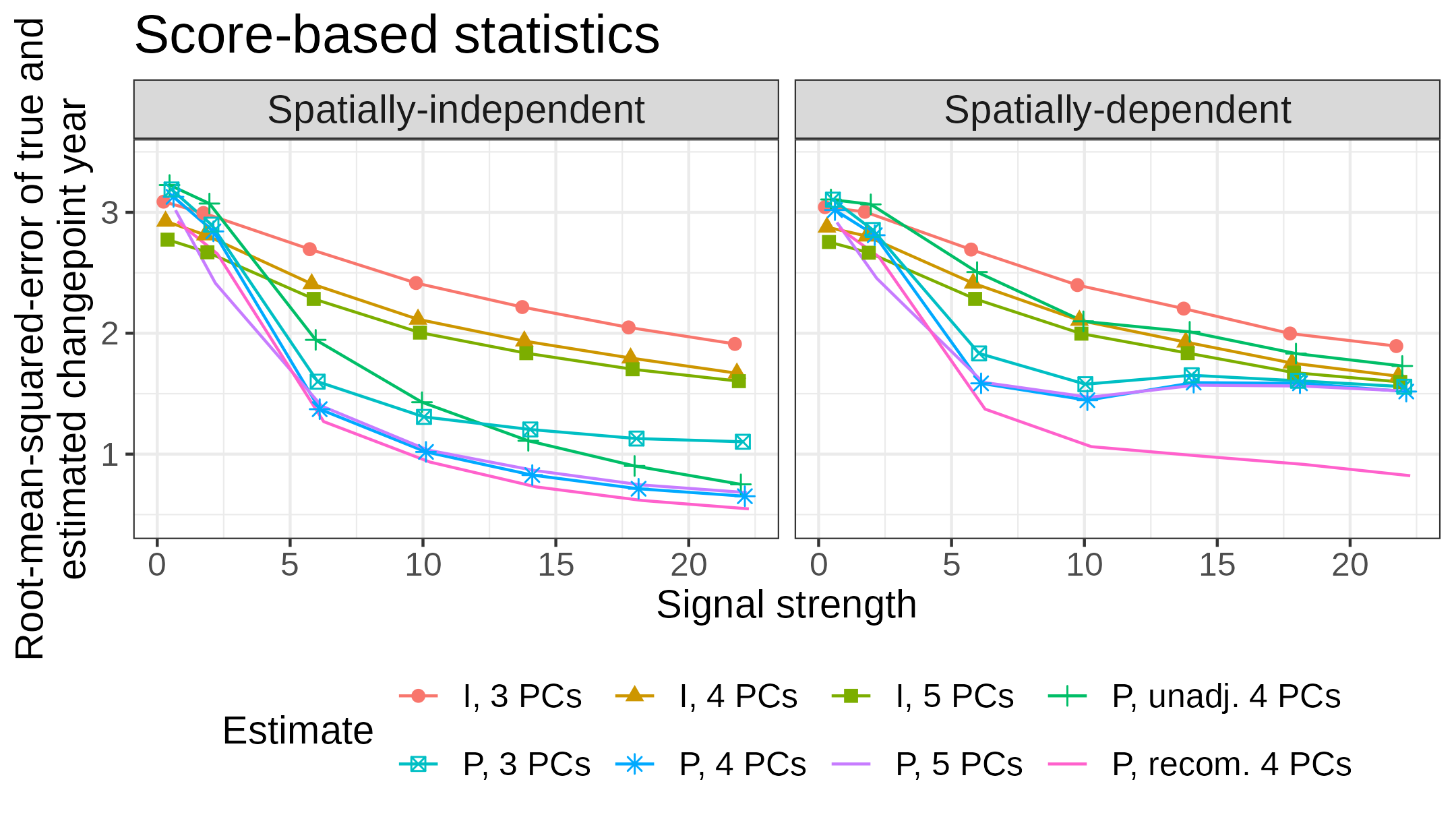}
\includegraphics[width = .48\textwidth]{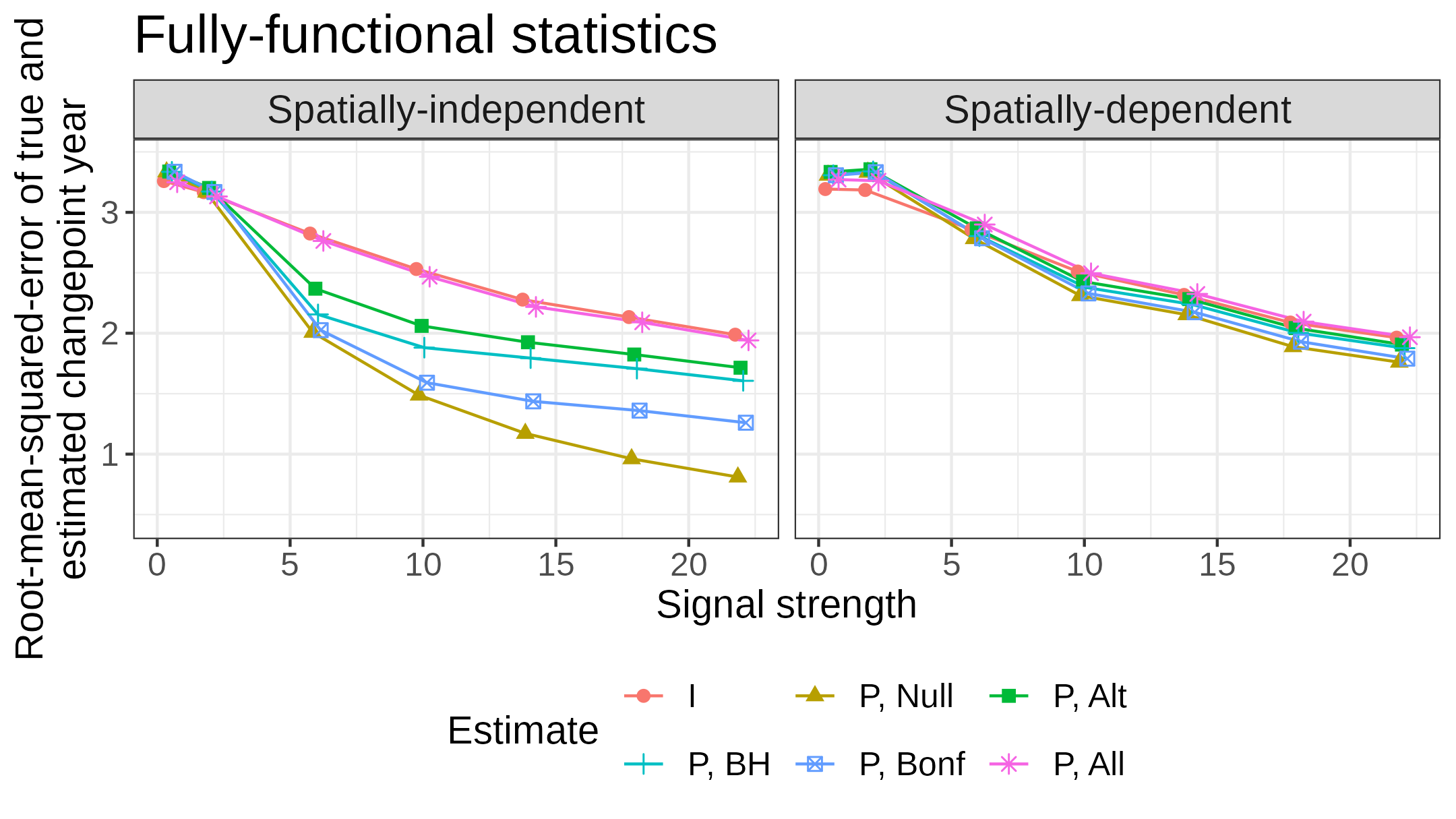}
\caption{Estimation year performance across simulation setups (dependence structure and signal strength). (Left) Score-based test statistics. (Right) Fully-functional test statistics. The scale of the plots are the same to ease comparison. }\label{fig:sim_year_est_all}
\end{figure}

In addition to improving the test performance of the estimators, we can also dramatically improve the estimation of the time of change. 
We first examine performance across dependence structures and simulation strength in Figure \ref{fig:sim_year_est_all} in terms of root-mean-squared-error between the true and estimated changepoint. 
Here, we compare results for all locations with a true changepoint, regardless of whether a changepoint was detected, so that we can evaluate this performance straightforwardly and separately from testing.

Generally, the spatially-predicted test statistics improve upon test statistics computed individually at each location.
While there is no improvement for lower signal strengths, at mid-to-high signal strengths, the predicted test statistics can cut RMSE in half in the spatially-independent setting.  
In the spatially-dependent setting, there are more modest but substantial improvements. 
As signal strength increases from high to higher, the gap between spatial and individual test statistics decreases, as the change time is more easily estimated using data at each location. 

Comparing results across the independent and dependent settings, the test statistics computed at individual locations perform the same with or without spatial dependence. 
This result is expected since the marginal distribution at each location is the same, with spatial dependence only affecting how different locations relate to each other.
For the predicted test statistics, adding spatial dependence makes the problem fundamentally harder by adding spatially-correlated noise. 

We also compare the performance of the score-based and fully-functional test statistics. 
When computed individually at each location, the score-based test statistics with 4 or 5 principal components perform better than the fully-functional statistics. 
This result mostly holds for spatially-predicted test statistics also. 
We see that predictions under alternative hypotheses do not improve the test statistics much compared to other predicted test statistics. 
Only using three principal components is substantially worse than using four or five, since the change may be filtered into the covariance when estimating the model under the null. 
An example of the changepoint estimation is given in Section \checklater{S3}.


\section{Data analysis} \label{sec:data_analysis}
\subsection{Application to HSW++ data}

For an initial test case for the methodology in the context of global climate data, we apply and evaluate the methodology on a simplified atmospheric model branded as HSW++. 
This model builds upon the framework of \cite{hsw_1994} and \cite{williamson_1998} in the thesis work of \cite{hollowed_2022}.
The primary goal is to evaluate the effects of a stratospheric aerosol injection of $\textrm{SO}_\textrm{2}$, sulfate, and ash, designed to simulate a volcanic eruption. 
While HSW++ is considerably simplified compared to Earth Systems Models (ESMs), it provides a middle ground between a statistical simulation study and an application to ESM output or climate reanalysis data. 
HSW++ assumes that land with no elevation covers the entire earth, there is no precipitation or water vapor, and there are no diurnal or seasonal cycles. 
See \cite{hollowed_2022} for a detailed description of the model. 

\begin{figure}
\includegraphics[width = .48\textwidth]{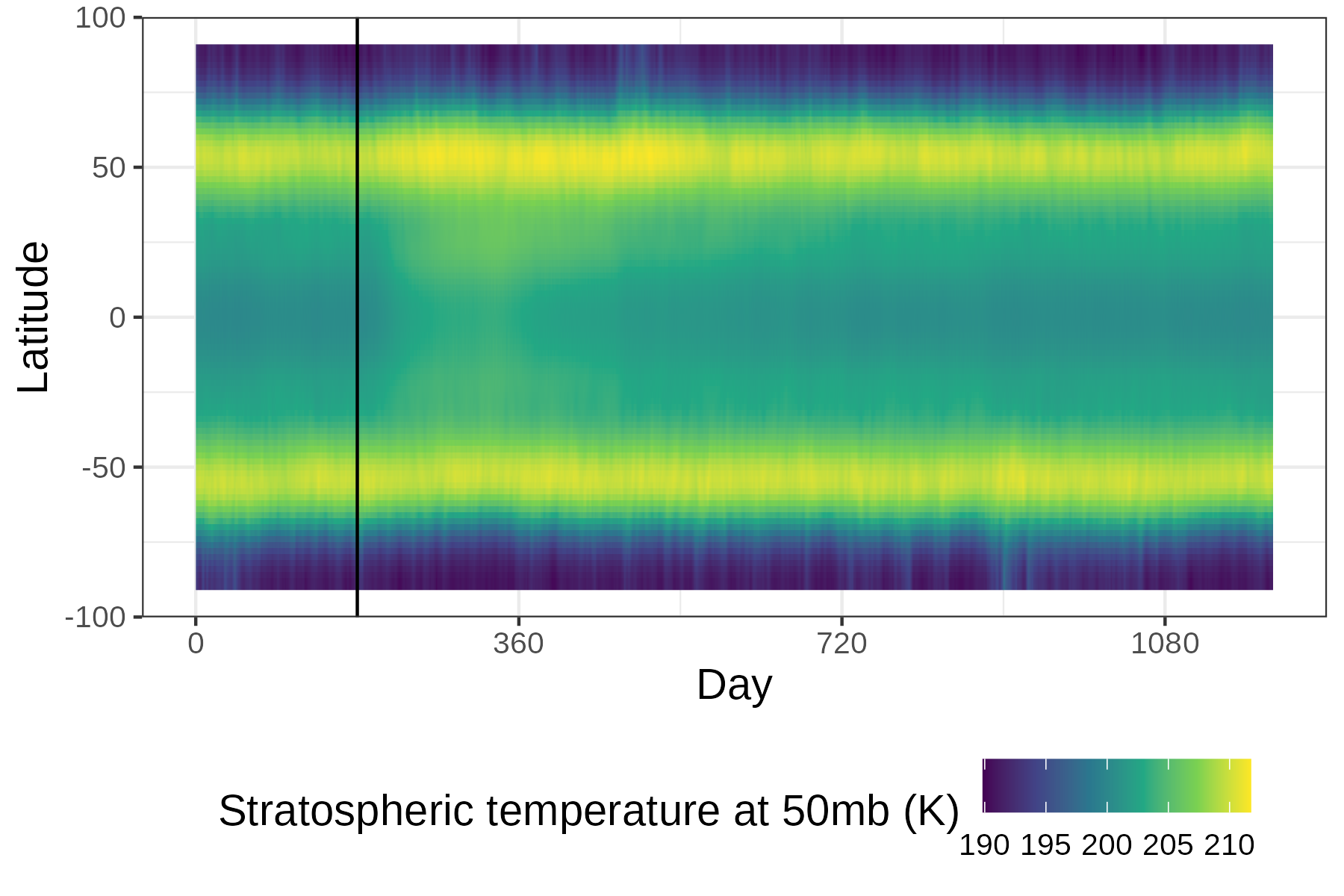}
\includegraphics[width = .48\textwidth]{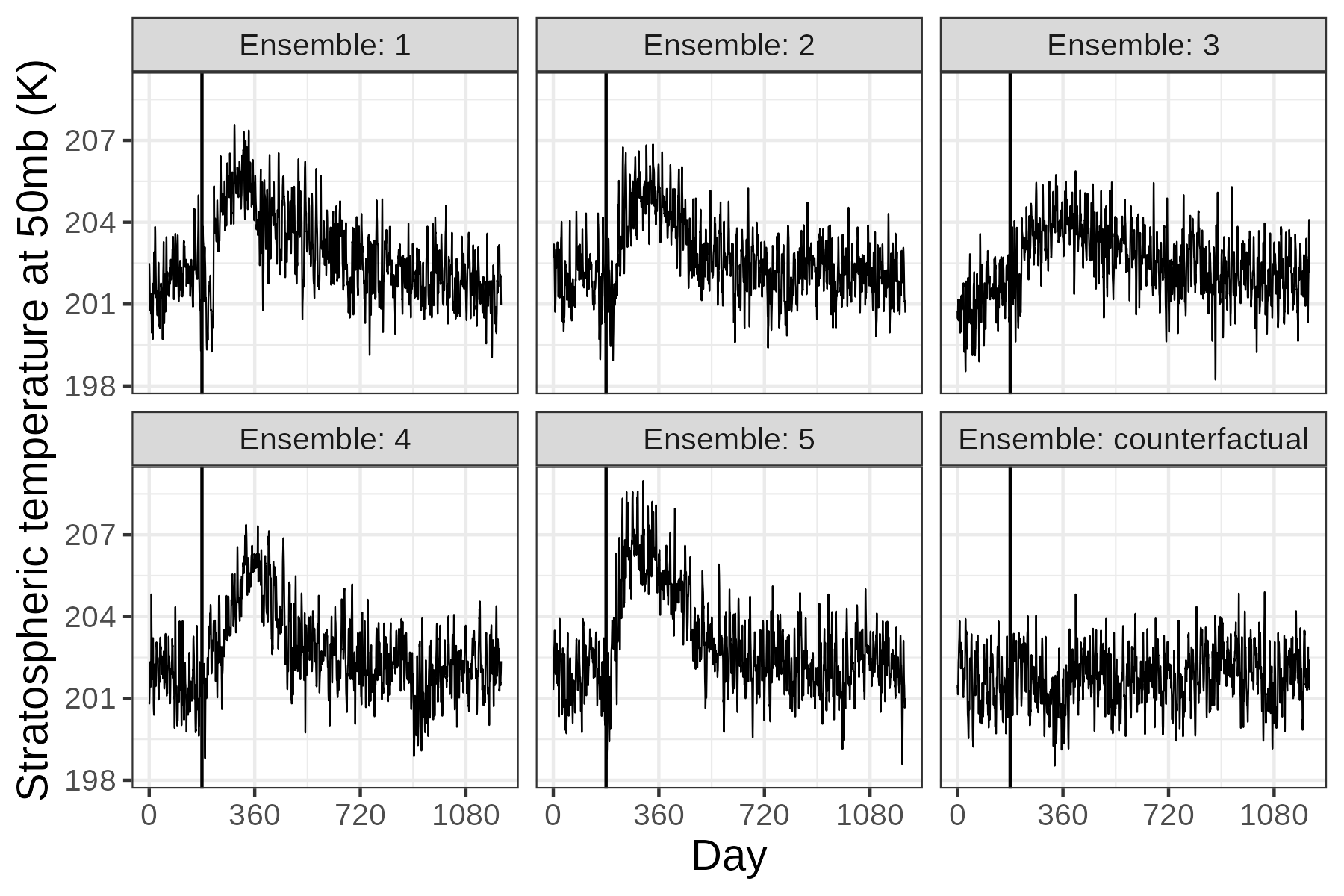}

\caption{(Left) Longitudinal and ensemble average of 5 HSW++ ensemble members with stratospheric injection. (Right) Example of HSW++ data for each ensemble member for $38^\circ \textrm{E}$ and $18^\circ \textrm{N}$. Unless otherwise noted, the vertical line represents the time of injection.  }\label{fig:HSW_data}

\end{figure}

We apply the developed methodology separately to a 5-member ensemble run of HSW++ where each member had a stratospheric aerosol injection event and different initial conditions, as well as a ``counterfactual'' member with no stratospheric aerosol injection event. 
Each ensemble member consists of data defined on a $2^\circ \times 2^\circ$ latitude-longitude grid (16,380 locations total) over the course of 1{,}200 days.
The stratospheric injection occurs on the 180th day. 
We use 6-hourly averaged data resulting in 4{,}800 total data points per location and ensemble.
Figure \ref{fig:HSW_data} shows a summary and example of the stratospheric (near-50mb) temperature data. 
From the lack of cyclical effects in the data, we can flexibly define $m$ and $N$ with $m=120$ and $N=40$, so that the ``years'' $k$ can now be naturally interpreted as a ``month'' of $30$ days. 
Throughout, we use $Q=10$ principal components, spherical harmonics up to degree $10$ resulting in $121$ spatial basis functions, a bandwidth of $h= 400$ kilometers, and $8$ neighbors for the Vecchia approximation. 
We use spatially-predicted test statistics under the null hypothesis and a Benjamini-Hochberg correction to the p-values.

\begin{figure}
\includegraphics[width = .32\textwidth]{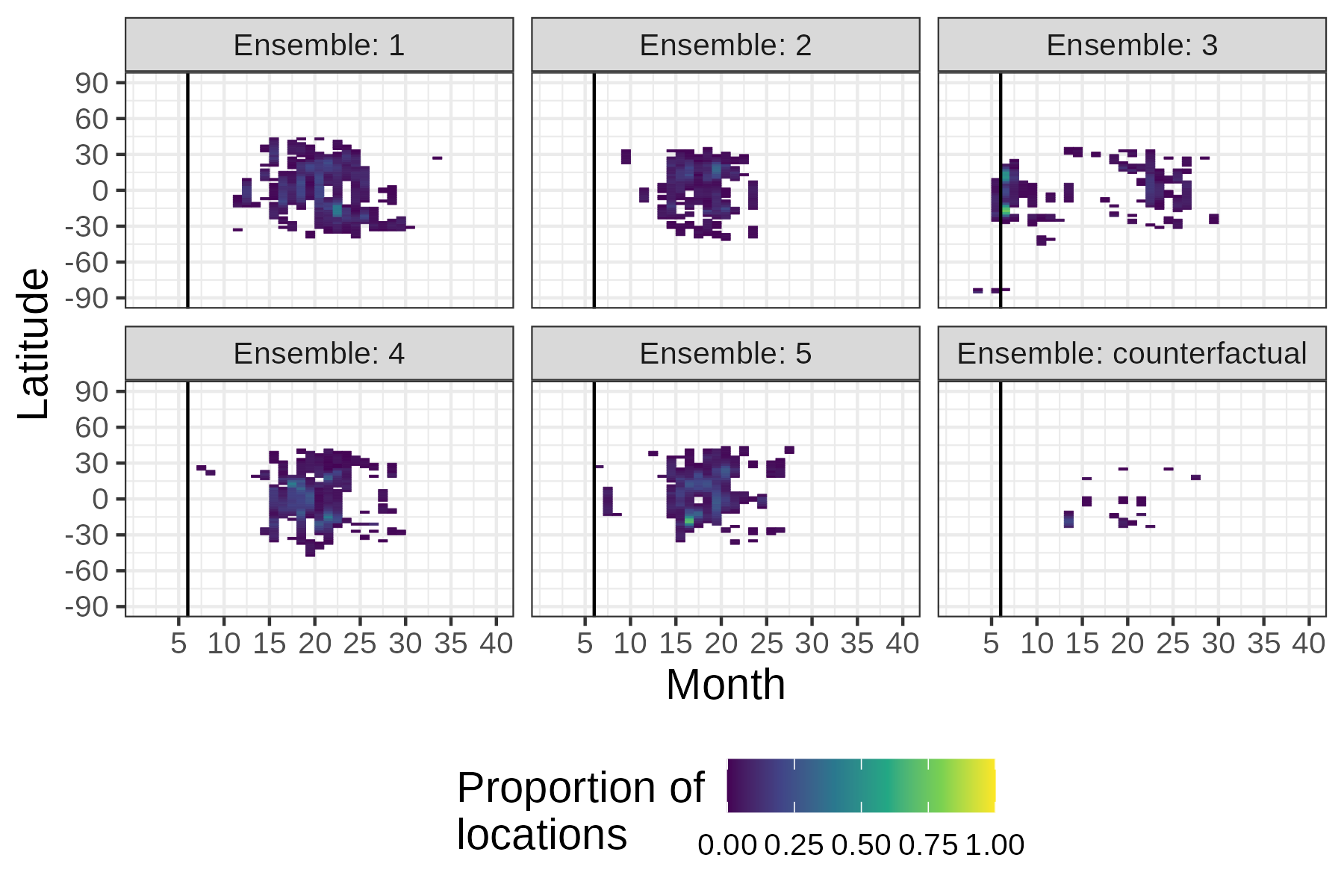}
\includegraphics[width = .32\textwidth]{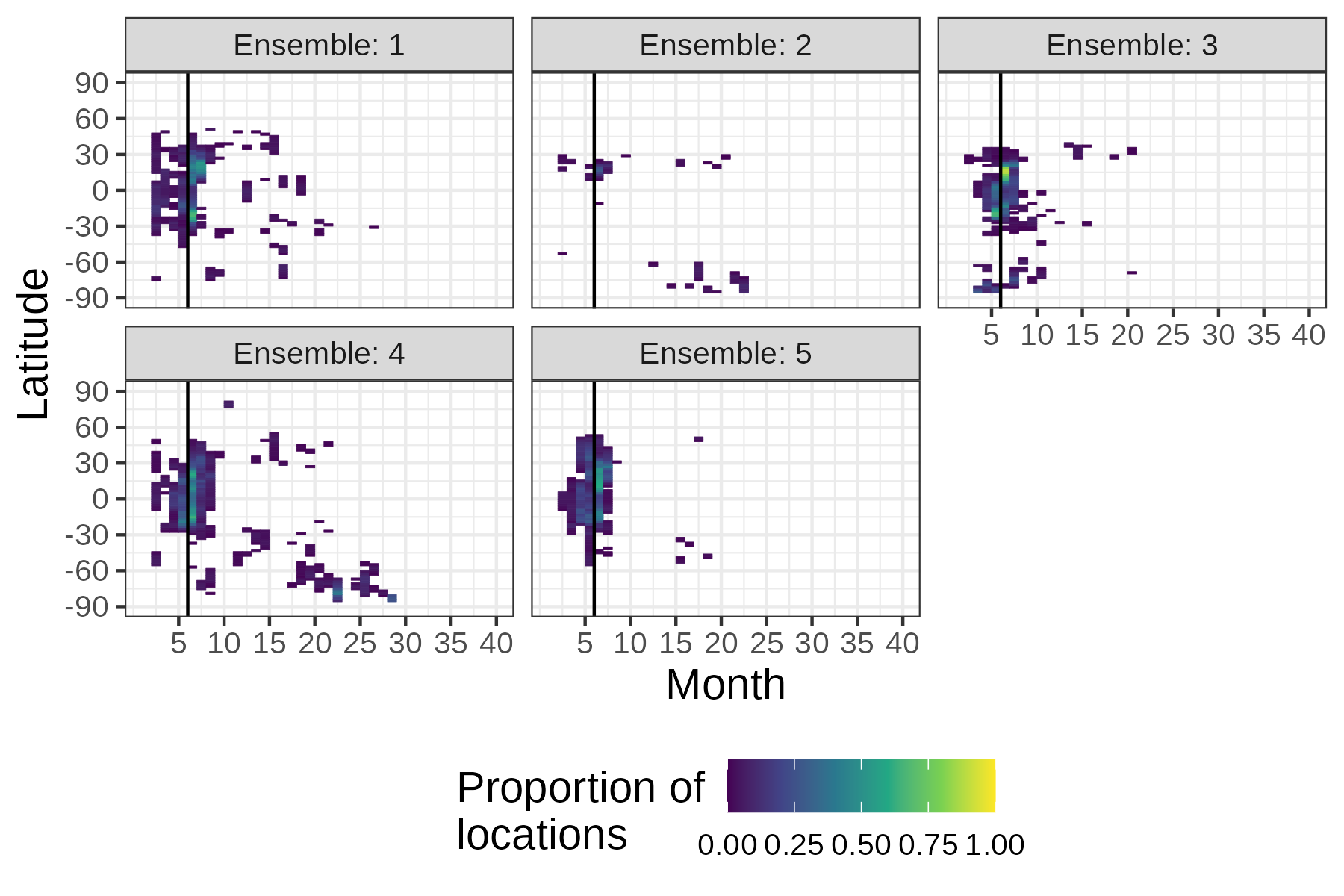}
\includegraphics[width = .32\textwidth]{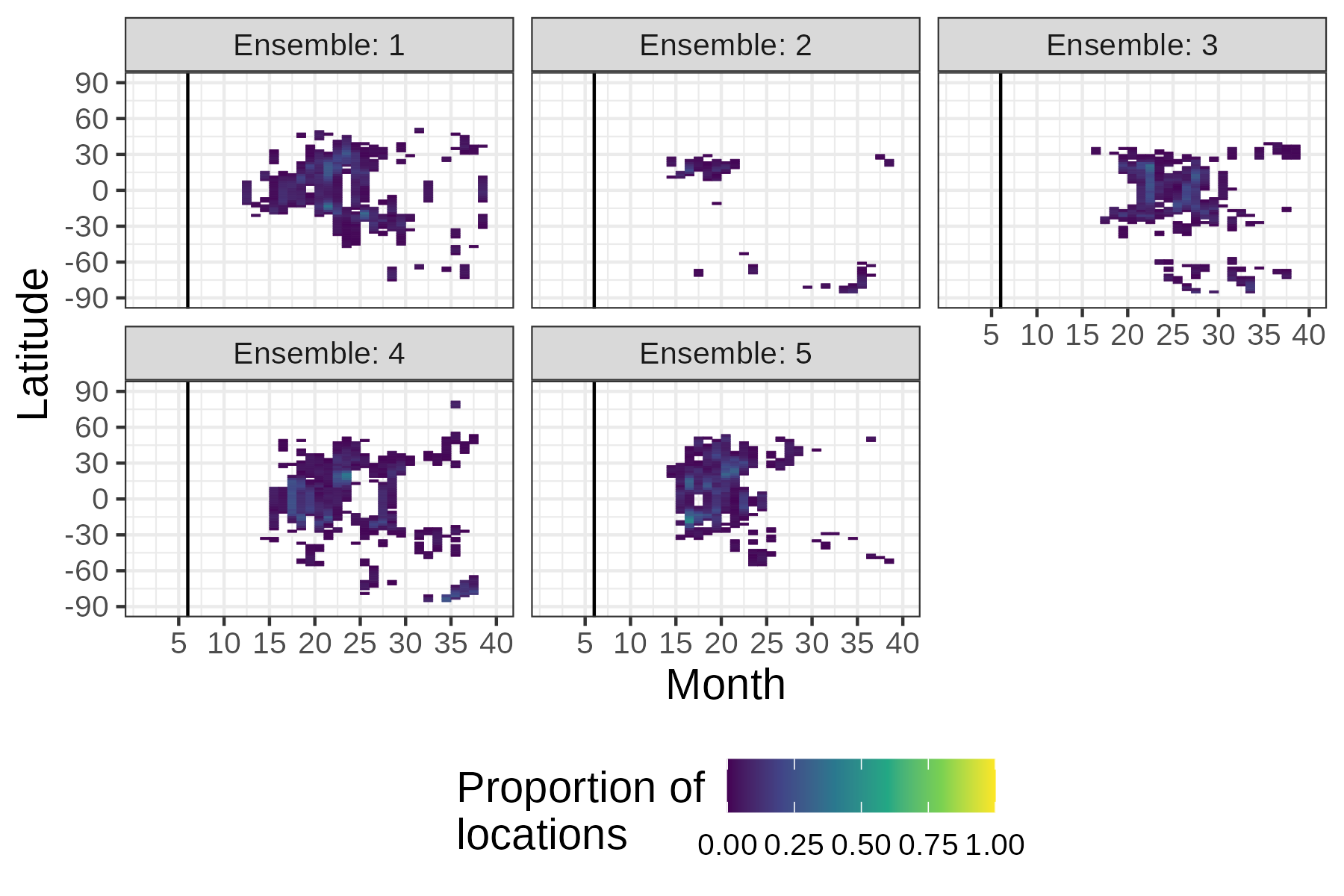}
%

\caption{Comparison of AMOC (Left) and epidemic (Middle and Right) changepoint model for HSW++. (Middle) For estimates of $\tau_1(\mb{s})$; (Right) for estimates of $\tau_2(\mb{s})$.  For each latitude, we plot the proportion of locations (of 180 different longitudes) that had a change detected during that month using Benjamini-Hochberg-corrected p-values.} \label{fig:HSW_change_results}

\end{figure}

At first, we focused on the ``at-most-one-change'' (AMOC) model described by \eqref{eq:orig_model}. 
The results for spatially-predicted test statistics are presented on the left of Figure \ref{fig:HSW_change_results}. 
While the third ensemble member had many changepoints detected around the time of injection (month 6), for other ensemble members many changes are detected near the middle of the time period (month 18). 
Comparing with Figure \ref{fig:HSW_data}, it appears that the AMOC changepoint model picks up a decrease in temperature associated with a ``return-to-normal'' rather than the immediate impact of the injection.
For this reason, we began to consider an ``epidemic''-type changepoint model studied in \cite{aston_2012}.
This model assumes replaces the indicator $\mathbb{I}(k> \tau(\mb{s}_i))$ in \eqref{eq:orig_model} with $\mathbb{I}(\tau_1(\mb{s}_i) < k \leq \tau_2(\mb{s}_i))$. 
Ideally, the change $\tau_1(\mb{s})$ represents the injection's initial impact, while $\tau_2(\mb{s})$ represents the ``return-to-normal'' previously detected. 
We use score-based statistics from \cite{aston_2012}: \begin{align*}
 T_{score, epi }^{\mb{s}_i}&= \frac{1}{N^3} \sum_{q = 1}^Q \frac{1}{\lambda_q^{\mb{s}_i}}  \sum_{t_1 = 1}^N \sum_{t_2 > t_1}^N\left( \sum_{k=t_1}^{t_2} Z_{qk}^{\mb{s}_i} - \frac{(t_2 - t_1 + 1)}{N}\sum_{k=1}^NZ_{qk}^{\mb{s}_i}\right)^2,\end{align*}
with convergence under the null hypothesis to $\displaystyle
T_{score, epi}^{\mb{s}_i} \overset{d}{\to} \sum_{q = 1}^Q \int_0^1\int_0^y (B_{q}(x) -  B_{q}(y))^2 dx \  dy.$
See \cite{aston_2012} for more information.
Our approach for including spatial information is directly applied by using predicted versions of principal component scores. 

The results for the epidemic-type changepoint are plotted on the right of Figure \ref{fig:HSW_change_results}. 
Estimates of $\tau_1(\mb{s})$ are concentrated around the stratospheric aerosol injection event, and estimates of $\tau_2(\mb{s})$ correspond more closely to the changepoints detected with the AMOC model.
There is variability in the results between ensembles. For ensemble 2, very few changepoints were detected after correction, and for ensemble 3, the estimates for $\tau_2(\mb{s})$ are later compared to other ensembles. 
For the counterfactual ensemble member, after Benjamini-Hochberg correction only few changepoints were detected using the AMOC model, while no changepoints were detected for the epidemic model.
This suggests that the changepoint detection procedures work well under a ``null hypothesis'' in this climate model. 

\begin{figure}

\includegraphics[width = .48\textwidth]{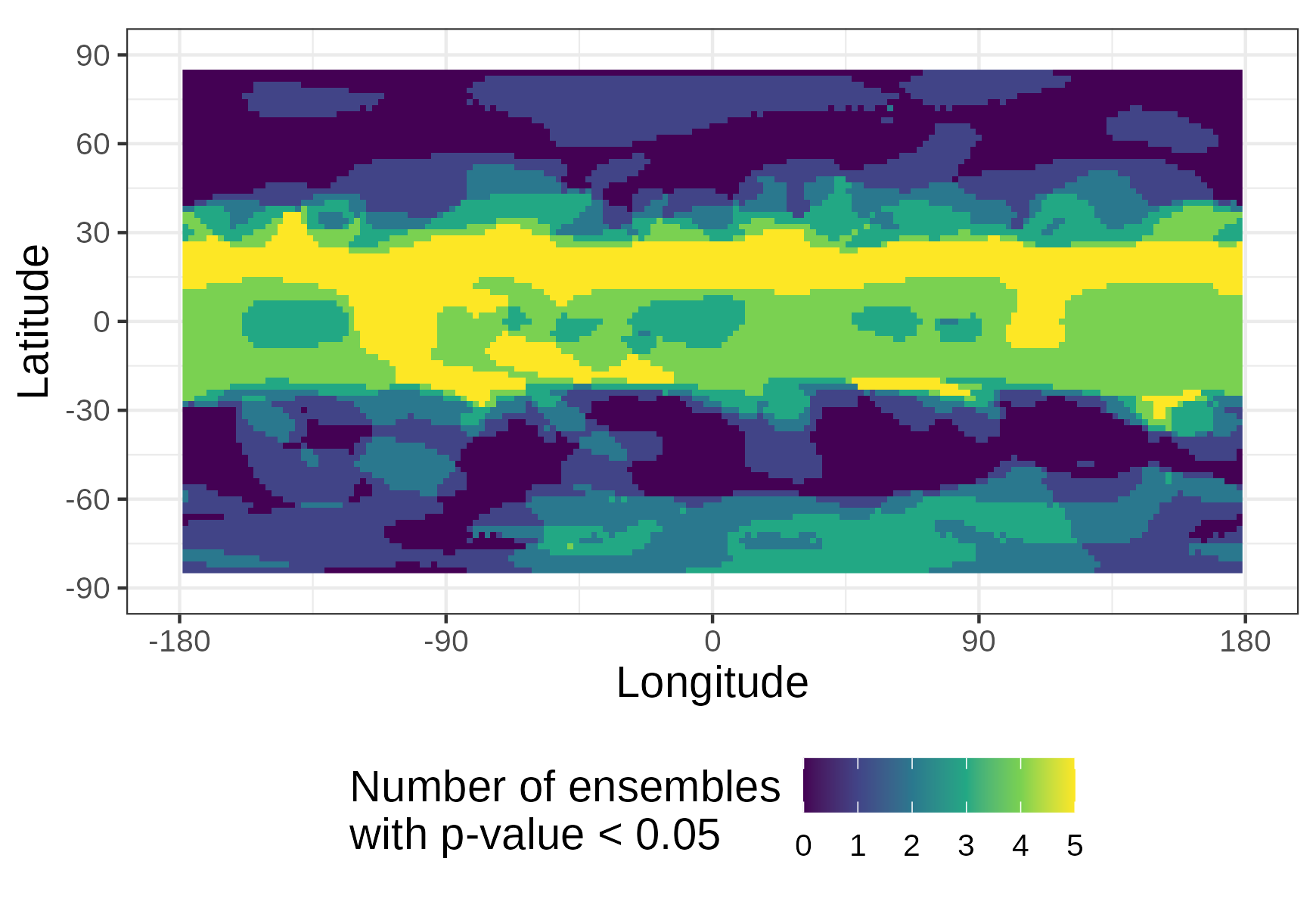}
\includegraphics[width = .48\textwidth]{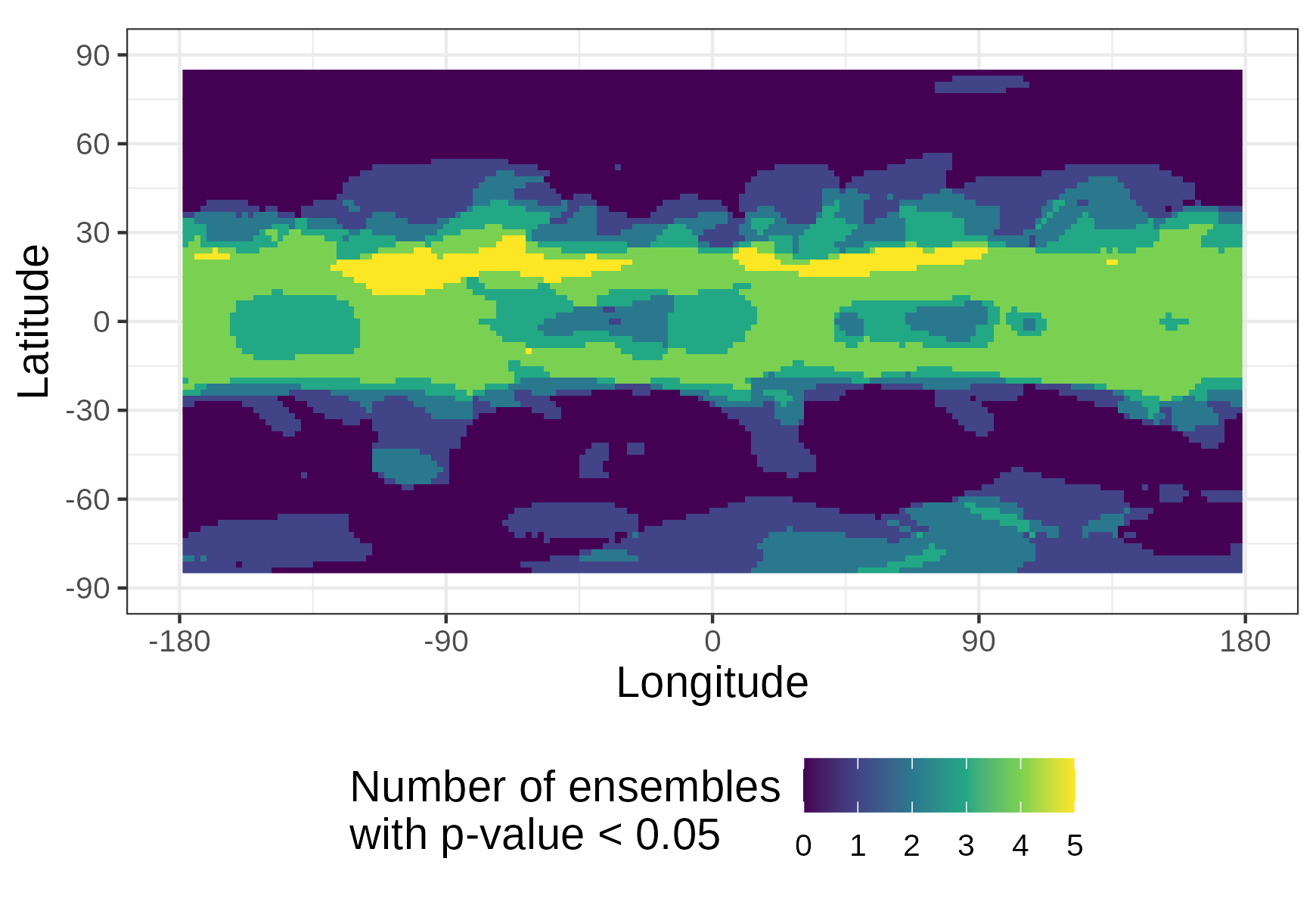}

\includegraphics[width = .48\textwidth]{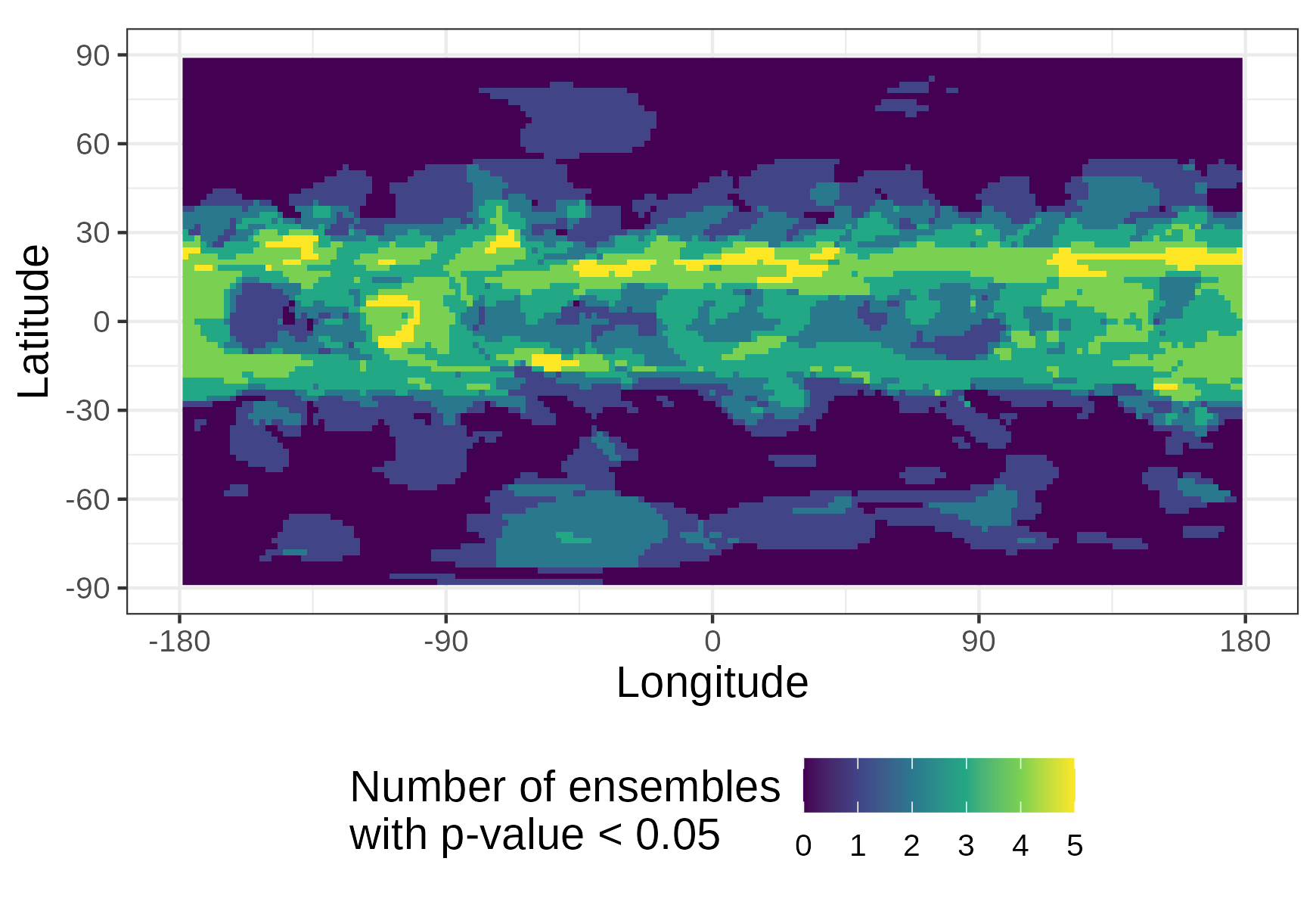}
\includegraphics[width = .48\textwidth]{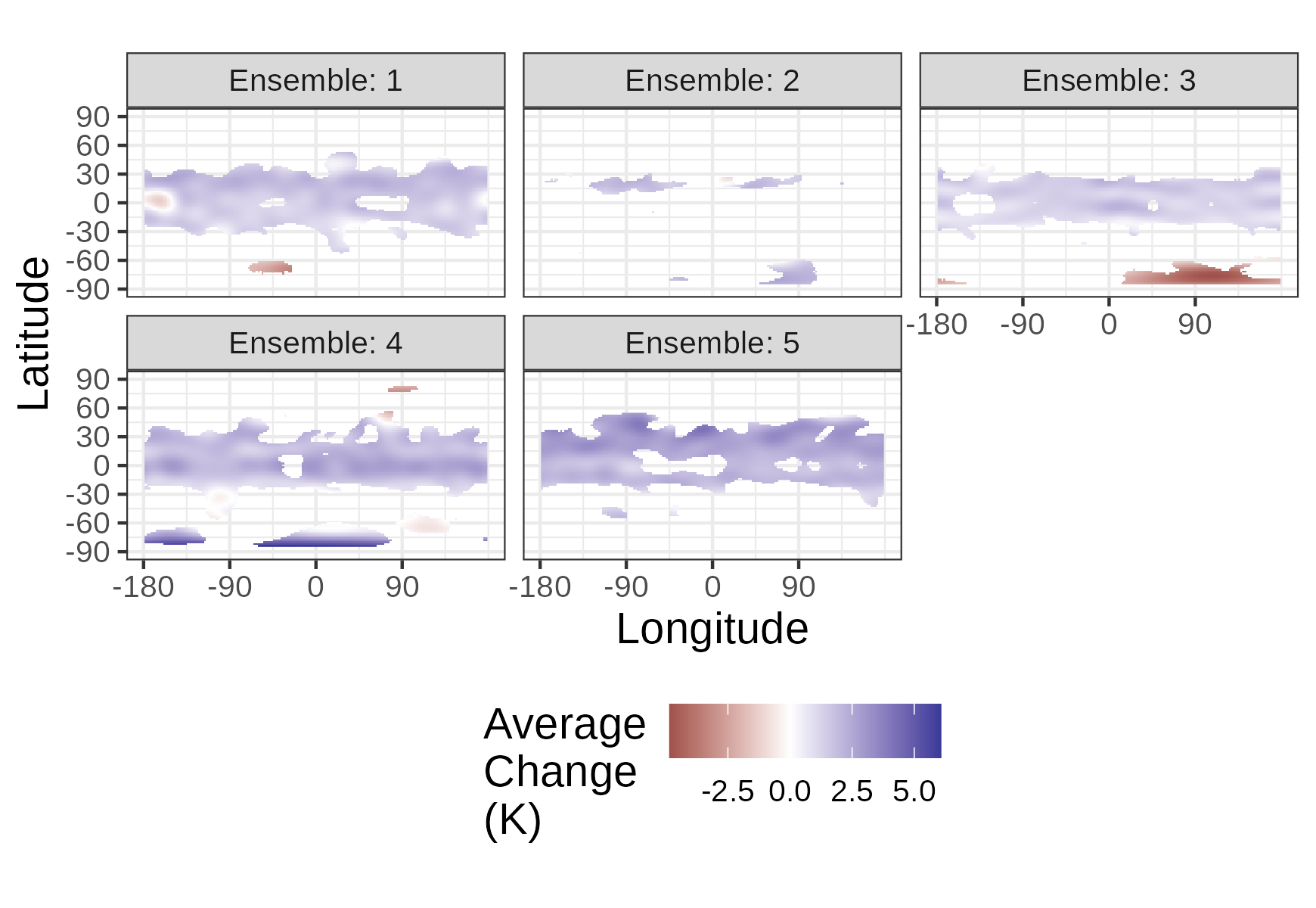}

\caption{HSW++ results. (Left) Number of ensembles with detected epidemic change after (Top) and before (Bottom) spatial prediction. (Top Right) Number of ensembles with detected epidemic change after Benjamini-Hochberg correction on spatially-predicted test statistics. (Bottom Right) Estimated average magnitude of change surface.}\label{fig:HSW_locs_results}

\end{figure}

In Figure \ref{fig:HSW_locs_results}, we plot results on the latitude-longitude grid. 
On the left, we compare the number of ensembles (of 5 with injection) that the locations had a detected epidemic changepoint using individual test statistics and test statistics based on spatial prediction, both without p-value correction at the $0.05$ level. 
We see that the spatially-predicted test statistics increase power with more detected changepoints. 
On the top right, we give the same plot for Benjamini-Hochberg-corrected p-values based on spatial prediction. 
P-value correction reduces the number of times changepoints were detected, yet more changepoints are detected compared to only using data at each location. 
Furthermore, the areas with changepoints detected for all five ensemble correspond to the approximate latitude of the injection site ($15^\circ$N).
On the bottom right of Figure \ref{fig:HSW_locs_results}, we plot an estimate of $\int_0^{1}\delta(\mb{s}, u) du$, the averaged amount of change associated with the epidemic time period. 
Grid cells were plotted if a changepoint was detected using a Benjamini-Hochberg-corrected p-value. 
There is generally an increase in stratospheric temperature across the ensembles between 1-4 degrees K for detected locations in the tropics, in line with expectations. 

In general, the HSW++ data reinforces the validity of our approach to spatially predict test statistics by detecting expected changes, and the form of the changepoint model (AMOC or epidemic) can largely affect results and interpretation of observed changes.

\subsection{Application to MERRA-2 reanalysis}

\begin{figure}
\includegraphics[width = .48\textwidth]{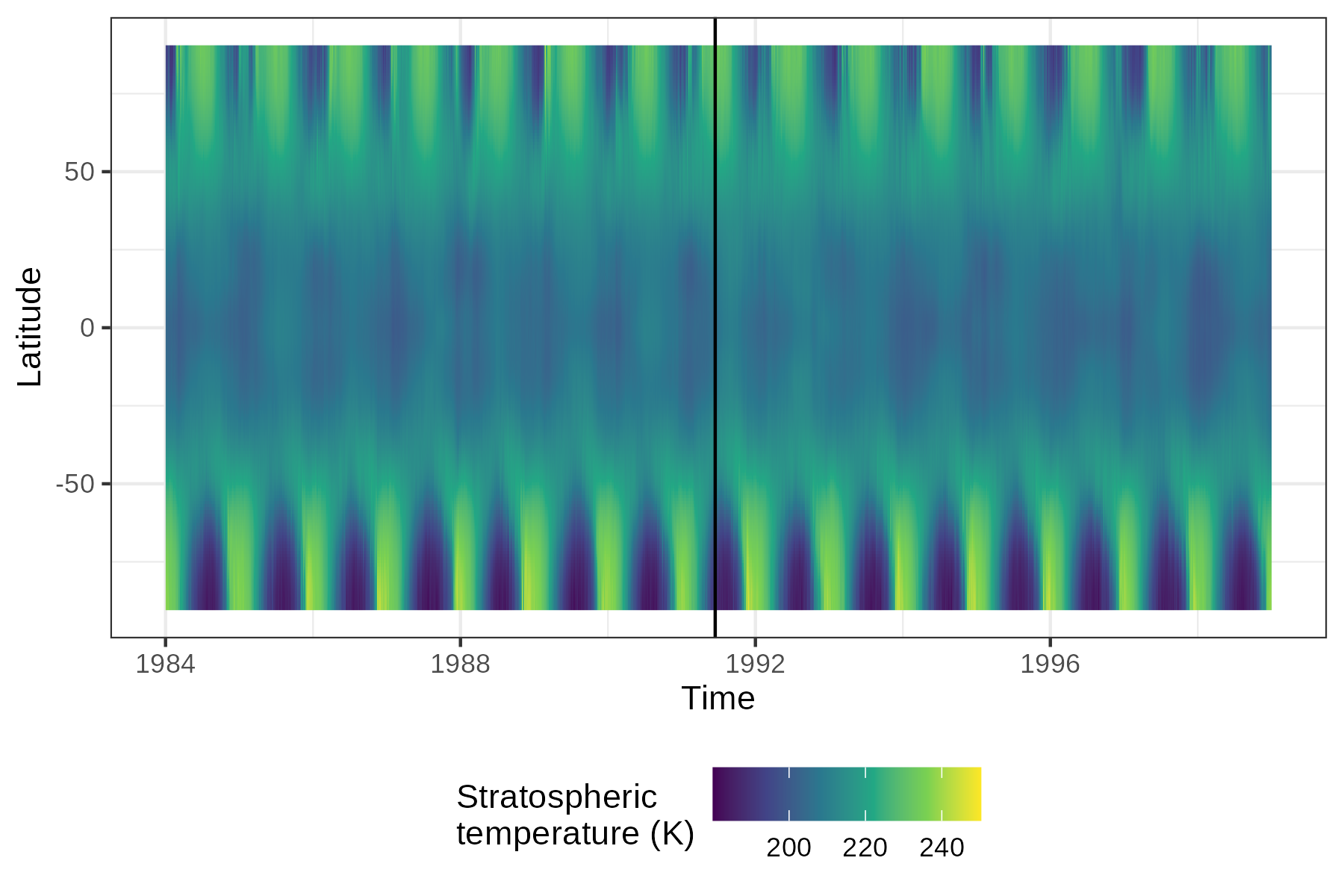}
\includegraphics[width = .48\textwidth]{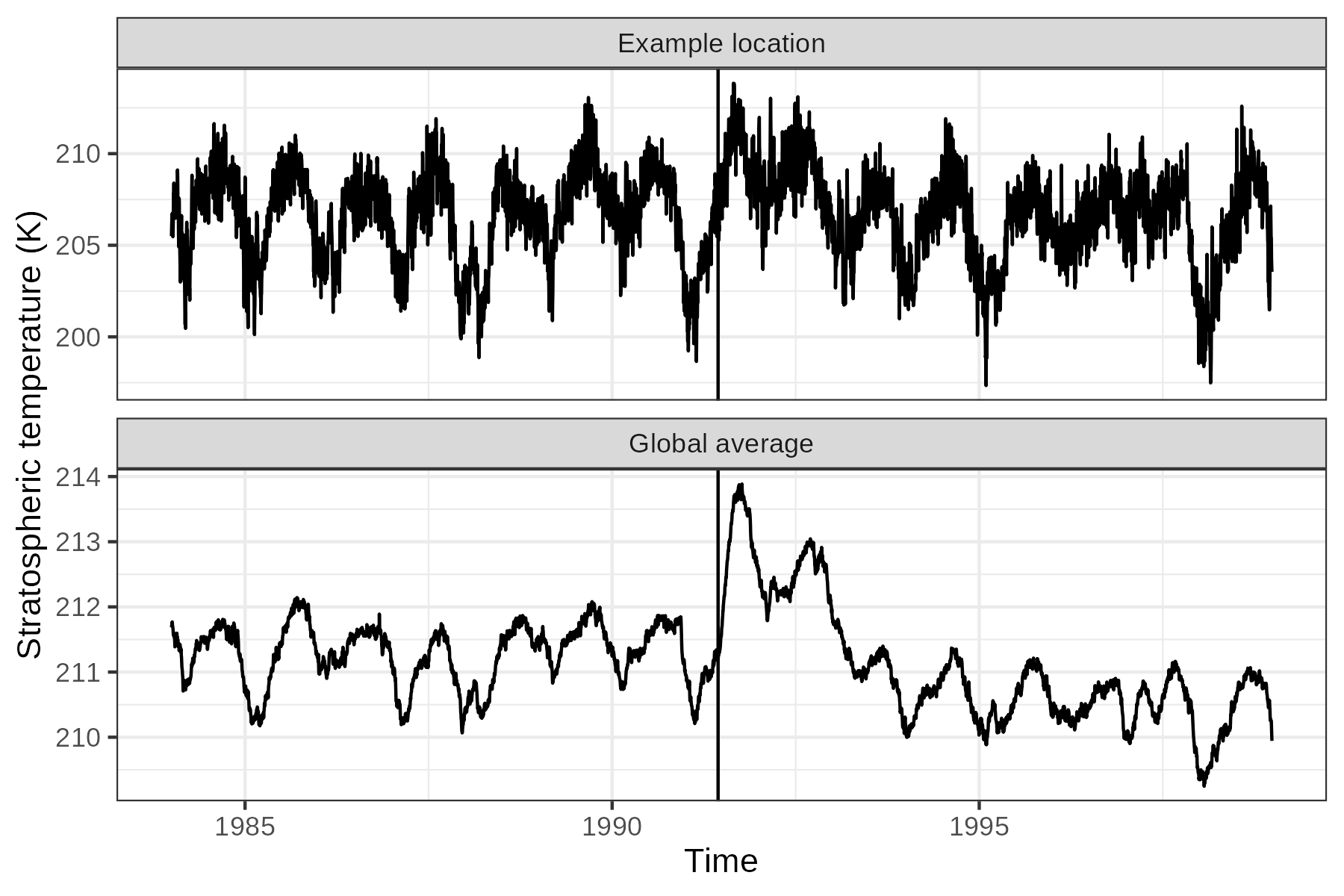}

\caption{(Left) Longitudinal average of MERRA-2 reanalysis of stratospheric temperature. (Top Right) Example of MERRA-2 reanalysis stratospheric temperature data for $37.5^\circ \textrm{E}$ and $18^\circ \textrm{N}$.  (Bottom Right) Global average of stratospheric temperature. }\label{fig:MERRA2_data}

\end{figure}

We next turn to the MERRA-2 climate reanalysis data \citep{merra2} which assimilates historical observations with a climate model to provide data at standardized locations and time points. 
We use stratospheric (near 50mb) temperature data from 1984-1998 ($N =15$) to provide substantial data around either side of the June 1991 eruption of Mt.\ Pinatubo.
We use daily-averaged data, wrangled into $m = 365$ observations per year. 
In Figure \ref{fig:MERRA2_data}, we plot an overview of the MERRA-2 data. 
As expected, we see strong seasonal cycles that dominates other sources of variability. 
In contrast to the HSW++ data, the change in stratospheric temperature around the time of the aerosol injection is less discernible. 
We use the same number of basis functions and principal components as the HSW++ analysis.

\begin{figure}[ht]
\includegraphics[width = .48\textwidth]{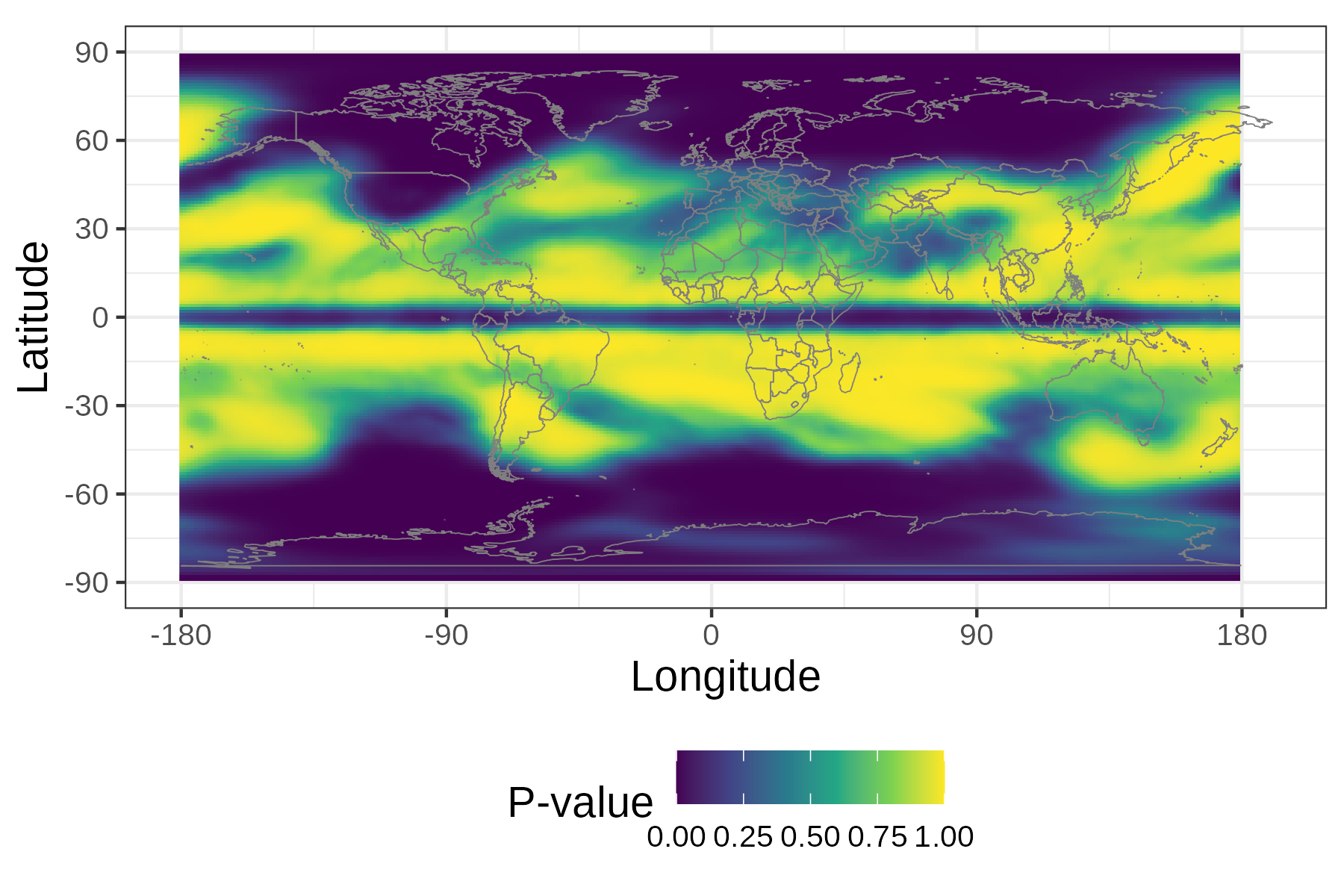}
\includegraphics[width = .48\textwidth]{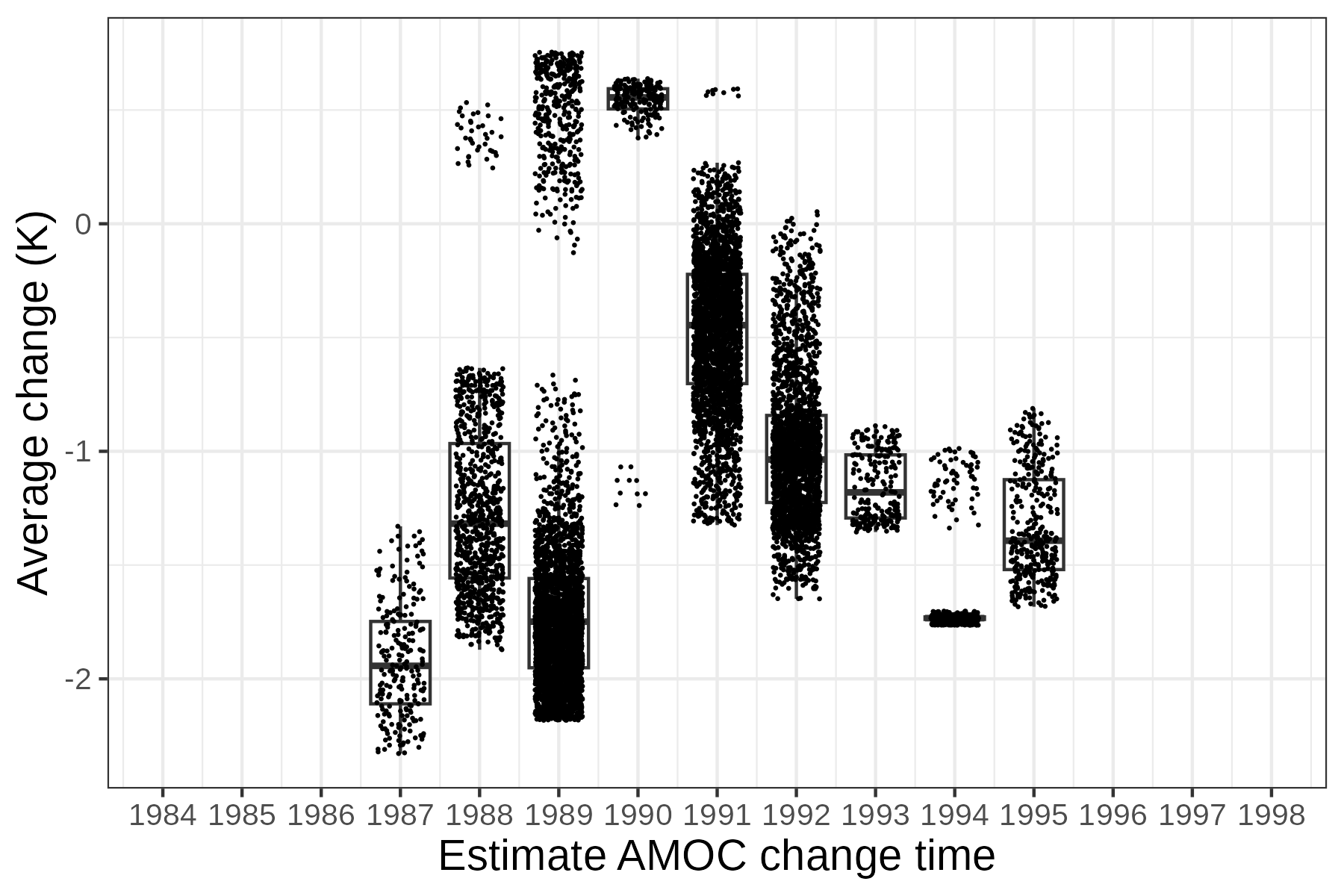}

\includegraphics[width = .48\textwidth]{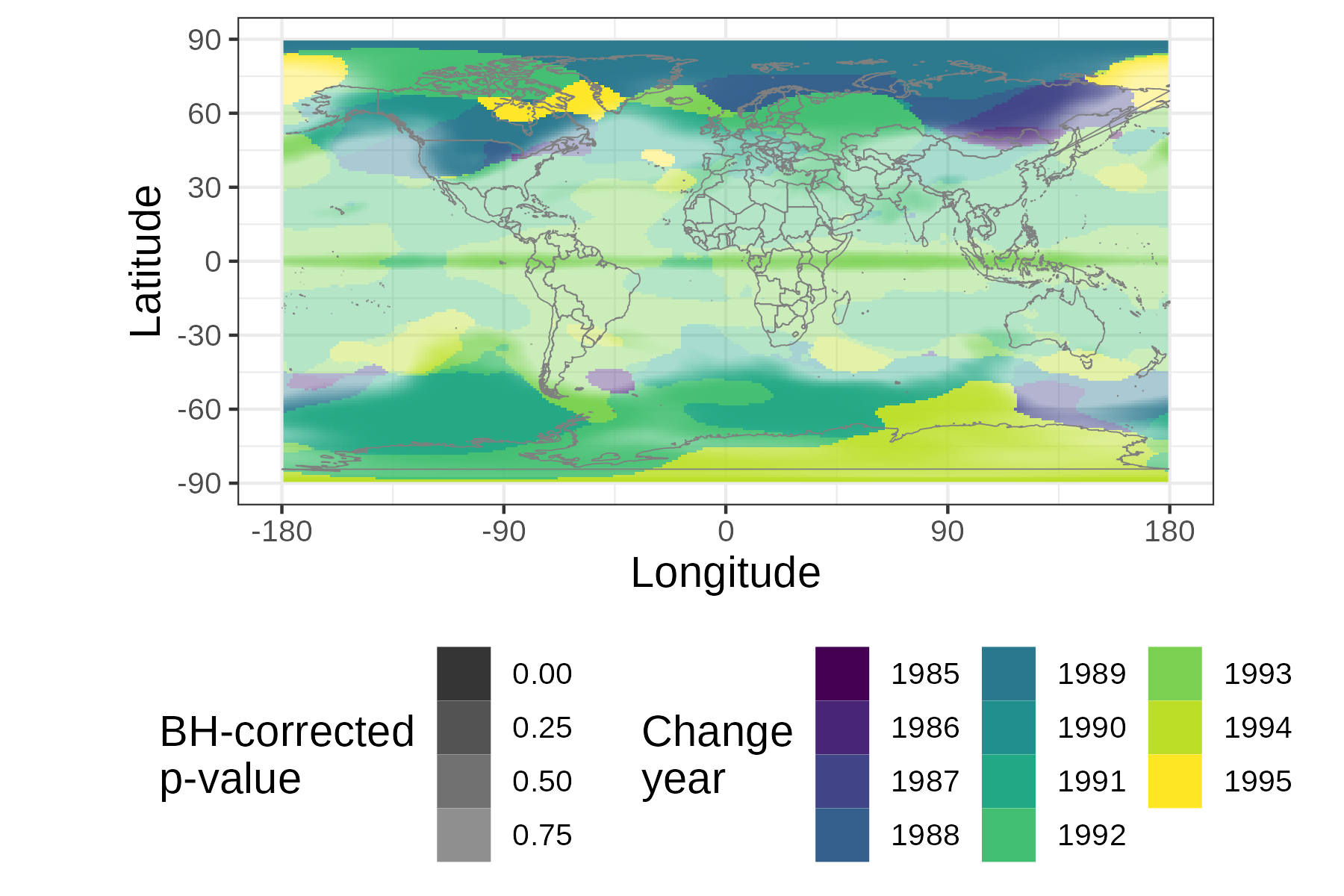}
\includegraphics[width = .48\textwidth]{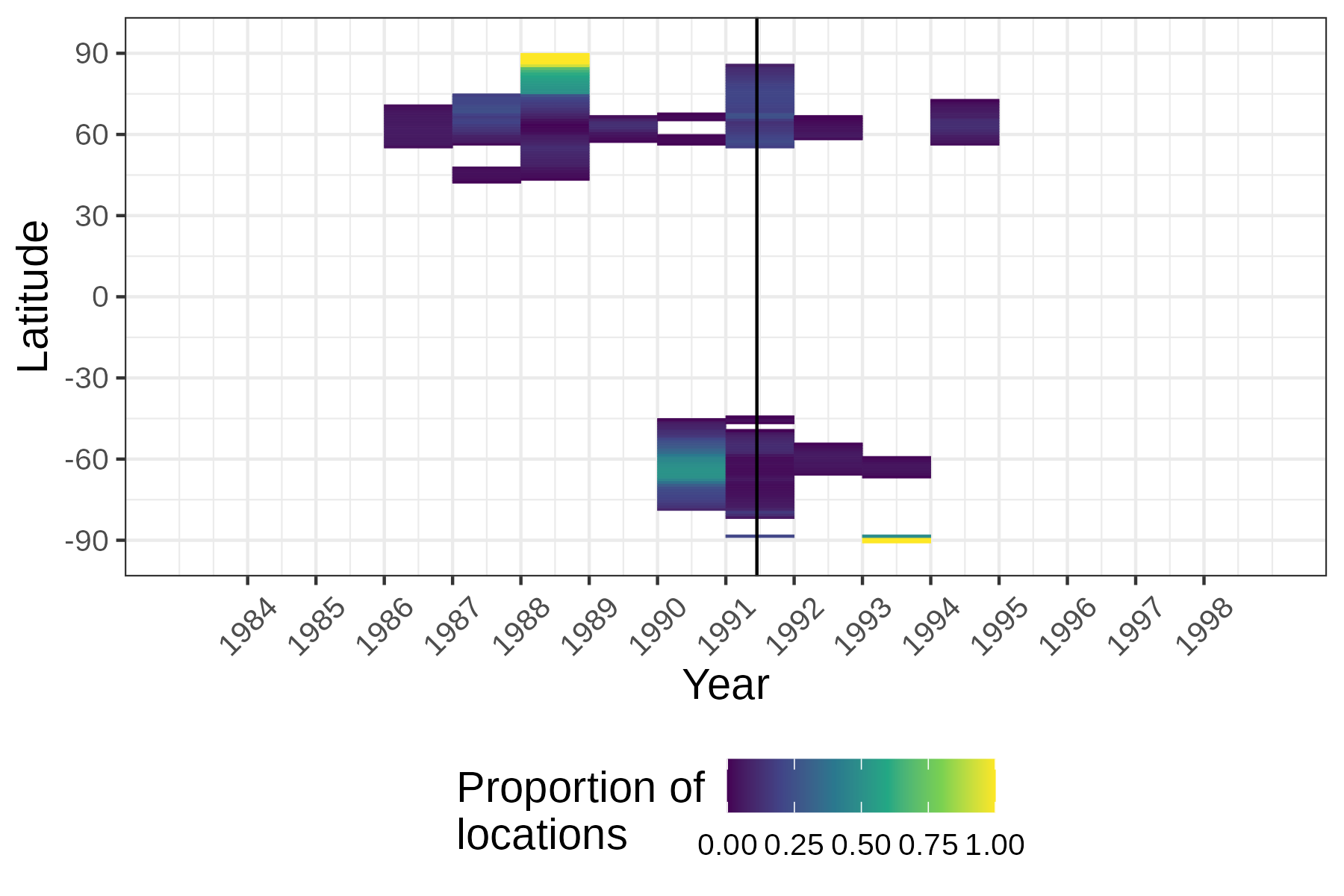}

\caption{MERRA-2 stratospheric temperature results for AMOC model. (Top Left) Unadjusted p-values. (Top Right)  Boxplot and points of the average change estimate by $\hat{\tau}(\mb{s})$ for locations with significant BH-adjusted p-value. (Bottom Left) Estimated $\hat{\tau}(\mb{s})$.
(Bottom Right) For each latitude, the proportion of locations (of 288 different longitudes) with $\hat{\tau}(\mb{s})$ for each year and significant BH-corrected p-values. }\label{fig:merra2_amoc}

\end{figure}

We present results first from the AMOC model in Figure \ref{fig:merra2_amoc}. 
There are low p-values in the poles as well as around the equator. 
While detected changepoints in the tropics (approximately $30^\circ$S to $30^\circ$N) may be a direct result of the eruption of Mt.\ Pinatubo, changes near the poles would be more likely associated with other changes or variability in the climate. 
In the top right of Figure \ref{fig:merra2_amoc}, we plot a boxplot of $\hat{\Delta}(\mb{s}_i) = \int_0^{1}  \hat{\delta}(\mb{s}_i, u) du$ for different $\mb{s}_i$ with a detected changepoint. 
The results are mostly consistent with expected changes in stratospheric temperature based on the eruption of Mt.\ Pinatubo. 
For example, for detected changes with the change period beginning for 1992 or later ($\hat{\tau}(\mb{s})\geq 1991$), there is an expected decrease in stratospheric temperature on the order of 1 to 2 K. 
This suggests the detection procedure detects a ``return-to-normal'' type change in the data. 
Similarly, there are detected changepoints with a later increase in stratospheric temperature in 1987 and 1990.
Many changes may also be associated with other sources of variability in the climate data. 
In the bottom of Figure \ref{fig:merra2_amoc}, we provide summaries of $\hat{\tau}(\mb{s})$. 
Many changepoints are concentrated around the time of the eruption, potentially detecting either a ``return-to-normal'' type change or a change based on the original eruption. 

\begin{figure}
\includegraphics[width = .48\textwidth]{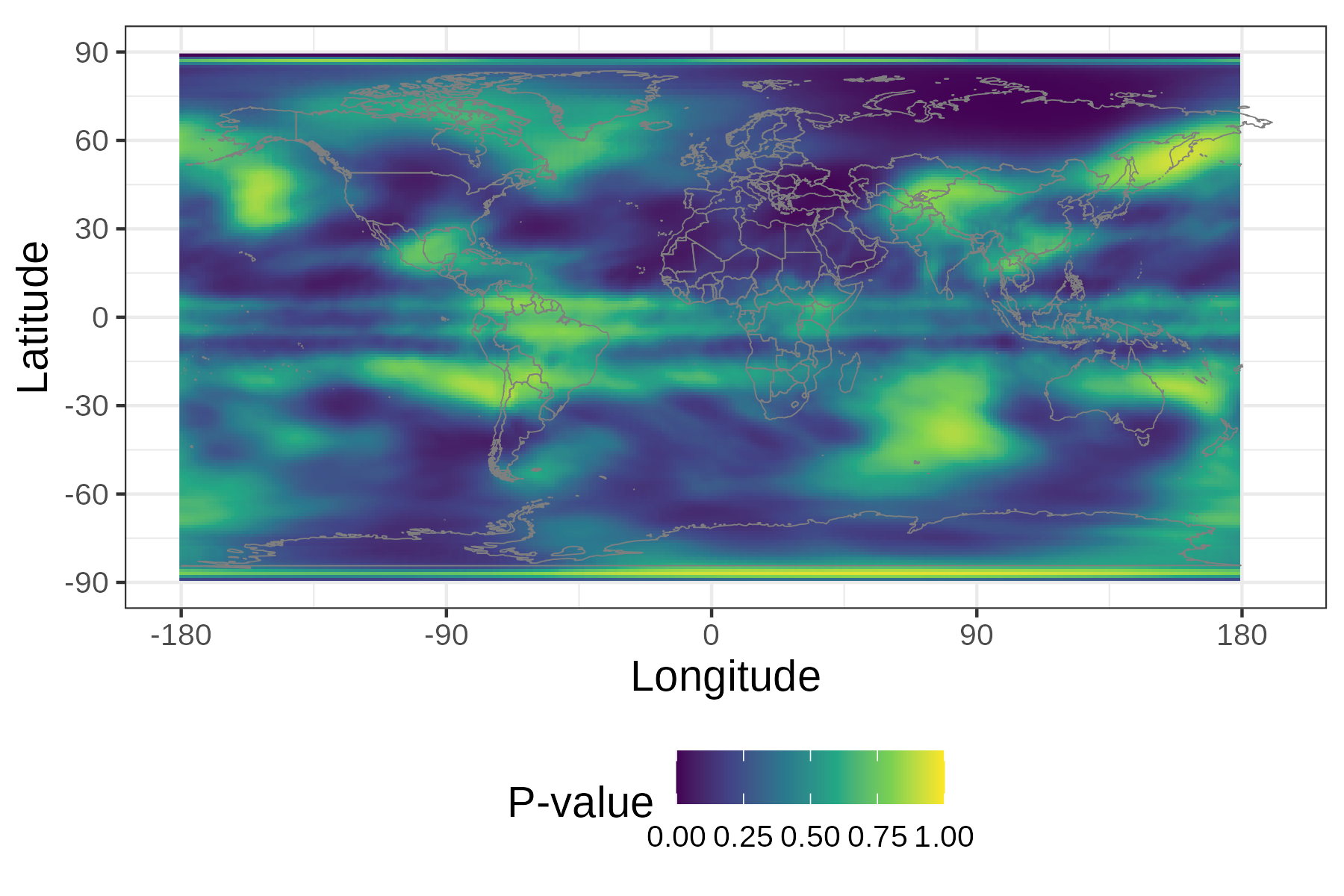}
\includegraphics[width = .48\textwidth]{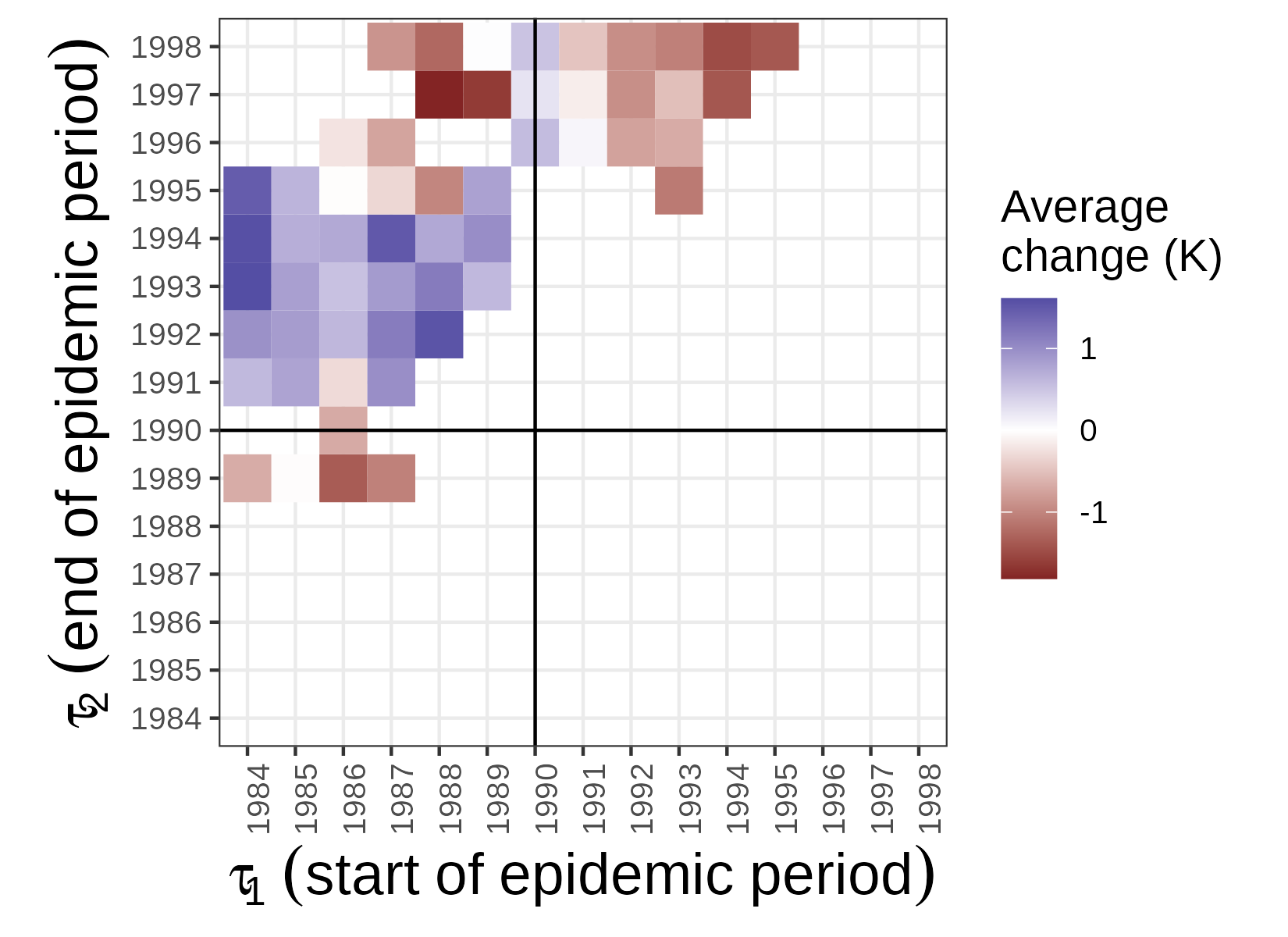}

\caption{(Left) Unadjusted epidemic p-values. (Right) Mean average change estimate for the epidemic changepoint model for all locations by year of detected change. 
}\label{fig:MERRA2_boxplot}

\end{figure}

In Figure \ref{fig:MERRA2_boxplot}, we give results for the epidemic changepoint model. 
Since the epidemic changepoint model is more complex, we have less power, and after Benjamini-Hochberg correction to the p-values, no detections are made at the $\alpha = 0.05$ level. 
On the other hand, there are locations with significant changes when not using p-value adjustment, and there are substantially lower p-values in the tropics compared to the p-values from the AMOC change model.
As with the HSW++ data, the epidemic change model may be more appropriate. 
In the right panel of Figure \ref{fig:MERRA2_boxplot}, we plot $\textrm{mean}(\hat{\Delta}(\mb{s}))$ when grouping by both the estimated $\hat{\tau}_1(\mb{s})$ and $\hat{\tau}_2(\mb{s})$. 
The estimated changes again are in accordance with the expected effects of Mt.\ Pinatubo on stratospheric temperature. 
When $\hat{\tau}_1(\mb{s}) \leq 1989$, an average increase in stratospheric temperature on the order of $0.25-1.6$ K is estimated, with these epidemic periods spanning the immediate aftermath of the eruption (1991-1994). 
Alternately, when $\hat{\tau}_1(\mb{s}) \geq 1991$, an average decrease in stratospheric temperature is estimated, suggesting a return-to-normal-type changepoint is estimated using $\hat{\tau}_1(\mb{s})$.
More results for the epidemic model are given in the supplement Section \checklater{S5}. 

In Figure \ref{fig:MERRA2_example}, we plot two examples of locations in our changepoint analysis. 
We plot the MERRA-2 stratospheric temperature data and our estimates. 
Even though the p-values are relatively large, the estimated changes use structure in the data. 
On the left panel, the years 1989-1992, and especially the year after the eruption of Mt.\ Pinatubo in June 1991, have a higher stratospheric temperatures than other years. 
The AMOC changepoint model groups this 4-year period with 1984-1988, leading to an estimated decrease in temperature after the change in 1992. 
The epidemic changepoint selects 1989-1992 corresponding to an expected increase in temperature.  
At the other location on the right panel, both models group the higher temperatures immediately after the eruption with 1984 through 1990, which had higher temperature compared to the years 1993-1998. 
As with the HSW++ data, the changepoint detection procedures may pick up unexpected events, and visualization and evaluation of the estimated changes can inform resulting interpretation. 
 
\begin{figure}[ht]
\centering
\includegraphics[width = .48\textwidth]{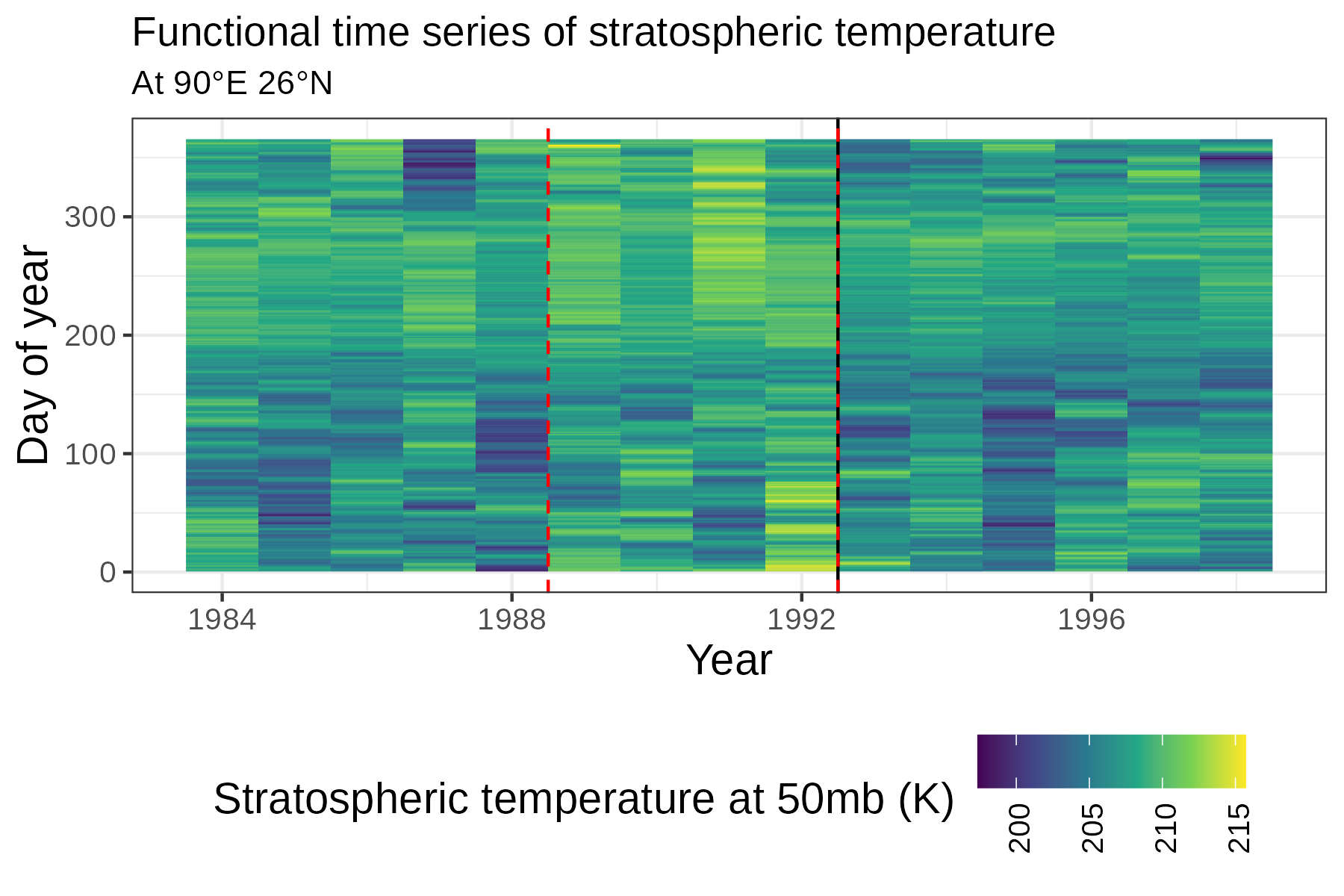}
\includegraphics[width = .48\textwidth]{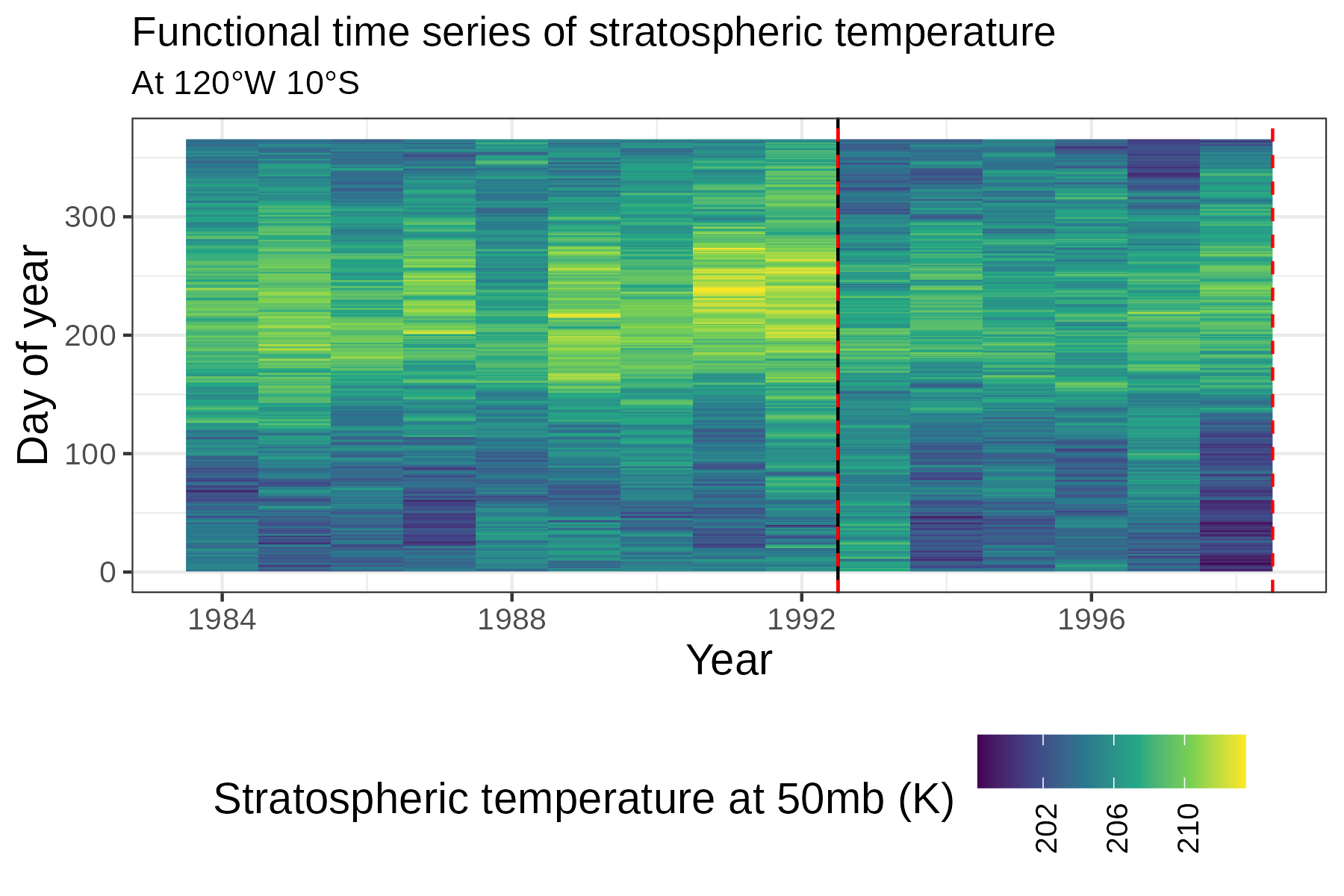}

\caption{Two examples of stratospheric temperature and estimated change points. 
The solid black line is the change detected using the AMOC model, with p-values of $0.472$ (Left) and $ 0.996$ (Right). 
The two dashed red lines are the change period detected using the epidemic model, with p-values of $0.428$ (Left) and $0.259$ (Right).}\label{fig:MERRA2_example}

\end{figure}

In the supplement Section \checklater{S6}, we also give the results on another atmospheric variable from MERRA-2, aerosol optical depth (AOD). 
Overall, the results are mostly similar to the results for stratospheric temperature. 
While power continues to be a challenge for detecting changepoints, the estimated amount of changes largely give expected changes associated with Mt.\ Pinatubo, while describing complex spatial heterogeneity in the climate. 

\section{Discussion}\label{sec:discussion}

In future efforts to understand changes in climate due to volcanic eruptions or climate geoengineering events, the ability of statistical procedures to detect and estimate such changes will play an important role in the interpretation and response to such events. 
Such statistical procedures could also play a role in the planning of climate interventions.
We introduce an approach using functional time series for such situations, and we comprehensively evaluate its abilities and limitations for this purpose.
In this work, we provide a comprehensive framework for detecting changepoints in globally-indexed functional time series that handles spatial heterogeneity in the measured variable and its changepoints.  
We demonstrate that one can use basic approaches to improve both changepoint detection power and the estimation of the changepoint year using spatial information.
We iteratively apply the methodology on increasingly complex source data, demonstrating its use on a statistical simulation study, a simplified climate model, and climate reanalysis data. 
Moreover, we use the simulation study to evaluate different test statistics and approaches for using spatial information, finding that estimating the model under the null hypothesis and using score-based test statistics consistently improve estimates.

The full-variability analysis on MERRA-2 reanalysis data is challenging due to two main reasons.
First, there are complex sources of variability.
While the functional time series approach handles seasonality naturally, other variability like long-term trends are ignored. 
The second challenge is a lack of power due to a reduced sample size. 
In our analysis on climate reanalysis data, we use $N= 15$ years; in this setting, it is challenging to detect changes based on asymptotic results that take $N \to \infty$.
However, this problem does not have a simple solution.
One could aim to increase the number of years of data, yet over longer periods of time different events may be influential.
For example, the 1982 volcanic eruption of El Chinch\'on in Mexico likely influenced the climate in similar ways to the eruption of Mt.\ Pinatubo in 1991 \citep{marshall_2022}. 
One could also aim to use time periods more granular than years for the functional time series, yet one would need an alternate approach to handle seasonality. 
One could also only infer changepoints at a reduced set of locations, reducing the challenge of the multiple testing problem, or apply more sophisticated spatial testing approaches \citep{risser2019spatially, yun2022detection}. 
This challenge is especially pronounced for the epidemic changepoint model, as the epidemic change period may only last one or two years. 
Therefore, considering the problem under the framework of anomaly detection may be useful. 
To evaluate the test statistics' sensitivity to the reduced sample size, we repeated the HSW++ analysis for the first ensemble member using $N = 12$ and $m = 400$ instead of $N = 40$ and $m = 120$, with results presented in the supplement Section \checklater{S4}. 
As expected, this decreased the power for detecting changepoints, suggesting that the reduced $N$ substantially challenged the full-variability analysis.

One other limitation of this work is the lack of dependence assumed across $k$. 
As detailed in \cite{lund_2022}, ignoring such dependence may lead to conservative or anticonservative testing results. 
However, since, the functional time series are partitioned into relatively long time periods of years, dependence may be weaker between years compared to other settings (where $k$ might index weeks or days). 
As written, the spatial approaches developed here should be especially useful when the variable varies relatively smoothly without hard boundaries between regions of change and no change. 
Otherwise, the spatial model may unduly extrapolate changes in the mean to areas where there is no true change in the mean. 
Using spatial approaches like clustering would be useful for this setting with a variable like surface temperature, which is strongly influenced by hard topographic boundaries.

\section*{Acknowledgment}
This paper describes objective technical results and analysis. Any subjective views or opinions that might be expressed in the paper do not necessarily represent the views of the U.S. Department of Energy or the United States Government. This work was supported by the Laboratory Directed Research and Development program at Sandia National Laboratories; a multi-mission laboratory managed and operated by National Technology and Engineering Solutions of Sandia, LLC, a wholly owned subsidiary of Honeywell International, Inc., for the U.S. Department of Energy's National Nuclear Security Administration under contract DE-NA0003525.


%
%
%
%

\bibliographystyle{apalike2}

\bibliography{main.bib}

\end{document}


\def\spacingset#1{\renewcommand{\baselinestretch}%
{#1}\small\normalsize} \spacingset{1}


\if0\blind
{
  \title{\bf Supplement to ``Detecting changepoints\\ in globally-indexed functional time series''}
  \author{Drew Yarger and J. Derek Tucker\hspace{.2cm}\\
    Statistical Sciences, Sandia National Laboratories}
  \maketitle
} \fi

\if1\blind
{
  \bigskip
  \bigskip
  \bigskip
  \begin{center}
    {\LARGE\bf Supplement to ``Detecting changepoints\\ in globally-indexed functional time series''}
\end{center}
  \medskip
} \fi

\bigskip


\spacingset{1.8} 

%
%
%
%


\section{Comparison of Assumptions}\label{app:assumptions}

We explain and remark upon the assumptions presented in Section \okfornow{2.2}. 

\begin{remark}[Discussion of independence assumptions]\normalfont
Across the varying indexes, dependence or independence may be assumed. 
For example, we assume independence of $\{Z_{qk}(\mb{s})\}$ across the principal component entry $q$.
This is referred to as ``weak separability'' for spatio-functional principal component models, and see \cite{liang_2022} who propose a test for this assumption. 
The term $U_{iqk}$ introduces a ``functional nugget'' process \citep[as described in][]{zhang2022unified} defined by $\sum_{q=1}^Q U_{iqk} \phi_q(u)$, representing error dependent in $u$ yet independent across $\mb{s}$.

In contrast to some functional time series literature, we will assume independence in $k$ rather than weak dependence in order to focus our effort on the spatial dimension.
Previous work has taken different approaches to modeling how data from different $k$ relates to each other: independence across $k$ \citep{gromenko:2017, berkes:2009, aue2009estimation}, independence for data on different sides of $\tau(\mb{s})$ \citep{zhao2019composite}, and weak dependence \citep{Aue_2018, li_changepoint_2022}, though in \cite{li_changepoint_2022} independence is assumed for the simulation study.
This separates our model into ``spatially-coherent variability'' $\sum_{q=1}^Q \tilde{Z}_{qk}(\mb{s}) \phi_q(u)$ and ``nuisance variability'' of  $\sum_{q=1}^Q U_{iqk} \phi_q(u) + W_{ijk}$. 
Since they represent different sources of error, we assume that the components $\{W_{ijk}\}$, $\{U_{iqk}\}$, and $\{\tilde{Z}_{qk}(\mb{s}_i)\}$ are independent from each other.
\end{remark}

\begin{remark}[Discussion of spatially-varying components of $\epsilon_{k}(\mb{s}, u)$]\normalfont \label{rem:lr}

Throughout, we take the spatially-varying variances $\sigma_{q}^2(\mb{s})$ and $\gamma_q^2(\mb{s})$ to have general, nonparametric smooth form. 
We use local regression to estimate these parameters \citep{fan1996local} which we describe in more detail here.

Let $K(\mb{s})$ be a kernel function; we use the Epanechnikov kernel $K(\mb{s}) = (3/4) \left(1 - \left\lVert \mb{s}\right\rVert^2\right)_+$ where $(x)_+ = x$ if $x > 0$ and $0$ otherwise. 
If we let $K_h(\mb{s}) = K(\mb{s}/h)/h$ for a bandwidth $h$ that we choose later, a locally-constant or Nadaraya-Watson estimator of $\sigma_q^2(\mb{s}_0)$ for a fixed $\mb{s}_0$ is: \begin{align*}
\hat{\sigma}^2_q(\mb{s}_0)&= \argmin_{\beta \in \mathbb{R}}\sum_{i_1=1}^n \sum_{i_2=1}^n K_h(\mb{s}_{i_1} - \mb{s}_0)  K_h(\mb{s}_{i_2} - \mb{s}_0) \sum_{k=1}^N (Z_{qk}(\mb{s}_{i_1})Z_{qk}(\mb{s}_{i_2}) - \beta)^2
\end{align*}
This estimate uses the spatial correlation of data points that are near $\mb{s}_0$ to estimate $\hat{\sigma}^2_q(\mb{s}_0)$ and uses data from all years in the estimate. 
Notice that the optimization problem may also be written up to a constant as \begin{align*}
\hat{\sigma}^2_q(\mb{s}_0)&= \argmin_{\beta \in \mathbb{R}}\sum_{i_1=1}^n \sum_{i_2=1}^n K_h(\mb{s}_{i_1} - \mb{s}_0)  K_h(\mb{s}_{i_2} - \mb{s}_0) (C_{i_1, i_2, q} - \beta)^2
\end{align*}where $C_{i_1, i_2, q} = \frac{1}{N}\sum_{k=1}^N Z_{qk}(\mb{s}_{i_1})Z_{qk}(\mb{s}_{i_2})$ which simplifies computations. 

To estimate $\gamma_q^2(\mb{s})$, we similarly estimate $\hat{\gamma}^2_q(\mb{s}_0) = \hat{\lambda}_{q}(\mb{s}_0)  -  \hat{\sigma}^2_q(\mb{s}_0)$, where \begin{align}
\hat{\lambda}_{q}(\mb{s}_0)&= \argmin_{\beta \in \mathbb{R}}\sum_{i=1}^n K_h(\mb{s}_{i} - \mb{s}_0) \sum_{k=1}^N \left(Z_{qk}(\mb{s}_{i})^2 - \beta\right)^2.\label{eq:lr_var}
\end{align}
In summary, $\hat{\lambda}_{q}(\mb{s}_0)$ is an estimate of $ \sigma_q^2(\mb{s}_0) + \gamma_q^2(\mb{s}_0)$, so we subtract off $\hat{\sigma}_q^2(\mb{s}_0)$ to estimate $\gamma_q^2(\mb{s}_0)$. 
Like with the estimate $\hat{\sigma}^2_q(\mb{s}_0)$, one can summarize over $k$ first to reduce the size of the resulting weighted least-squares problem. 
 
Here, the principal components $\phi_q(u)$ are assumed to be spatially constant as in \cite{li_changepoint_2022}. 
Developing a cohesive spatially-varying principal component framework is a challenging area for future work.
\end{remark}

\begin{remark}[Covariance function flexibility]\normalfont
For our simulation study, we will simulate from a covariance model like \okfornow{(4)}, yet aim to estimate the smoothness parameter of the Mat\'ern process when estimating it. 
For our data application, a covariance function on a sphere is used that has considerably more complexity (see Section \okfornow{3}).  
The smoothness parameter of the Mat\'ern parameter is estimated, and a spatially-varying anisotropic parameterization replaces $\alpha_q$ in \okfornow{(4)}. 
\end{remark}

\begin{remark}[Normality]\normalfont
While much of the changepoint literature does not assume normality of the original data, we assume joint normality for all random components $\{W_{ijk}\}$, $\{U_{iqk}\}$, and $\{\tilde{Z}_{qk}(\mb{s}_i)\}$. 
We do this primarily so that conditional distributions that define spatial predictions are straightforward under the framework of Gaussian processes. 
\end{remark}

\section{Estimation details for penalized least squares}\label{sec:est_details}

We briefly outline the approach to compute the solution to the minimization problem.
Explicitly, we use $
\beta(\mb{s}, u) = \sum_{\ell_s = 1}^{L_s} \sum_{\ell_u = 1}^{L_u} \Theta_{\ell_s, \ell_u} \mathcal{S}_{\ell_s} (\mb{s}) \psi_{\ell_u} (u)$, where $\{\Theta_{\ell_s, \ell_u}\}$ are coefficients to be estimated, $\{ \mathcal{S}_{\ell_s} (\mb{s})\}_{\ell_s = 1}^{L_s}$ are the spherical harmonic functions up to some degree, and $\{ \psi_{\ell_u} (u)\}_{\ell_u = 1}^{L_u}$ is a cubic B-spline basis with knots evenly spaced on $[0,1]$. 
Let $\mb{\Psi}_s= [\mathcal{S}_{\ell_s}(\mb{s}_i)]_{i=1, \ell_s = 1}^{n, L_s} \in \mathbb{R}^{n \times L_s}$ and $\mb{\Psi}_u= [\psi_{\ell_u}(u_j)]_{j=1, \ell_u = 1}^{m, L_u} \in \mathbb{R}^{m \times L_u}$ be basis matrices for the data observed on the sphere and on the functional domain, respectively. 
Let $\mb{\Theta}$ denote $\textrm{vec}(\{\Theta_{\ell_s, \ell_u}\})$ and $\overline{\mb{Y}} = \textrm{vec}(\{\overline{Y}^\mu (\mb{s}_i, u_j)\}_{i=1, j= 1}^{n,m})$. 
As suggested in \cite{wood_2006}, we take the penalty as $\zeta \cdot \textrm{vec}(\{\Theta_{\ell_s, \ell_u}\})^\top (\mb{I}_{L_s}\otimes \mb{\Omega}) \textrm{vec}(\{\Theta_{\ell_s, \ell_u}\})$, where $\mb{I}_{L_s}$ is the identity matrix of dimension $L_s$, $\mb{\Omega}$ is a $L_u \times L_u$ matrix with entries defined by the inner product between the second derivatives $\int_0^1 \psi_{\ell_1}^{(2)}(u) \psi_{\ell_2}^{(2)}(u) du$, and $\otimes$ and vec$(\cdot)$ are the Kronecker product and vector operations, respectively. 
Then the optimization problem can be rewritten as optimizing over the coefficients $\mb{\Theta}$: \begin{align*}
\hat{\mb{\Theta}}
&= \argmin_{\mb{\Theta}}  \left( \overline{\mb{Y}} - (\mb{\Psi}_s \otimes \mb{\Psi}_u)\mb{\Theta}\right)^\top\left( \overline{\mb{Y}} - (\mb{\Psi}_s \otimes \mb{\Psi}_u)\mb{\Theta}\right) + \zeta \mb{\Theta}^\top (\mb{I}_{L_s} 
\otimes \mb{\Omega}) \mb{\Theta} \\ 
&= \left( \left(\mb{\Psi}_s^\top \mb{\Psi}_s\right) \otimes \left(\mb{\Psi}_u^\top \mb{\Psi}_u\right) + \zeta \mb{I}_{L_s} \otimes \mb{\Omega} \right)^{-1} (\mb{\Psi}_s \otimes \mb{\Psi}_u )^\top \overline{\mb{Y}}.
\end{align*}
The solution to the minimization problem can be straightforwardly computed, involving finding a Cholesky decomposition of the partially-sparse $L_sL_u \times L_sL_u$ matrix instead of directly inverting it. 
To compute $(\mb{\Psi}_s \otimes \mb{\Psi}_u )^\top \overline{\mb{Y}}$ quickly and efficiently, one may instead compute $\textrm{vec}(\mb{\Psi}_s^\top \{\overline{Y}^\mu (\mb{s}_i, u_j)\}_{i=1, j= 1}^{n,m} \mb{\Psi}_u )$.

\section{Additional simulation results}\label{app:simulation}

We first evaluate the changepoint detection procedures, when averaging over all simulations, signal strengths, and the two dependence structures. 
Primarily, we evaluate Type I and Type II error using false positive rate (FPR) and false negative rate (FNR).
We also consider results after adjusting p-values using a Benjamini-Hochberg adjustment or a Bonferroni adjustment for the $n=300$ locations. 
We evaluate false discovery rate (FDR) and family-wise error rate (FWER) for these adjustments, respectively. 

We focus on score-based test statistics, with Type I errors in Table \ref{tab:typeI_null}. 
\begin{table}\footnotesize\centering
\begin{tabular}{|c||c|c|c|c|c|c|c|c|}\hline
Type I Errors& I, 3 PCs & I, 4&  I, 5&  P, 3 & P, 4& P, 5 & P unadj., 4 & P recom., 4 \\ \hline
FPR, raw p-values & 0.037 &0.021& 0.010& {\color{red} 0.107} &  {\color{red} 0.119} &  {\color{red} 0.115} & {\color{orange} 0.065} & {\color{orange} 0.059}\\ \hline
FDR, using BH &0.000 &0.000&0.000  & 0.042 & 0.046 & 0.046 & 0.022 & 0.025\\ \hline
FWER, using Bonf &0.000 & 0.000 & 0.000 & {\color{red}0.271}& {\color{red}0.390} & {\color{red}0.397}& {\color{red}0.329}&  0.000 \\ \hline
\end{tabular}
\caption{Score-based Type I error rates over all simulations, with nominal level $\alpha = 0.05$. ``I'' refers to individual test statistics, ``P'' refers to predicted test statistics. Colors: {\color{black} Type I Error $\leq 0.05$}, {\color{orange} $0.05 <$Type I Error$\leq 0.10$}, {\color{red} Type I Error $>0.10$}. } \label{tab:typeI_null}
\vskip .5cm

\hskip .17cm\begin{tabular}{|c||c|c|c|c|c|c|c|c|}\hline\centering
Type II Errors& I, 3 PCs & I, 4 &  I, 5 &  P, 3 & P, 4 & P, 5 & P unadj., 4 & P recom., 4\\ \hline
FNR, raw p-values & 0.765 & 0.770 &0.857& {\color{red} 0.491} & {\color{red} 0.391} & {\color{red} 0.407} & {\color{orange} 0.497} & {\color{orange}0.420} \\ \hline
FNR, using BH & 0.998 & 1.000 & 1.000 & 0.568 & 0.446 & 0.462 & 0.547 & 0.624 \\ \hline
FNR, using Bonf & 1.000 & 1.000 & 1.000 &{\color{red} 0.749}& {\color{red}0.578} & {\color{red}0.608} & {\color{red}0.621} & 1.000\\ \hline
\end{tabular}
\caption{Score-based Type II error rates (FNR) over all simulations, with nominal level $\alpha = 0.05$. Colors: {\color{black} Type I Error $\leq 0.05$}, {\color{orange} $0.05 <$Type I Error$\leq 0.10$}, {\color{red} Type I Error $>0.10$}. }\label{tab:typeII_null}
\end{table}
The individual test statistics control Type I error, while their predicted versions give more mixed results. 
The predicted test statistics do not all have full control over Type I error based on the raw p-values, yet overall they perform well, especially with the recomputed version of the test statistic. 
In addition, all test statistics control for FDR when properly corrected. 
Using p-value correction on test statistics computed at each location gives virtually no power, while sharing information across space leads to detection of changepoints. 

Turning to the Type II error in Table \ref{tab:typeII_null}, we find that the predicted test statistics have dramatically more power. 
In particular, without p-value correction, the recomputed version of the test statistic correctly detects nearly $35\%$ more of locations with a changepoint. 
When only evidence at each location is taken into account, finding strong evidence necessary for multiple testing correction is more challenging. 
Overall, one can considerably improve power by using predicted test statistics at the expense of some increase in Type I error.

We briefly compare the testing results with \cite{gromenko:2017}.
Their FPCA test statistic without normalization had the most power, with results presented in Table \ref{tab:sim_comparison}. 
The row with \cite{gromenko:2017} is the proportion of {\it simulations} where a global changepoint was detected, while the other test statistics are the proportion of {\it locations} with a change that had a change detected.
Though the results are thus not directly comparable, \cite{gromenko:2017}'s test statistic is less powerful for lower signal strengths, yet overall it is well-behaved. 
In this setting, testing for a single changepoint is an oversimplification that provides no information about whether a specific location observed a change. 

\begin{table}\footnotesize\centering
\begin{tabular}{|c||c|c|c|c|c|c|c|}\hline
Signal strength& 0.5 & 2&  6 &  10& 14& 18 & 22 \\ \hline\hline
 \cite{gromenko:2017} & 1.000 &1.000& 0.995& 0.765 &  0.255&  0.105 & 0.060 \\ \hline
 I, 4 PCs & 0.971 &0.959 & 0.843& 0.684 &  0.577&  0.497 & 0.435 \\ \hline
 P, 4 PCs recom. & 0.971 &0.920 & 0.322& 0.174 &  0.134&  0.109 & 0.088 \\ \hline
\end{tabular}
\caption{Type II error rates by signal strength for \cite{gromenko:2017}, score-based individual test statistics, and score-based predicted test statistics.} \label{tab:sim_comparison}
\end{table}

We present testing results for fully-functional test statistics in Tables \ref{tab:typeI_alt} and \ref{tab:typeII_alt}. 
In contrast to the score-based test statistics, the predicted versions of fully-functional test statistics do not improve power compared to test statistics computed at each location.

\begin{figure}
\includegraphics[width = .48\textwidth]{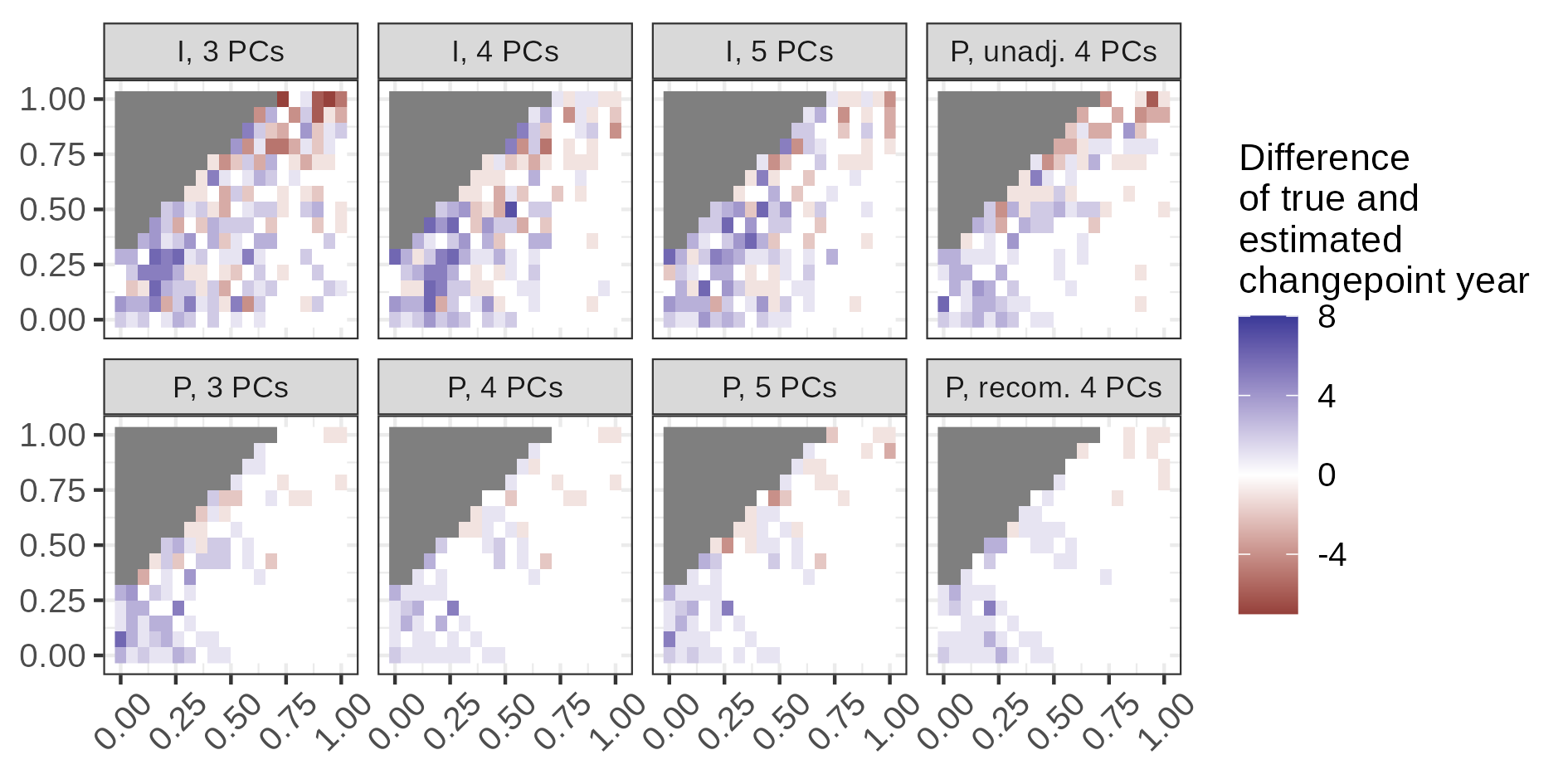}
\includegraphics[width = .48\textwidth]{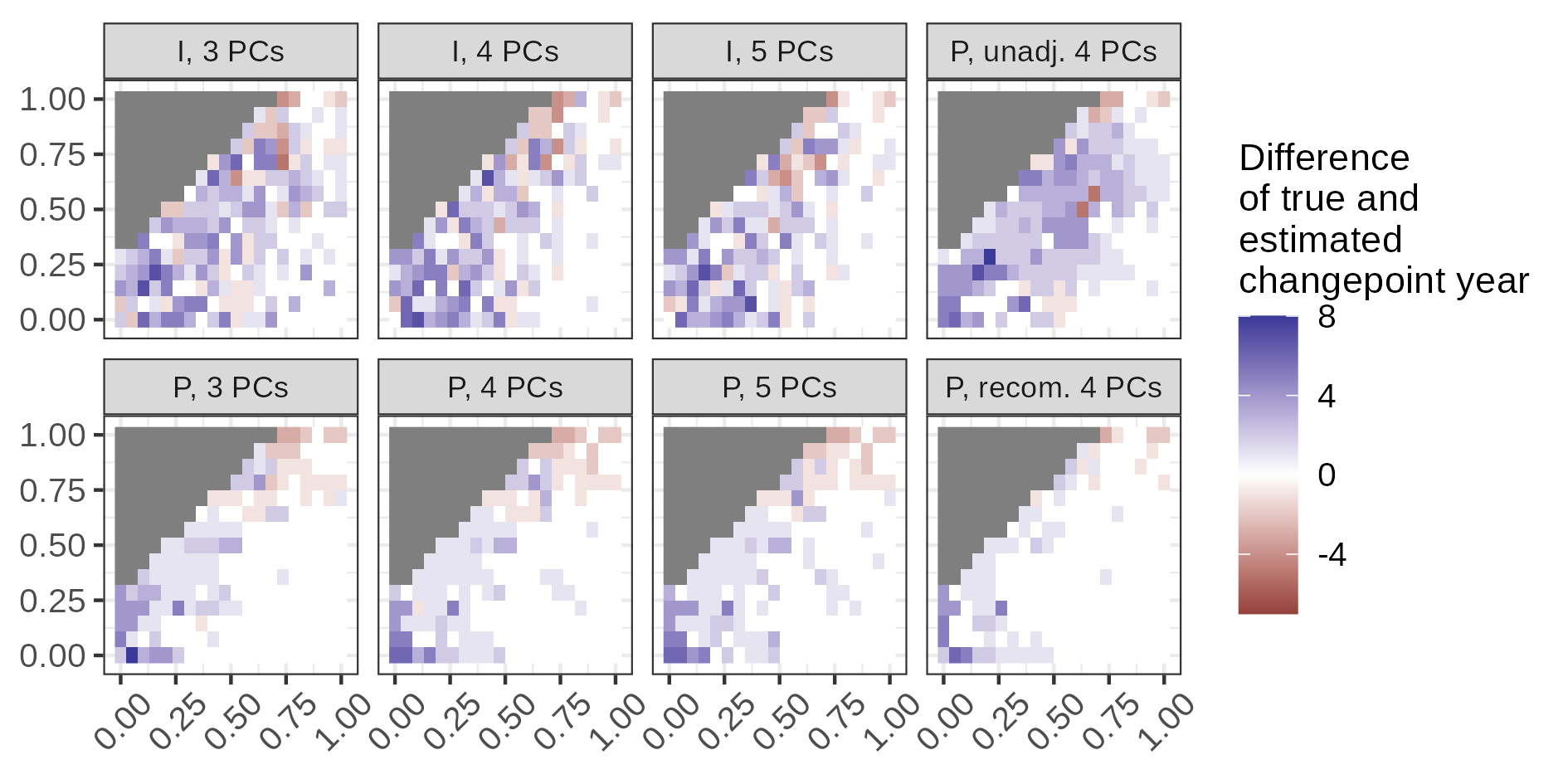}

\caption{Estimation year examples with $\eta = 10$ for score-based test statistics. (Left) A simulation without spatial dependence in $\epsilon_{k}(\mb{s}, u)$, (Right) A simulation with spatial dependence in $\epsilon_k(\mb{s},u )$. The color scale is shared across the two plots.  }\label{fig:sim_year_est_ex_score}
\end{figure}

We dive deeper into the results with two examples.
In Figure \ref{fig:sim_year_est_ex_score}, we plot the difference between estimated and true changepoints for score-based statistics for two different simulations with and without spatial dependence. 
The fully-functional results are plotted in Figure \ref{fig:sim_year_est_ex_ff} in Appendix \ref{app:simulation}. 
As expected, the highest differences between the true and estimated changepoints occur near the boundary between ``change'' and ``no change'' where the signal is weaker. 
Qualitatively, the predicted test statistics have different behavior under spatial independence versus dependence. 
In the spatially-independent case, predicting test statistics generally reduces the noise in the estimates and does not leave behind substantial spatial patterns.
In the spatially-dependent case, however, the predicted test statistic errors have more spatial coherency and patterns; this suggests that the spatial dependence in the errors is somewhat filtered into the estimates of the changepoint, reinforcing the challenge of separating changepoint-related spatial coherency and noise-related spatial coherency.

We also give an example of the estimated changepoint error for simulations with and without spatial dependence in Figure \ref{fig:sim_year_est_ex_ff} for fully-functional test statistics. 

\begin{table}\footnotesize\centering
\begin{tabular}{|c||c|c|c|c|c|c|}\hline
Type I Errors& I & P, Null  & P, Alt &  P, BH & P, Bonf & P, All  \\ \hline
FPR, raw p-values & {\color{red} 0.107}& 0.001 &{\color{orange} 0.052} &0.023& 0.006 & {\color{orange} 0.096}  \\ \hline
FDR, using BH & 0.043 & 0.000 &0.035  & 0.022&0.007 & 0.038  \\ \hline
FWER, using Bonf &{\color{red} 0.443} & 0.000 & {\color{red} 0.303}  & {\color{red} 0.226}  & {\color{red} 0.137}  & {\color{red} 0.276}   \\ \hline
\end{tabular}
\caption{Fully-functional Type I error rates over all simulations, signal strengths, and dependence structures, with nominal level $\alpha = 0.05$.  ``I'' refers to individual test statistics, and ``P'' columns are based on predicted test statistics under different hypotheses. Colors: {\color{black} Type I Error $\leq 0.05$}, {\color{orange} $0.05 <$Type I Error$\leq 0.10$}, {\color{red} Type I Error $>0.10$}. }\label{tab:typeI_alt}
\vskip .5cm

\hskip .17cm\begin{tabular}{|c||c|c|c|c|c|c|}\hline\centering
Type II Errors& I & P, Null  & P, Alt & P, BH & P, Bonf & P, All  \\ \hline
FNR, raw p-values & {\color{red} 0.613} &  0.810 & {\color{orange} 0.680} & 0.726 &  0.787 &  {\color{orange} 0.631}  \\ \hline
FNR, using BH & 0.792 &  0.906 & 0.719 &  0.743 &  0.797 &  0.710 \\ \hline
FNR, using Bonf & {\color{red} 0.781} & 0.989& {\color{red}0.772}&  {\color{red} 0.780} &  {\color{red}0.812} & {\color{red} 0.783}  \\ \hline
\end{tabular}
\caption{Fully-functional Type II error rates over all simulations, signal strengths, and dependence structures, with nominal level $\alpha = 0.05$. Colors: {\color{black} Type I Error $\leq 0.05$}, {\color{orange} $0.05 <$Type I Error$\leq 0.10$}, {\color{red} Type I Error $>0.10$}. }\label{tab:typeII_alt}
\end{table}

\begin{figure}
\includegraphics[width = .48\textwidth]{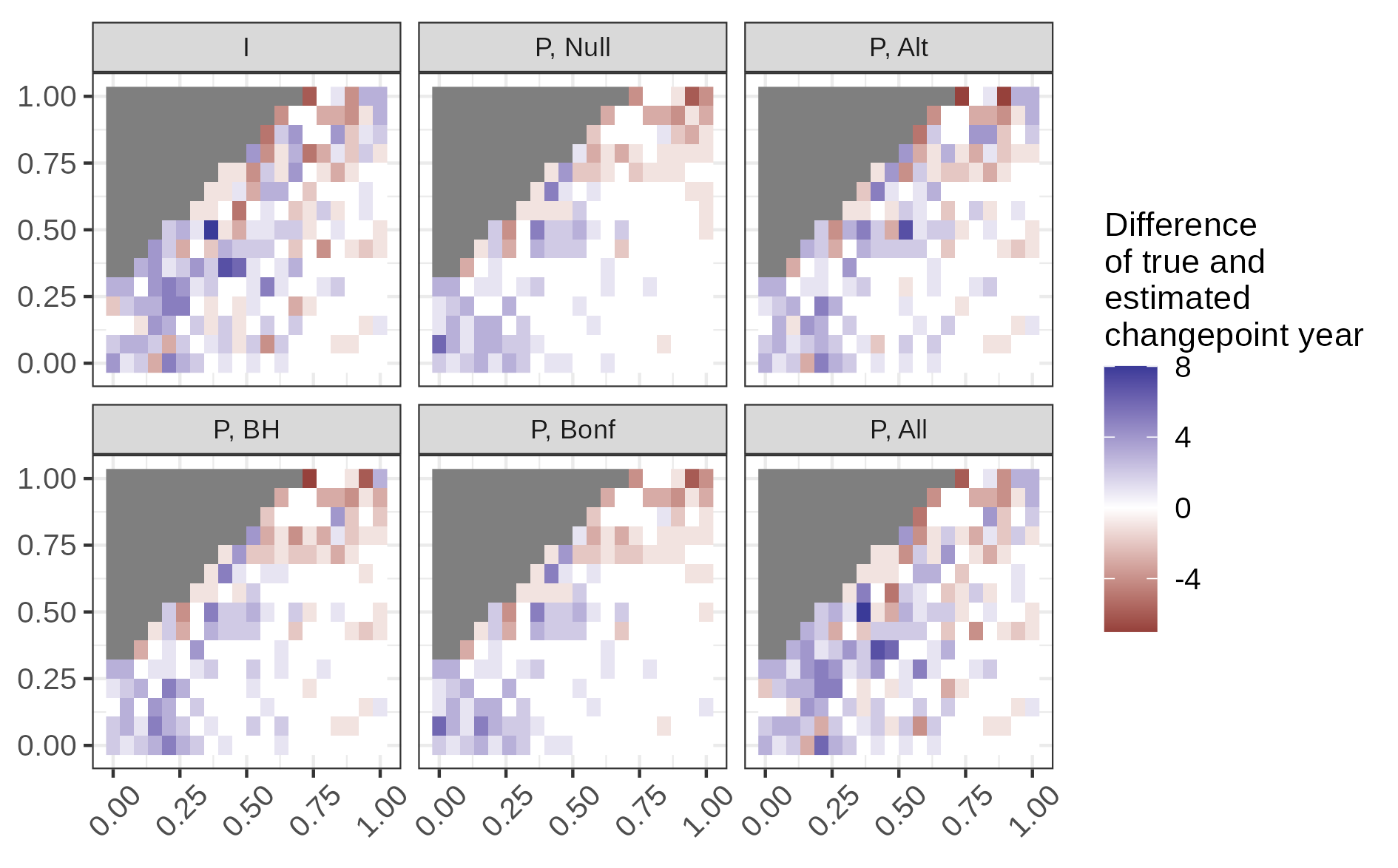}
\includegraphics[width = .48\textwidth]{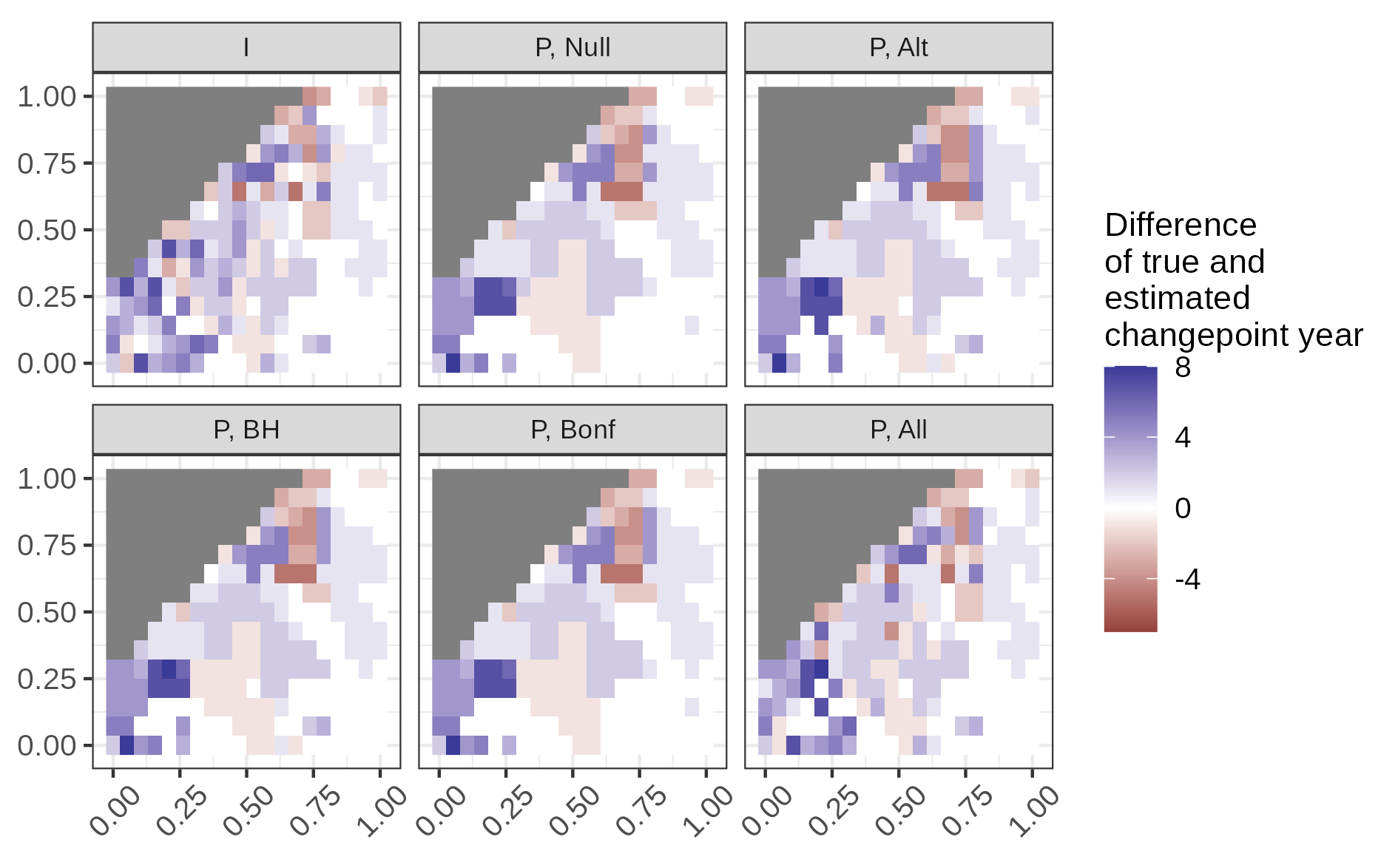}
\caption{Estimation year examples with $\eta = 10$ for fully-functional test statistics. (Left) A simulation without spatial dependence in $\epsilon_{k}(\mb{s}, u)$, (Right) A simulation with spatial dependence in $\epsilon_k(\mb{s},u )$. The color scale is shared across the two plots as well as across Figure \ref{fig:sim_year_est_ex_score}. }\label{fig:sim_year_est_ex_ff}
\end{figure}

\section{Additional HSW++ results}\label{app:hsw}

\begin{figure}
\includegraphics[width = .98\textwidth]{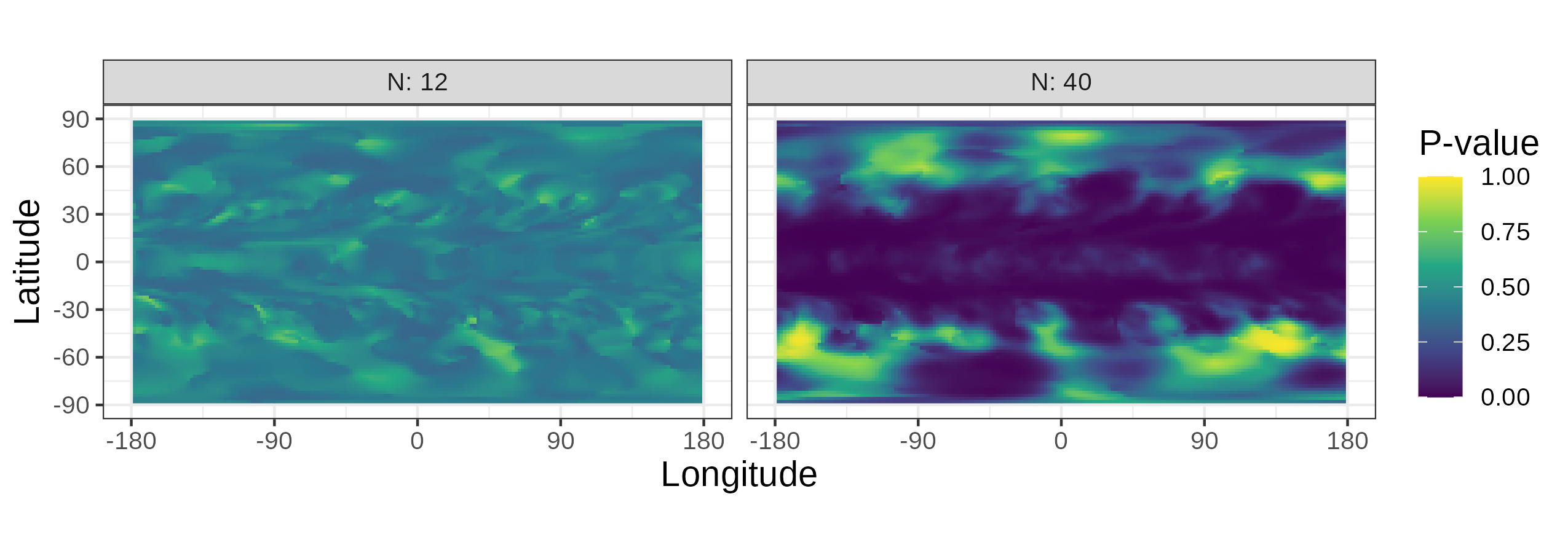}
\includegraphics[width = .98\textwidth]{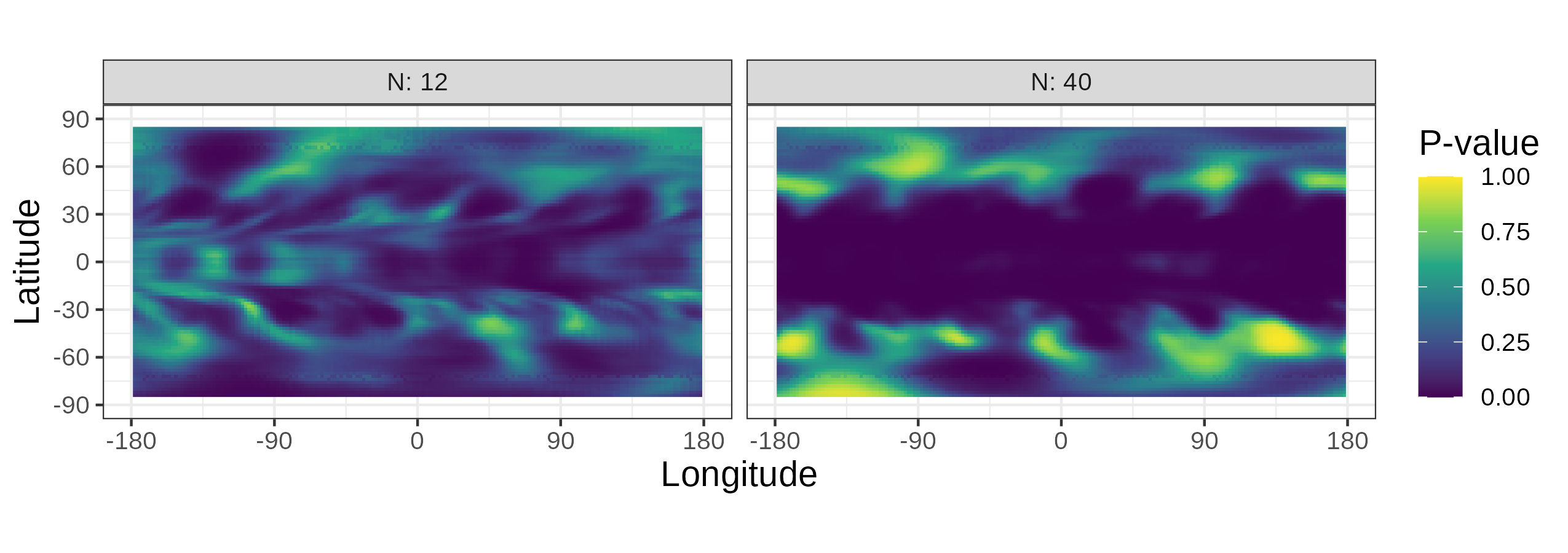}
\includegraphics[width = .98\textwidth]{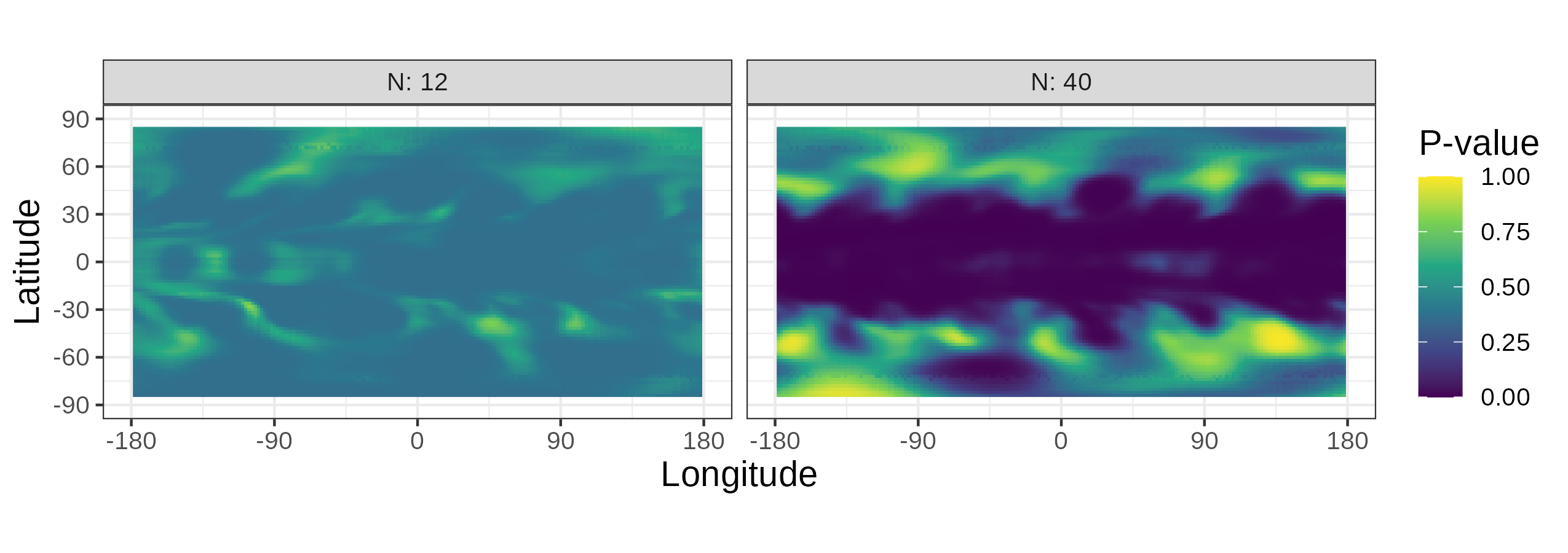}
\caption{Comparison of HSW++ results with different choices of $N$. (Top) Epidemic changepoint p-values using score-based tests statistics computed at each location. (Middle) Epidemic changepoint p-values using score-based tests statistics and spatially predicted scores. 
(Bottom) Epidemic changepoint p-values using score-based tests statistics, spatially predicted scores, and a Benjamini-Hochberg p-value correction. 
}\label{fig:HSW_N_comparison}
\end{figure}

In Figure \ref{fig:HSW_N_comparison} we compare HSW++ power results for the first ensemble member using a different value for $N$. 
We see that using individual test statistics has substantial power for $N=40$ but not $N=15$. 
For spatially-predicted test statistics, both values of $N$ have areas with smaller p-values. 
However, once one adjusts for multiple testing, only the $N=40$ case has power to detect changepoints. 

\section{Additional temperature MERRA-2 results}\label{app:merra2_strat}

\begin{figure}

\includegraphics[width = .48\textwidth]{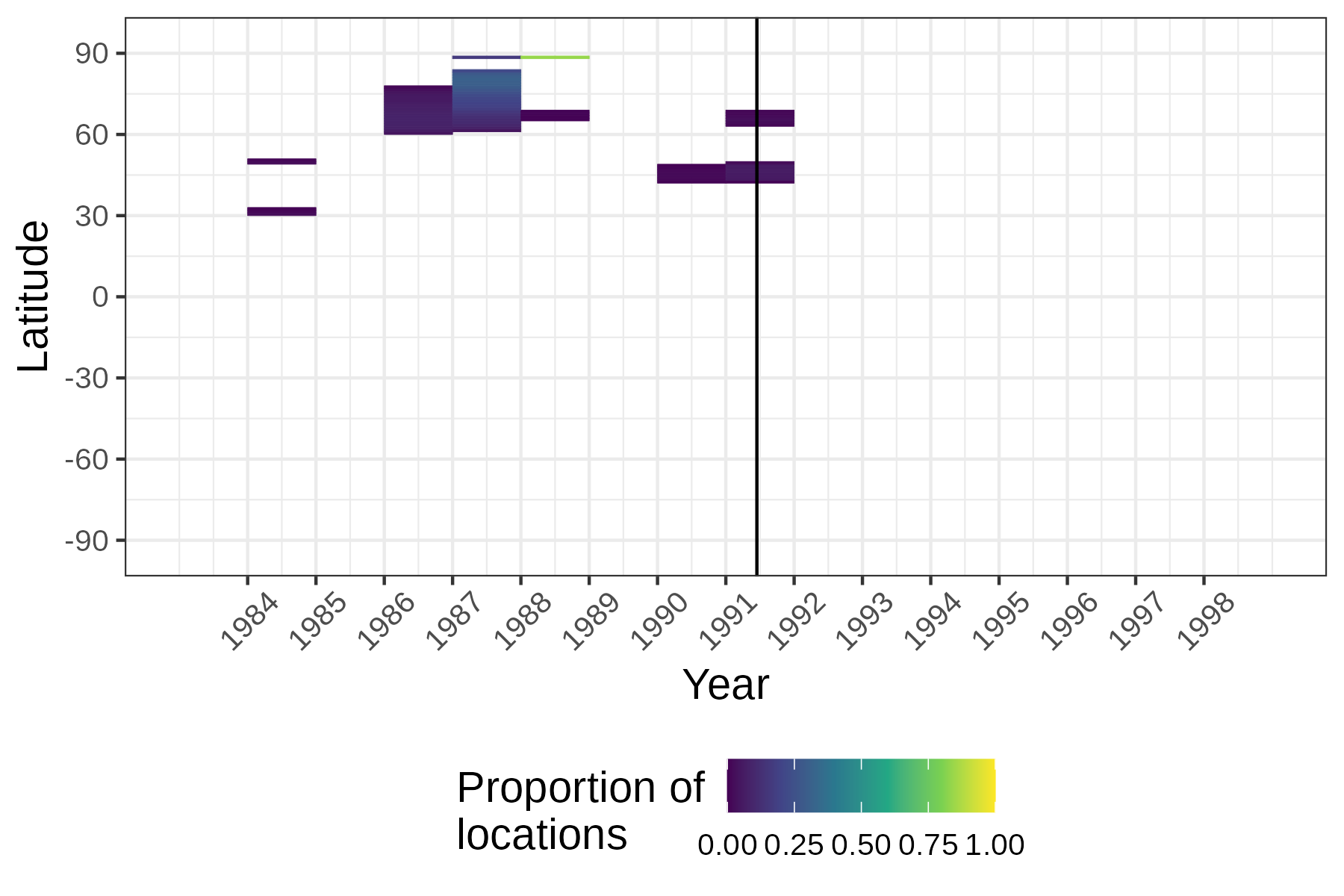}
\includegraphics[width = .48\textwidth]{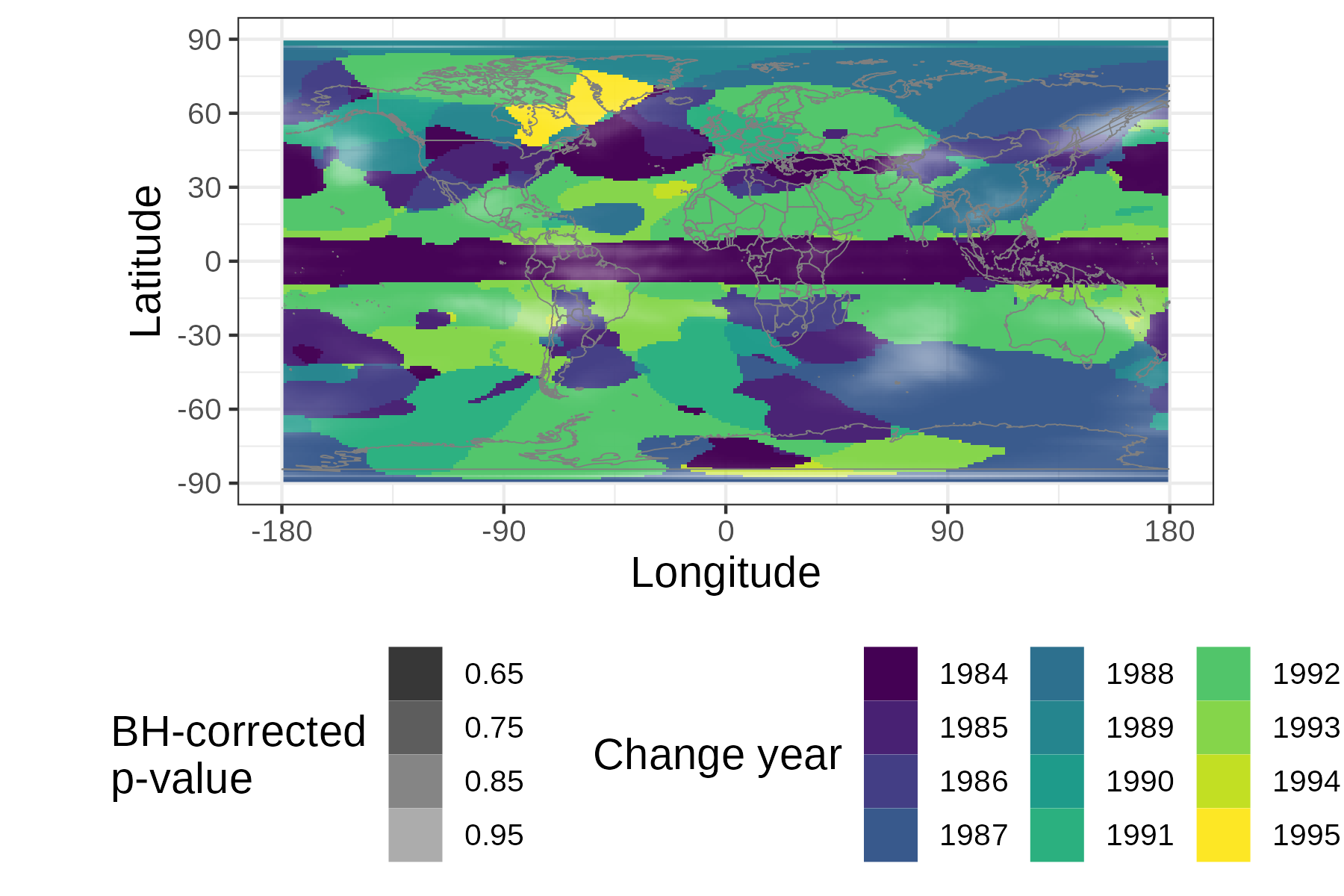}

\includegraphics[width = .48\textwidth]{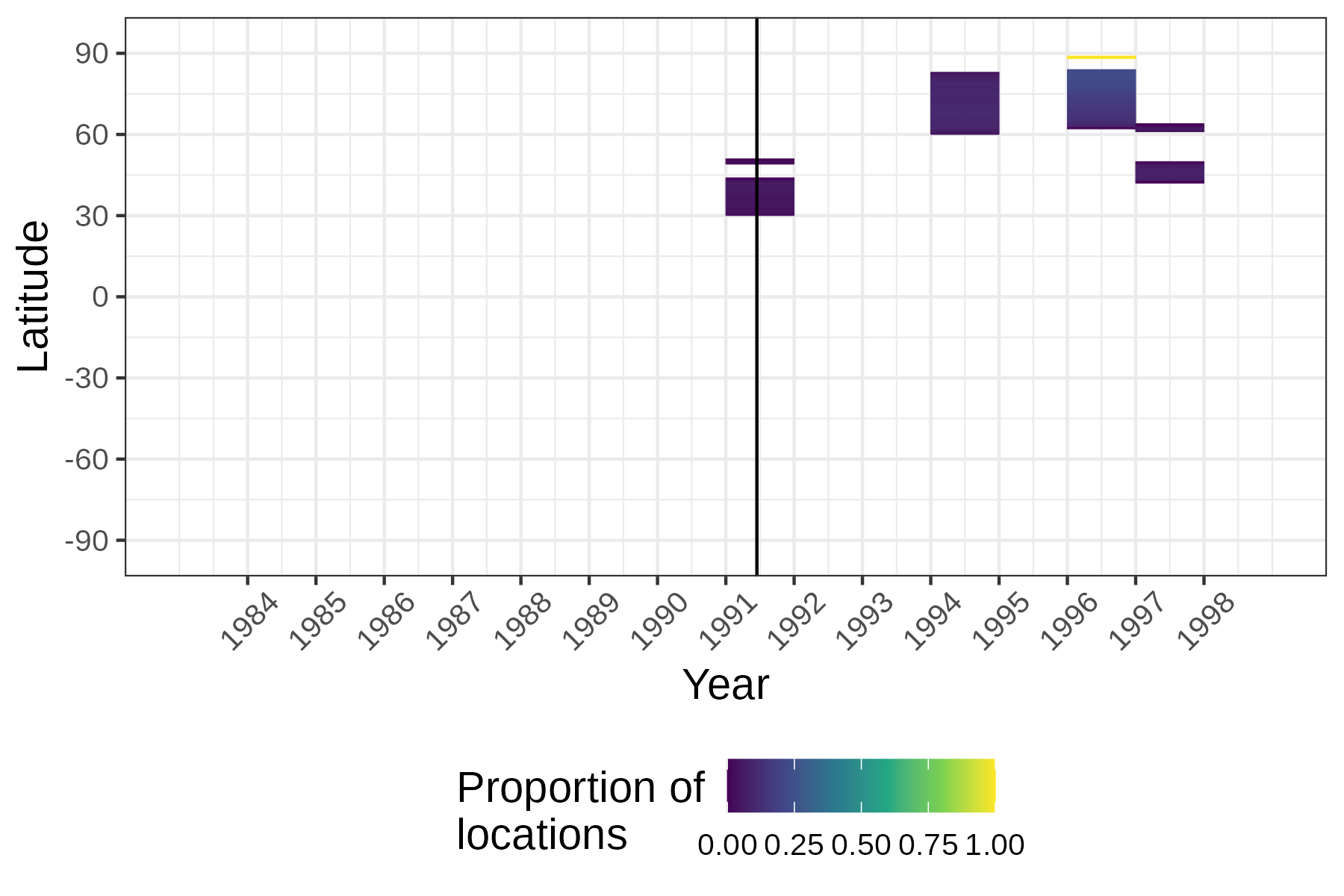}
\includegraphics[width = .48\textwidth]{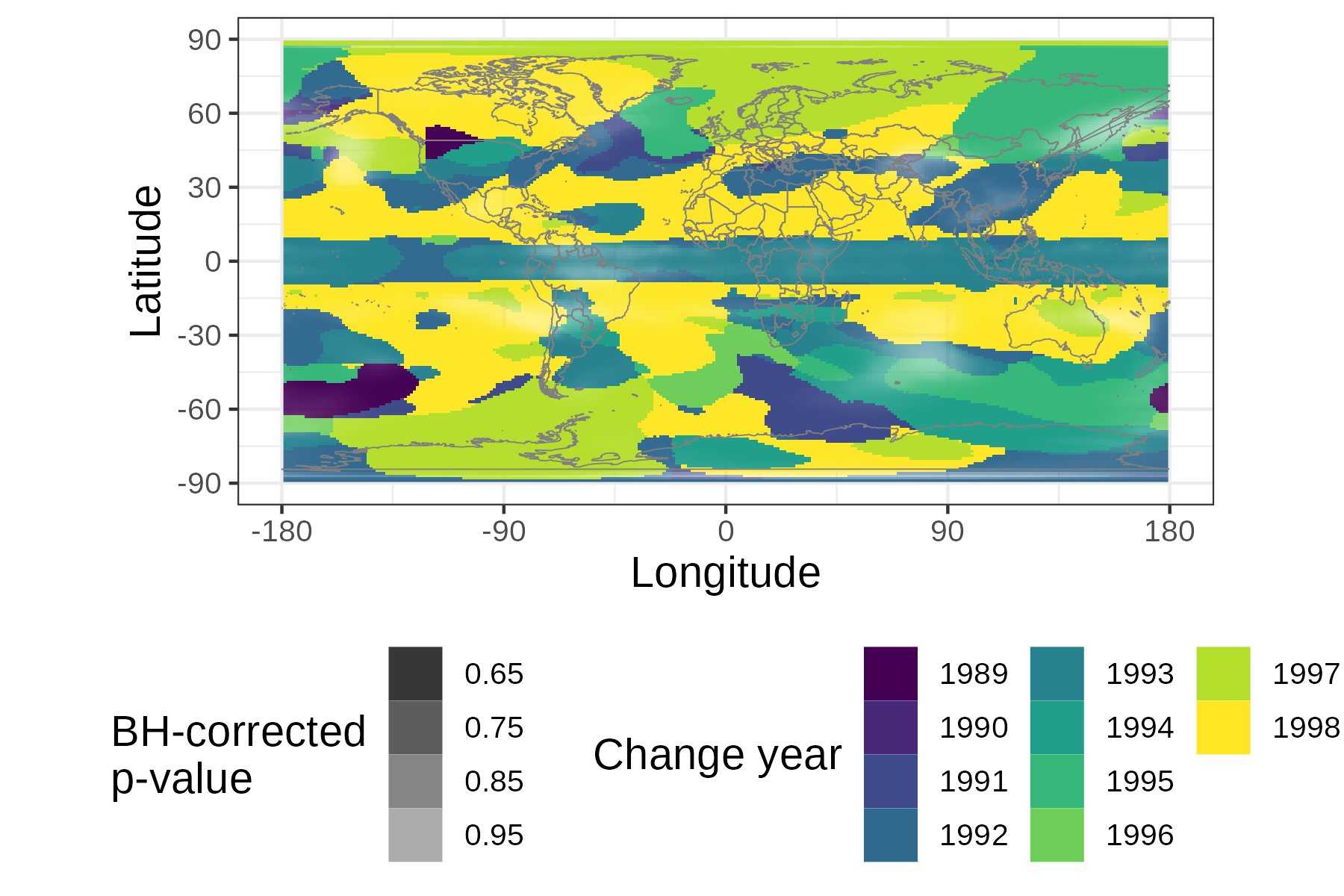}

\caption{MERRA-2 results under the epidemic model (Left) For each latitude, we plot the proportion of locations (of 288 different longitudes) that had a change detected during that month (Top) $\tau_1(\mb{s})$ and (Bottom) $\tau_2(\mb{s})$. (Right) Detected changepoint year for (Top) $\tau_1(\mb{s})$ and (Bottom) $\tau_2(\mb{s})$ using unadjusted p-values.
 The vertical line represents the time of the eruption of Mt.\ Pinatubo. }\label{fig:MERRA2_strat_more_epidemic}

\end{figure}

\begin{figure}

\includegraphics[width = .48\textwidth]{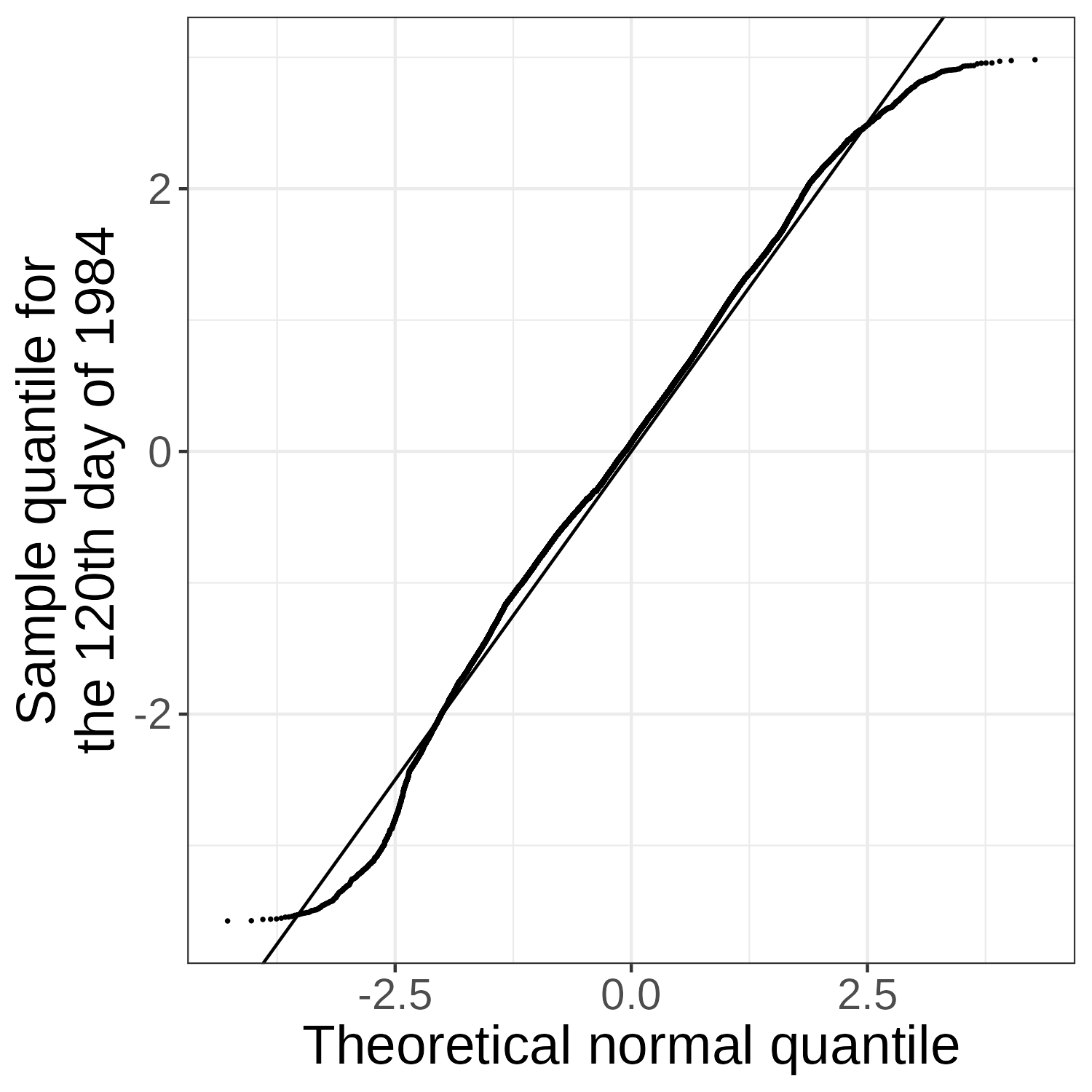}
\includegraphics[width = .48\textwidth]{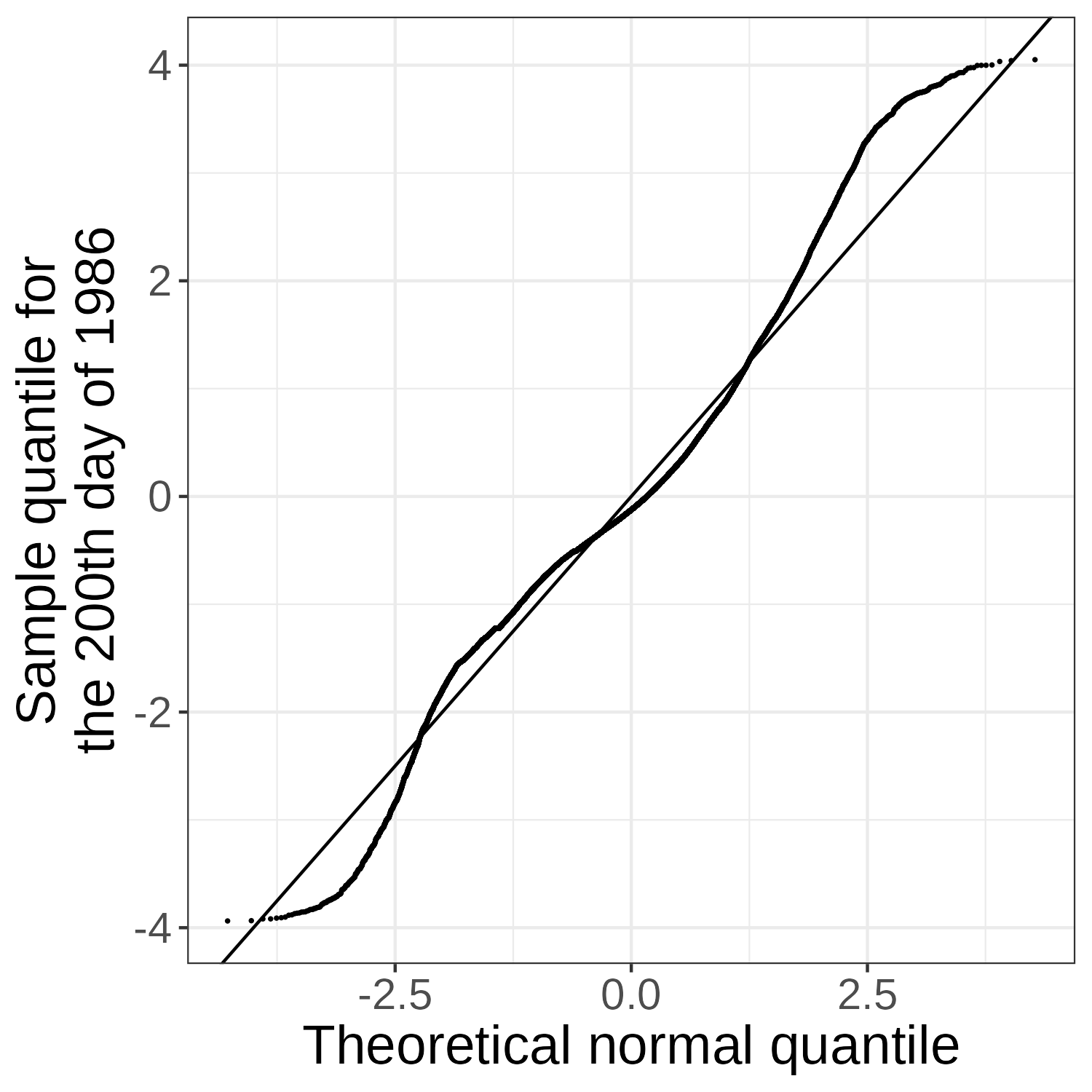}

\caption{Quantile-Quantile plot examples to evaluate the normality assumption for MERRA-2 stratospheric temperature during two different days at all locations. Even though we expect different locations to have different variances, the quantiles of the distributions match up relatively well with the normal distribution. There is some mismatch towards the tails of the distributions, yet this is expected when not adjusting for variances and working with quantiles more than 2.5 standard deviations away from the mean. }\label{fig:MERRA2_strat_qq}

\end{figure}

In Figure \ref{fig:MERRA2_strat_more_epidemic}, we plot more epidemic results; a relatively small number of changepoints are detected. 
In Figure \ref{fig:MERRA2_strat_qq}, we evaluate the normality assumption. 

%
%

\section{Aerosol optical depth MERRA-2 results}\label{app:merra2_AOD}

\begin{figure}
\includegraphics[width = .48\textwidth]{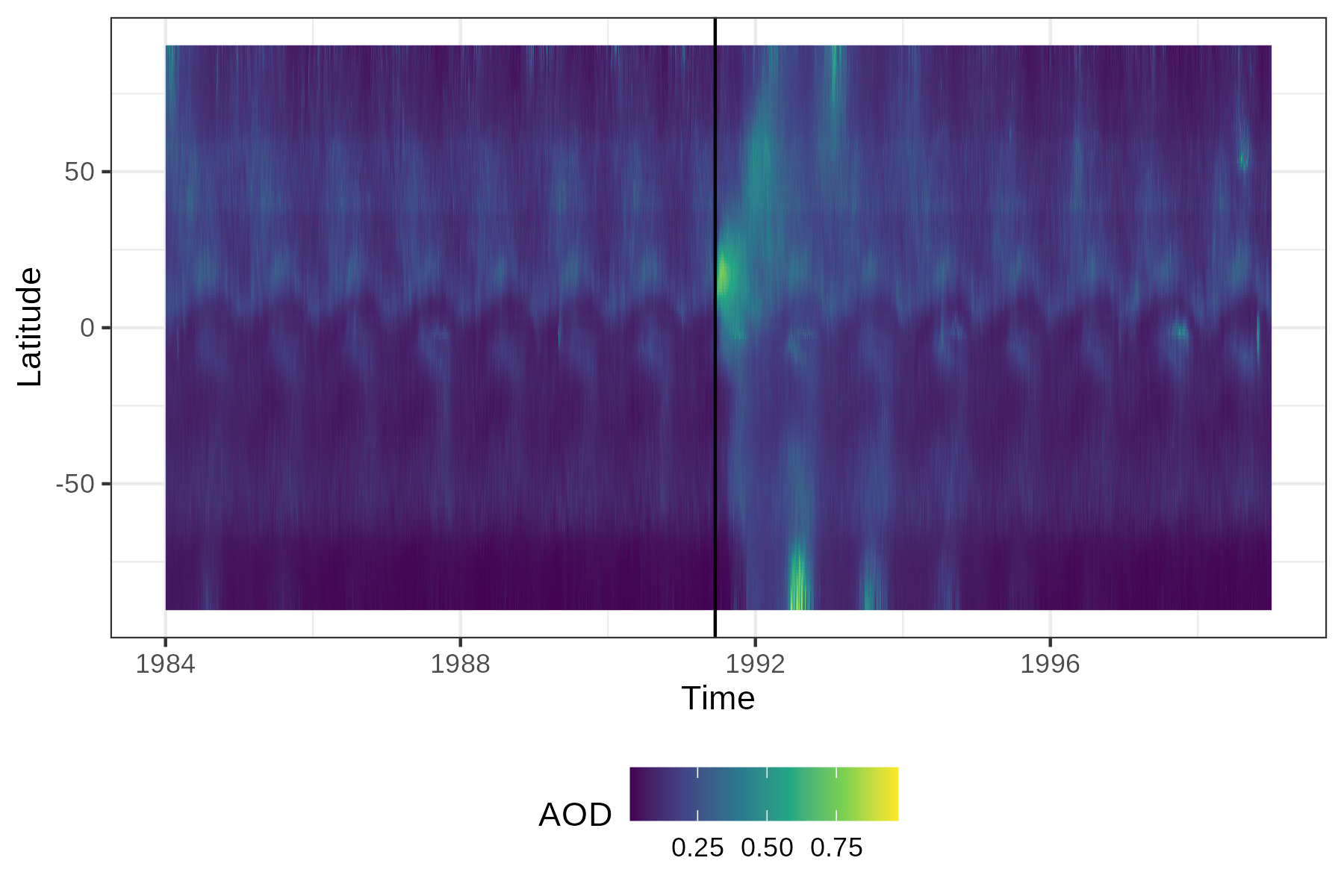}
\includegraphics[width = .48\textwidth]{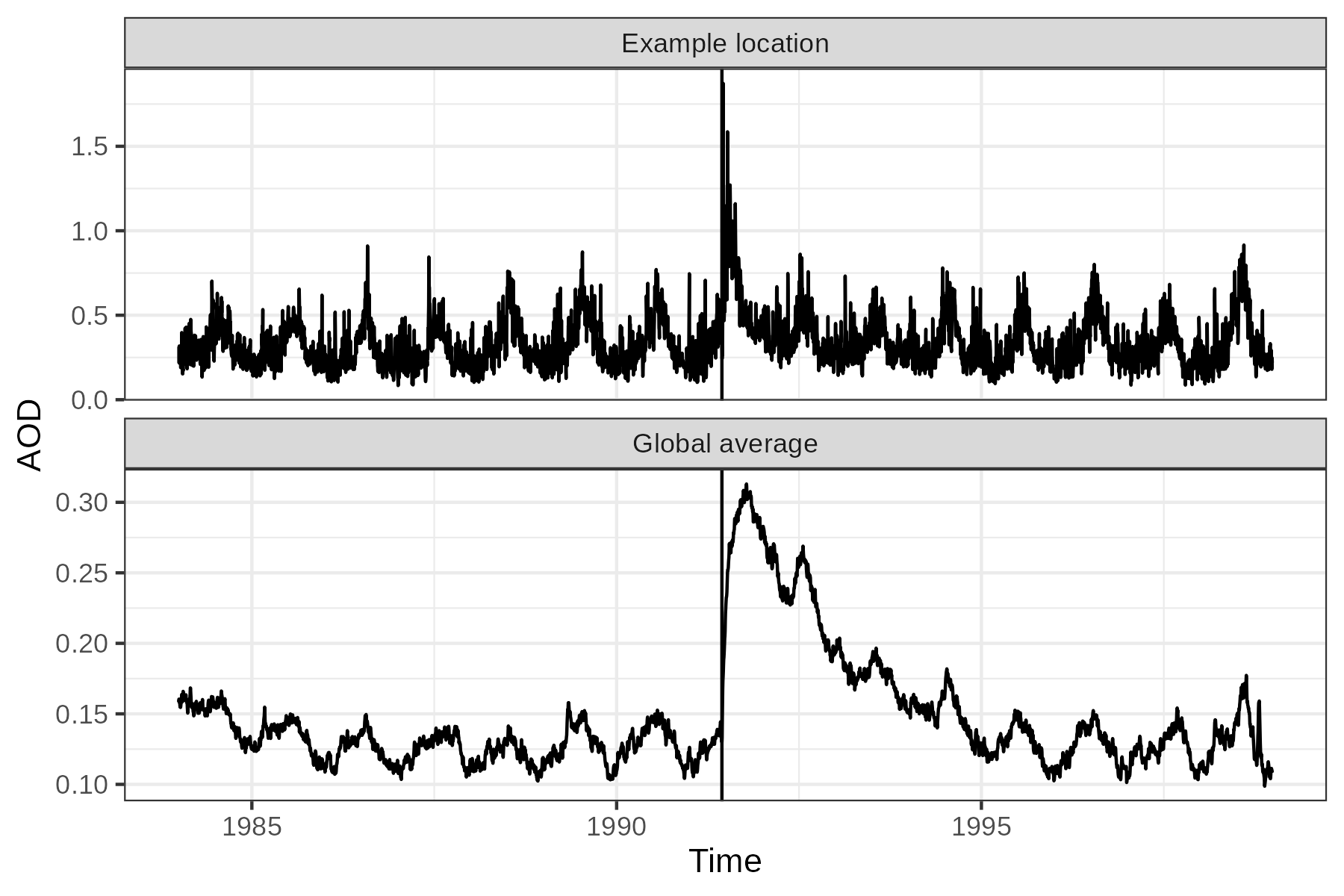}

\caption{AOD (Left) Longitudinal average of MERRA-2 reanalysis of aerosol optical depth. The vertical line represents the time of the eruption of Mt.\ Pinatubo. (Top Right) Example of MERRA-2 reanalysis aerosol optical depth data for $37.5^\circ \textrm{E}$ and $18^\circ \textrm{N}$. (Bottom Right) Globally-averaged AOD time series.  }\label{fig:MERRA2_data_AOD}

\end{figure}

\begin{figure}
\includegraphics[width = .48\textwidth]{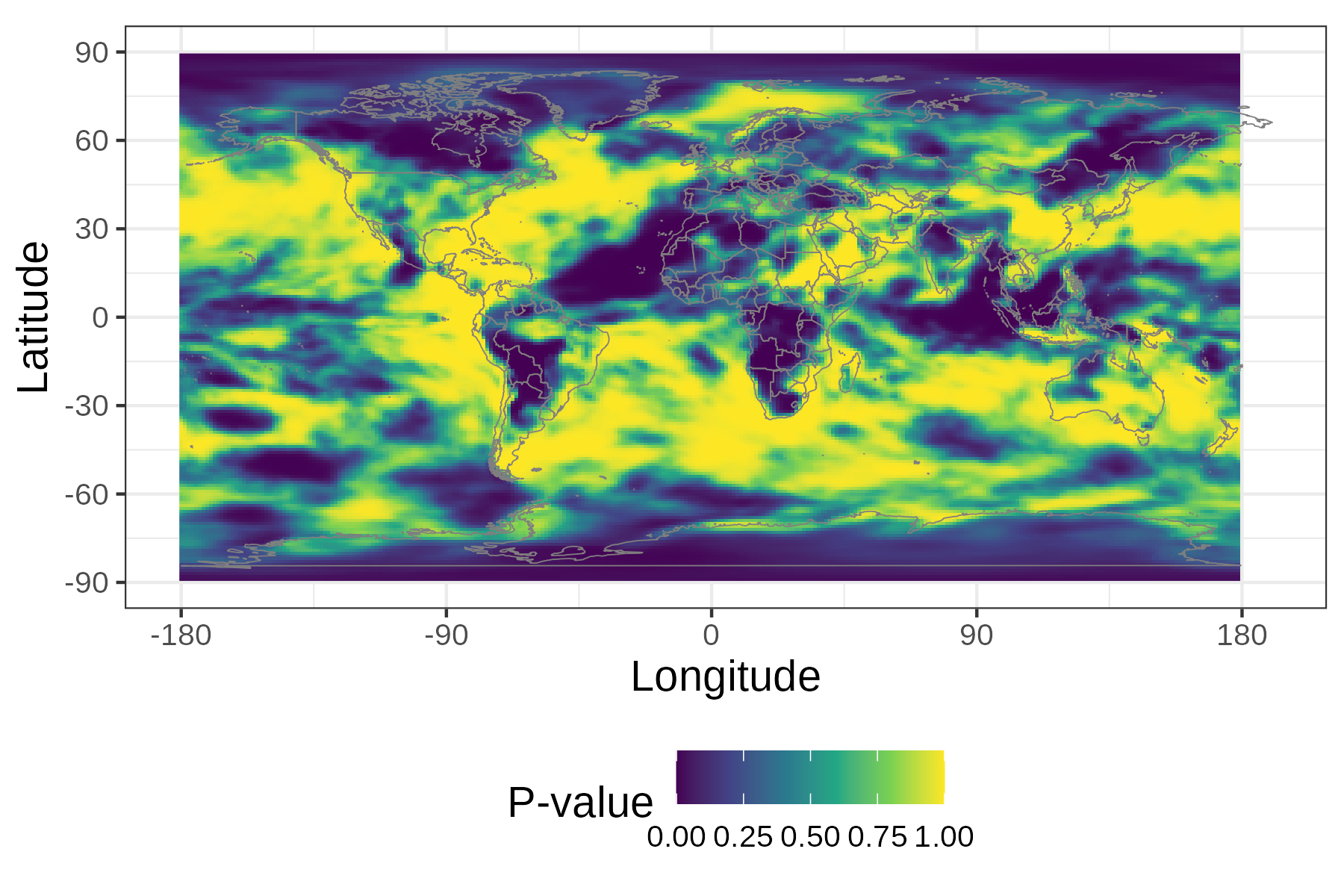}
\includegraphics[width = .48\textwidth]{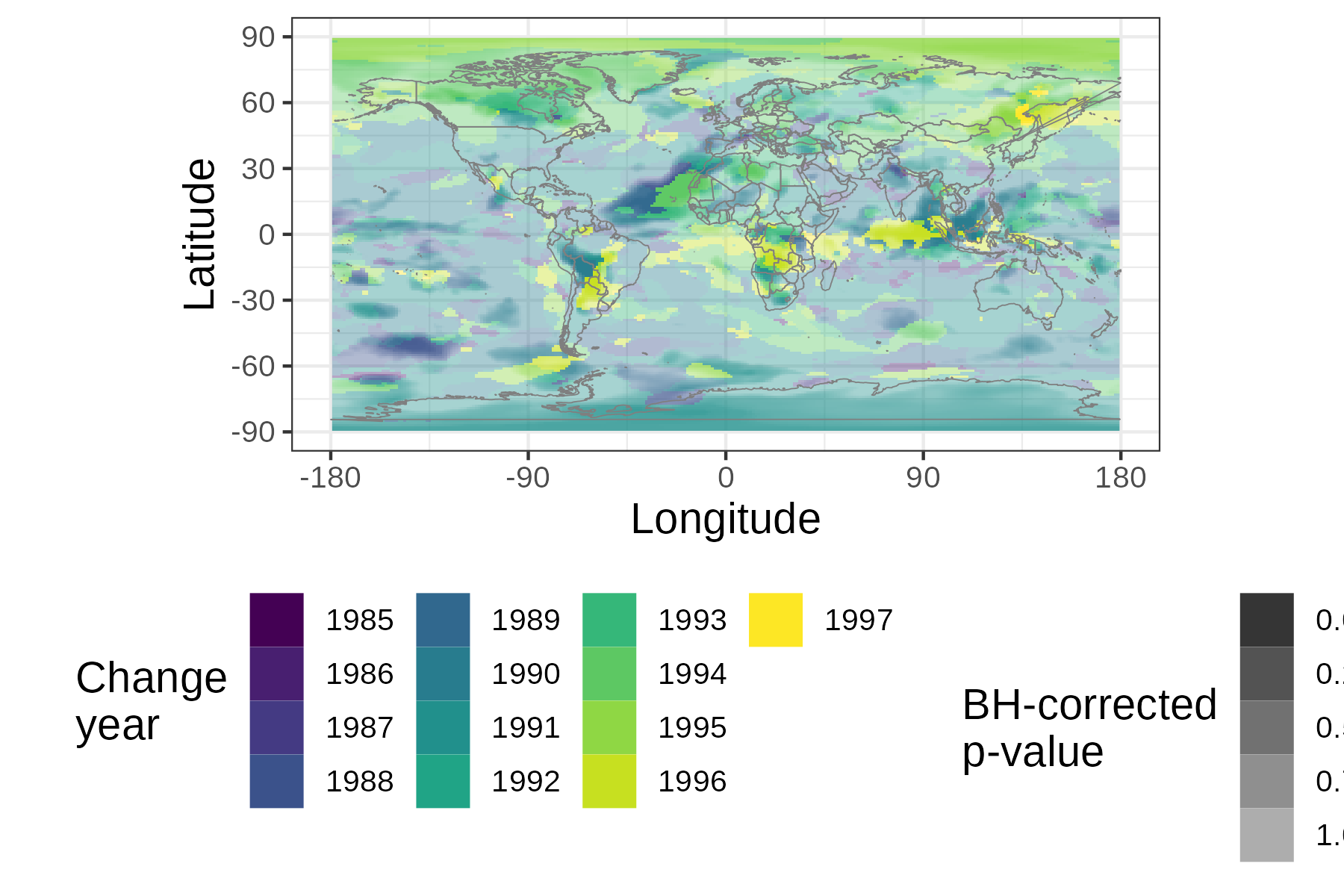}

\includegraphics[width = .48\textwidth]{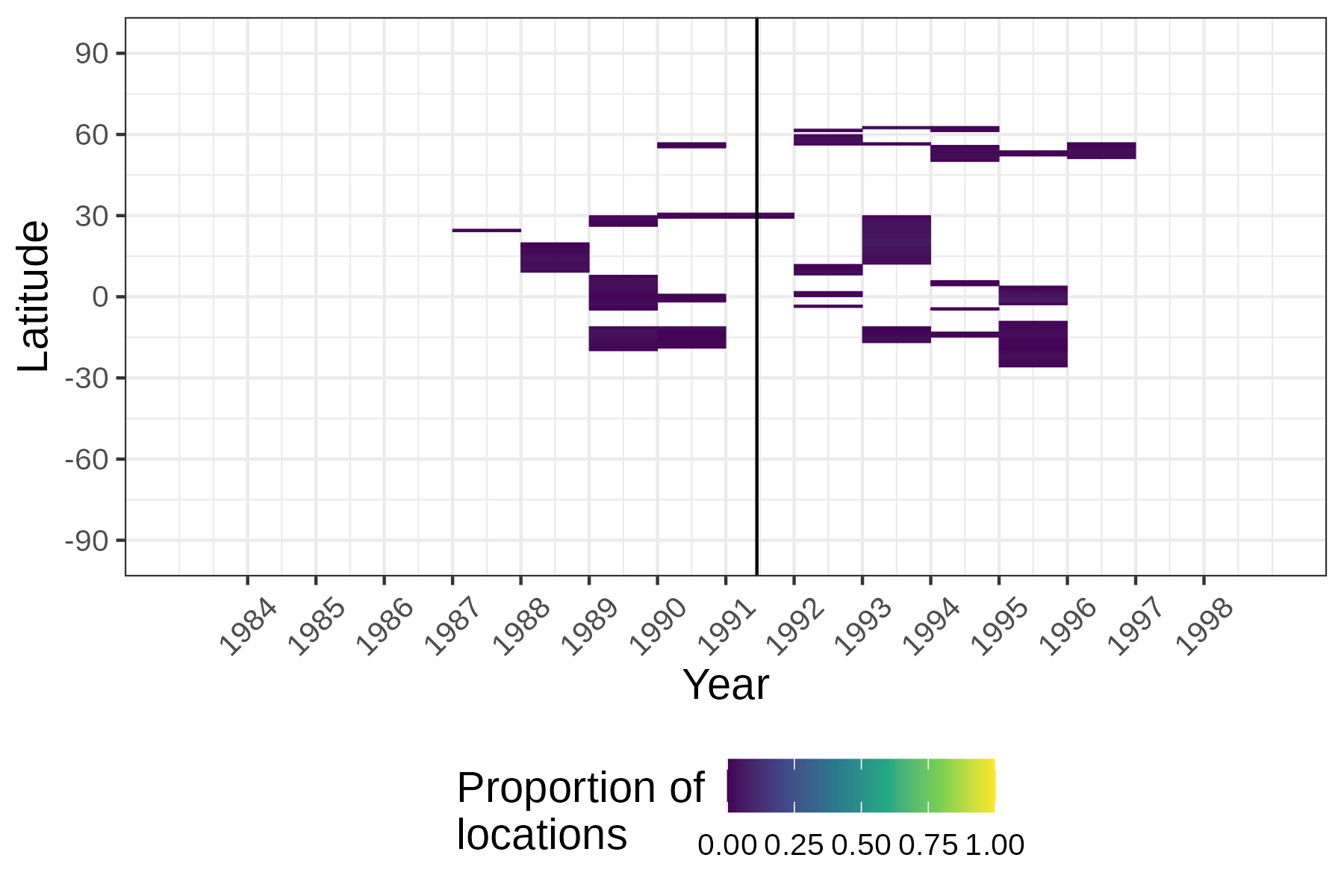}

\caption{AOD MERRA-2 results under the AMOC model (Top Left) Benjamini-Hochberg-corrected p-values (Top Right) Detected changepoint years using Benjamini-Hochberg-corrected p-values. 
(Bottom) For each latitude, we plot the proportion of locations (of 288 different longitudes) that had a change detected during that month using Benjamini-Hochberg-corrected p-values. The vertical line represents the time of the eruption of Mt.\ Pinatubo. }\label{fig:merra2_amoc_AOD}

\end{figure}

\begin{figure}
\includegraphics[width = .48\textwidth]{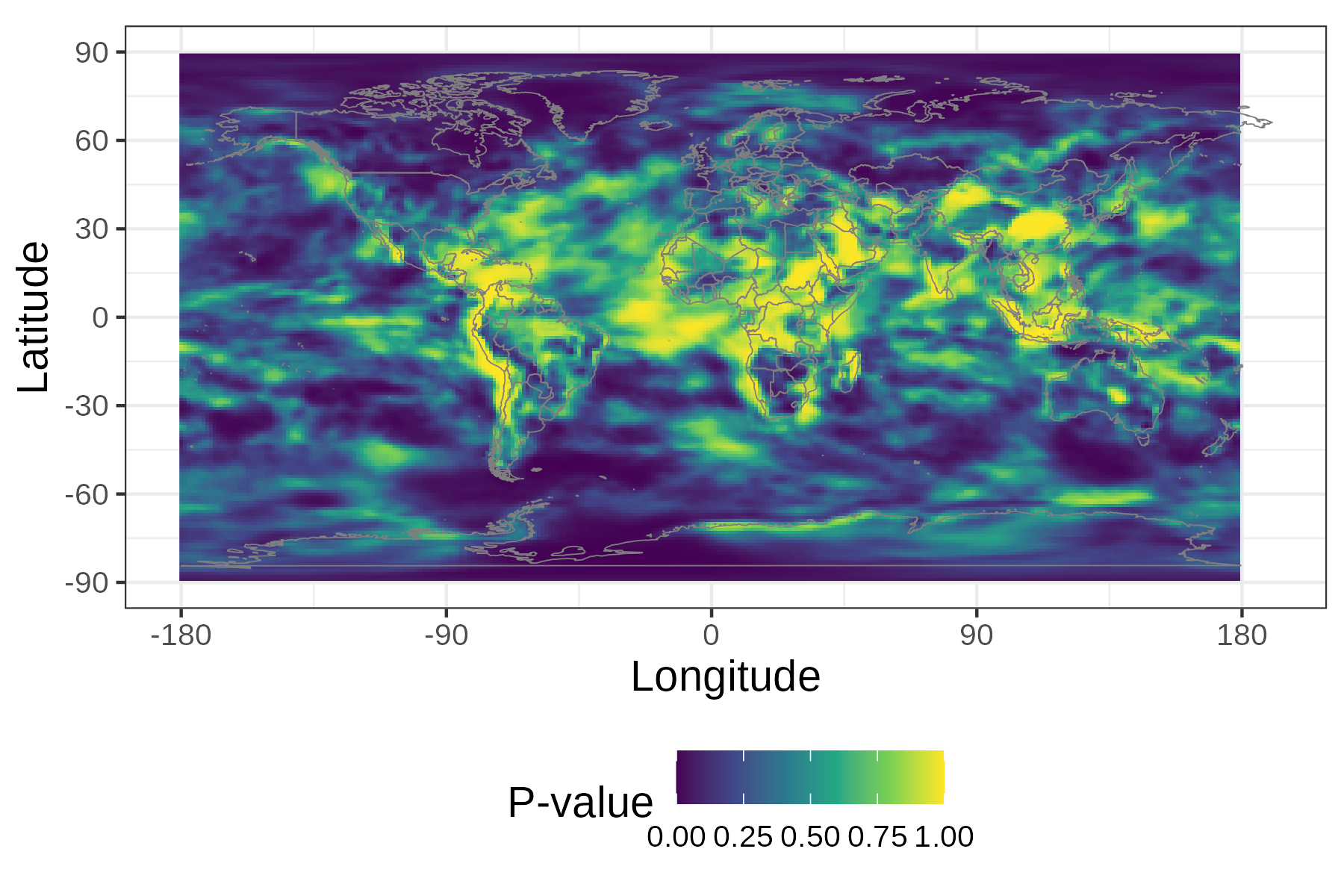}
\includegraphics[width = .48\textwidth]{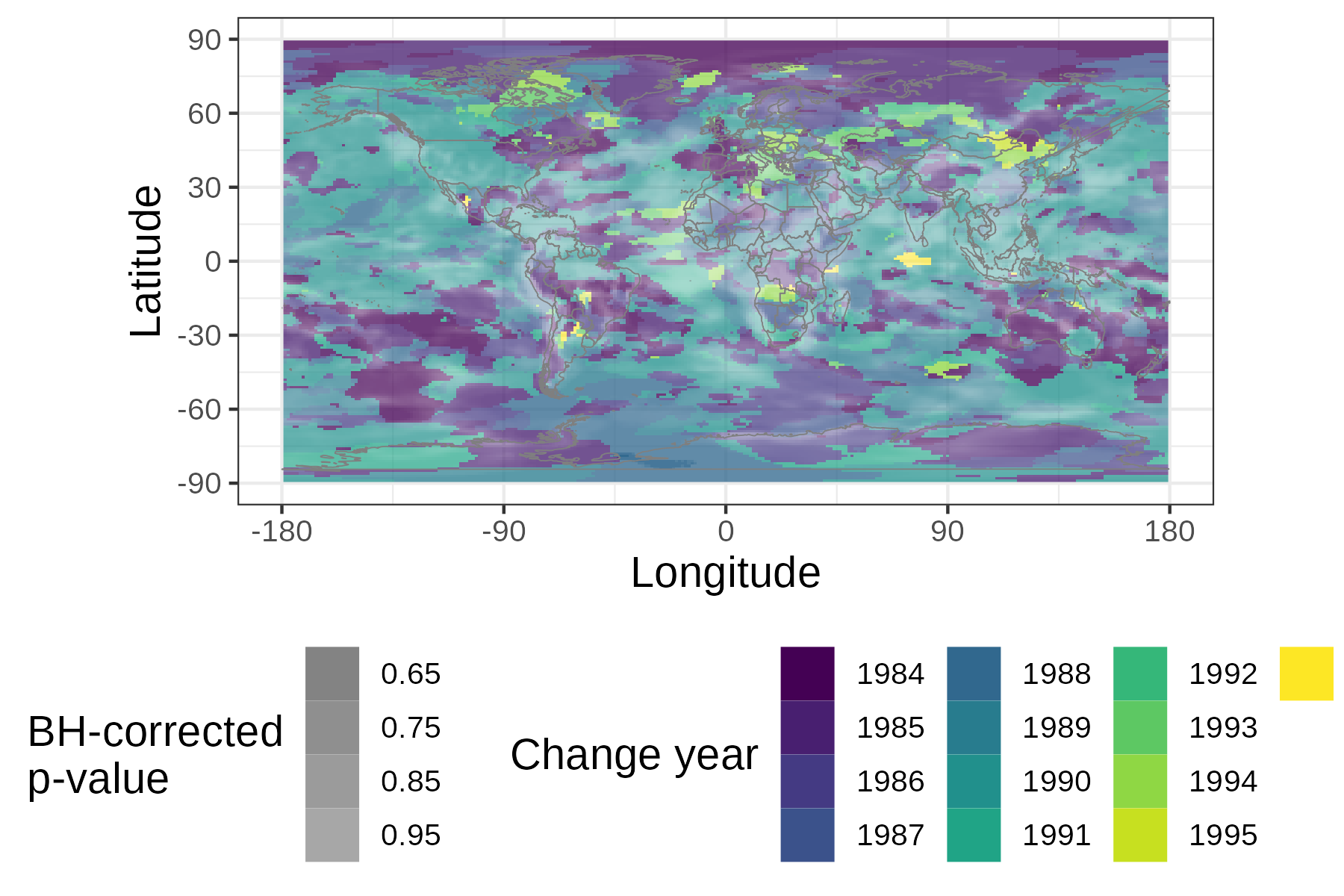}

\includegraphics[width = .48\textwidth]{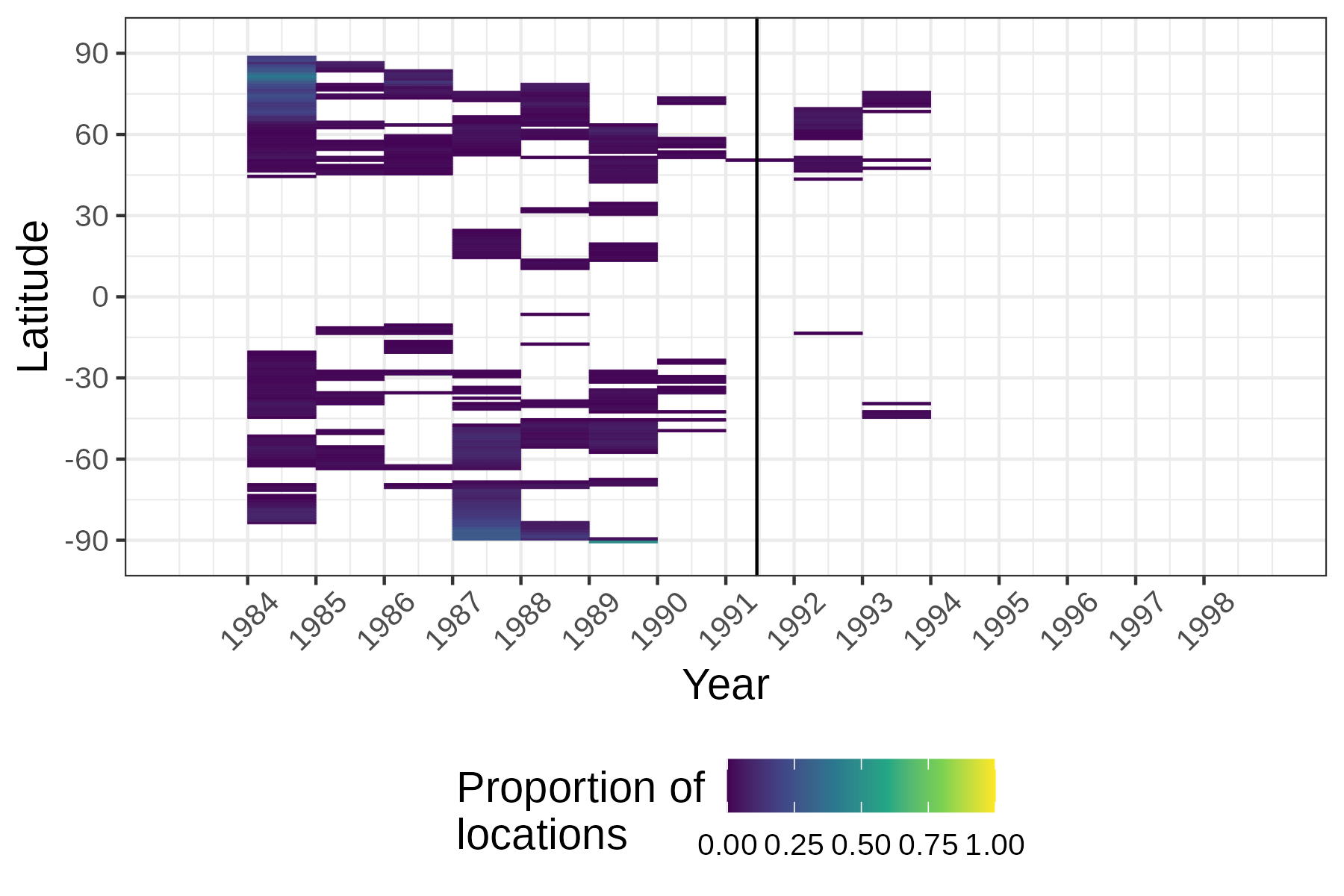}
\includegraphics[width = .48\textwidth]{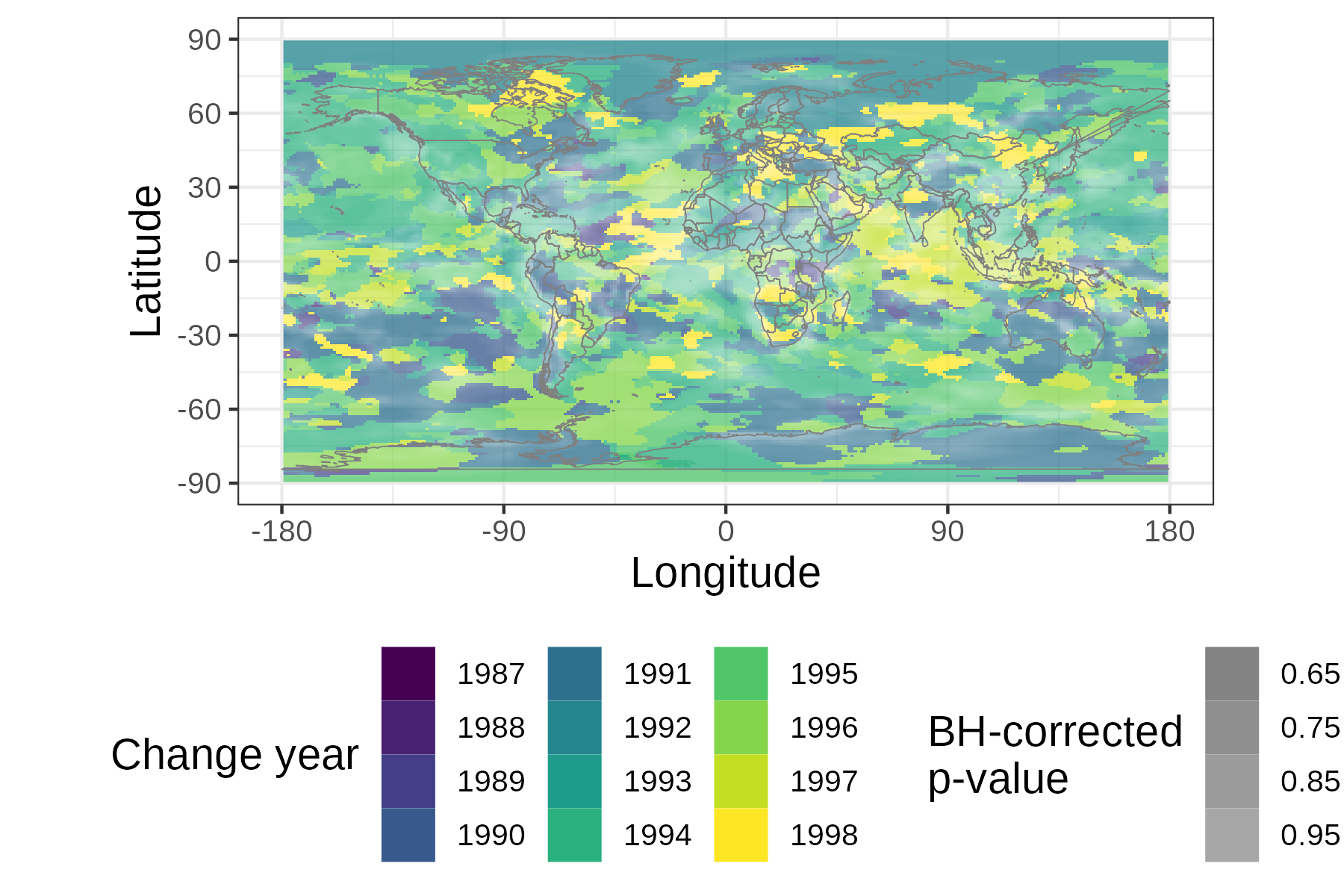}

\includegraphics[width = .48\textwidth]{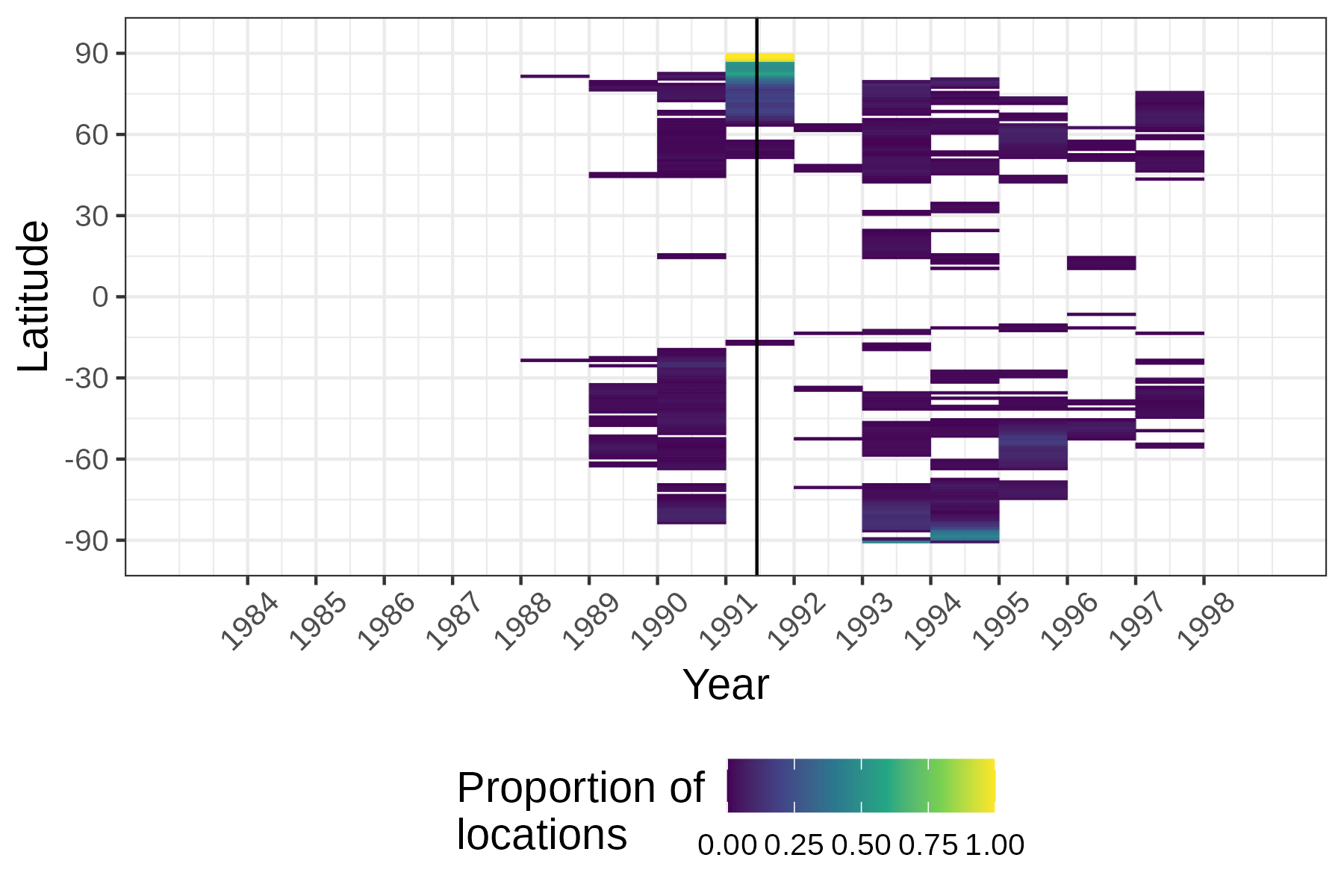}

\caption{AOD MERRA-2 results under the epidemic model (Top Left) Uncorrected p-values (Right) Detected changepoint year for (Top) $\tau_1(\mb{s})$ and (Bottom) $\tau_2(\mb{s})$.
(Middle and Bottom Left) For each latitude, we plot the proportion of locations (of 288 different longitudes) that had a change detected during that month (Middle) $\tau_1(\mb{s})$ and (Bottom) $\tau_2(\mb{s})$. The vertical line represents the time of the eruption of Mt.\ Pinatubo. }\label{fig:merra2_epidemic_AOD}

\end{figure}

\begin{figure}
\includegraphics[width = .48\textwidth]{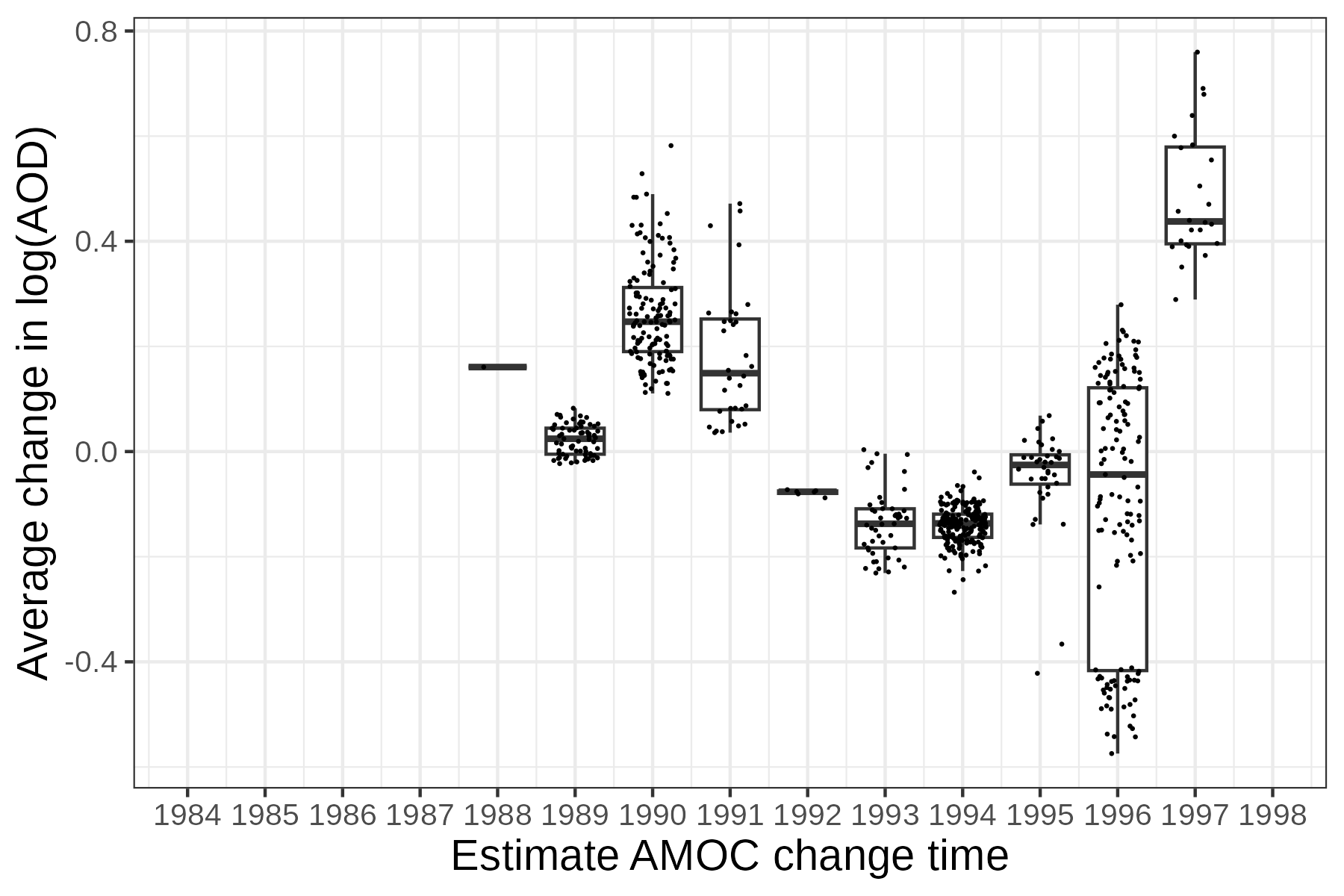}
\includegraphics[width = .48\textwidth]{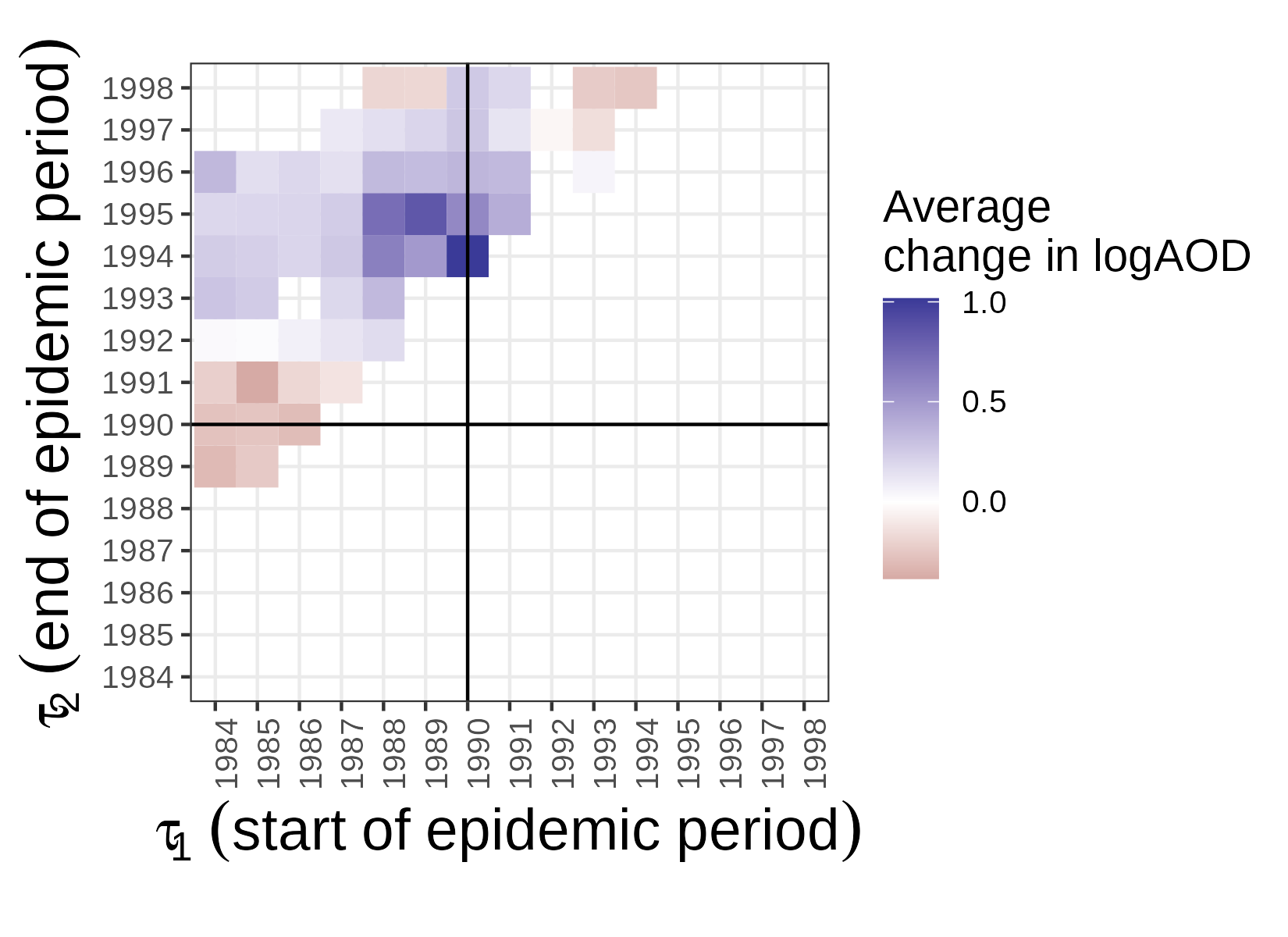}

\includegraphics[width = .48\textwidth]{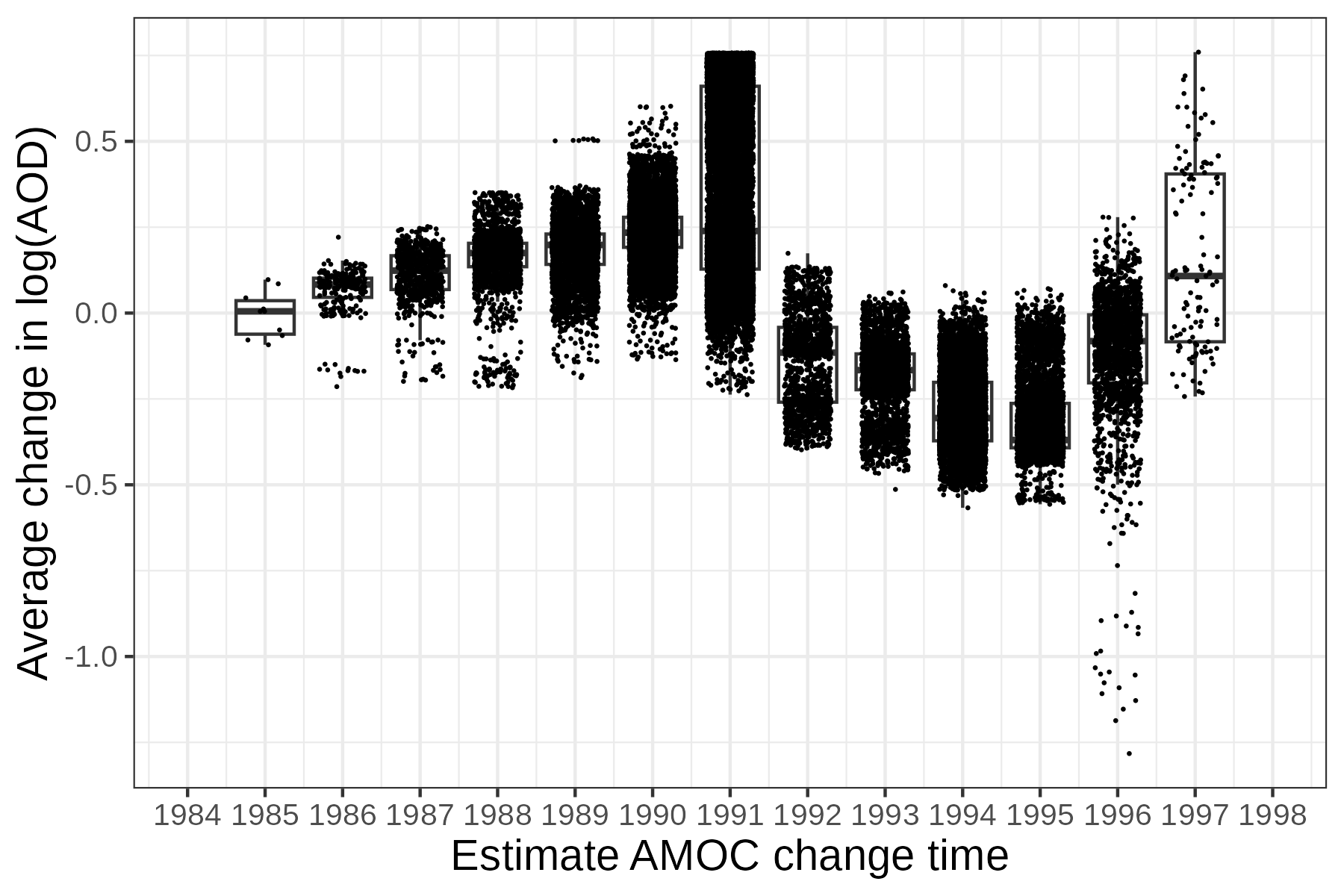}
\includegraphics[width = .48\textwidth]{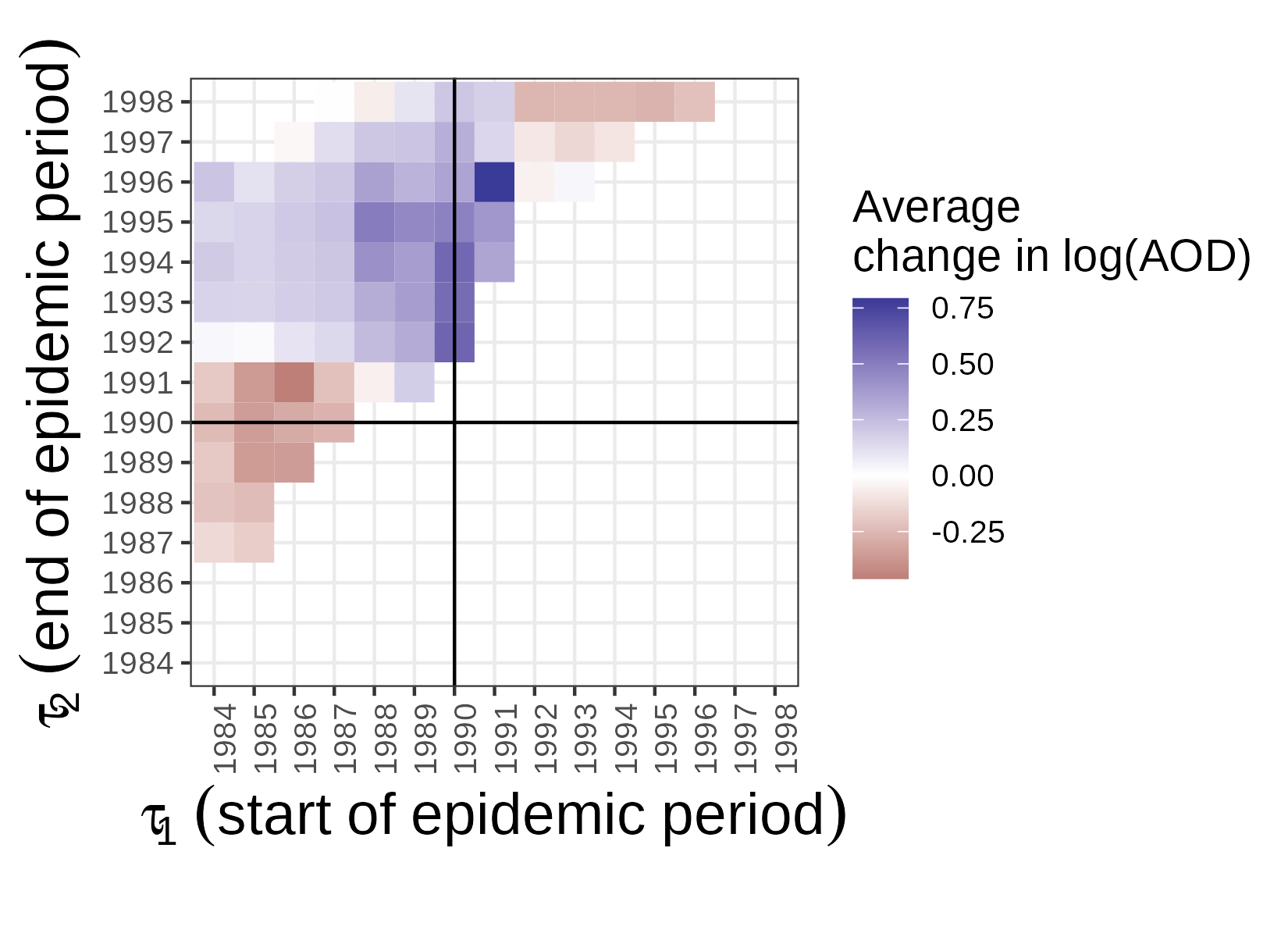}

\caption{AOD (Top Left) Boxplot of the average change estimate by the estimated AMOC change time for locations with significant Benjamini-Hochberg-adjusted p-value. Boxplots and jittered points are plotted for visualization. (Top Right) Mean average change estimate for the epidemic changepoint model for locations with significant uncorrected p-value. (Bottom) The same as above, but for all locations regardless of significance. }\label{fig:MERRA2_boxplot_AOD}

\end{figure}

\begin{figure}
\centering
\includegraphics[width = .48\textwidth]{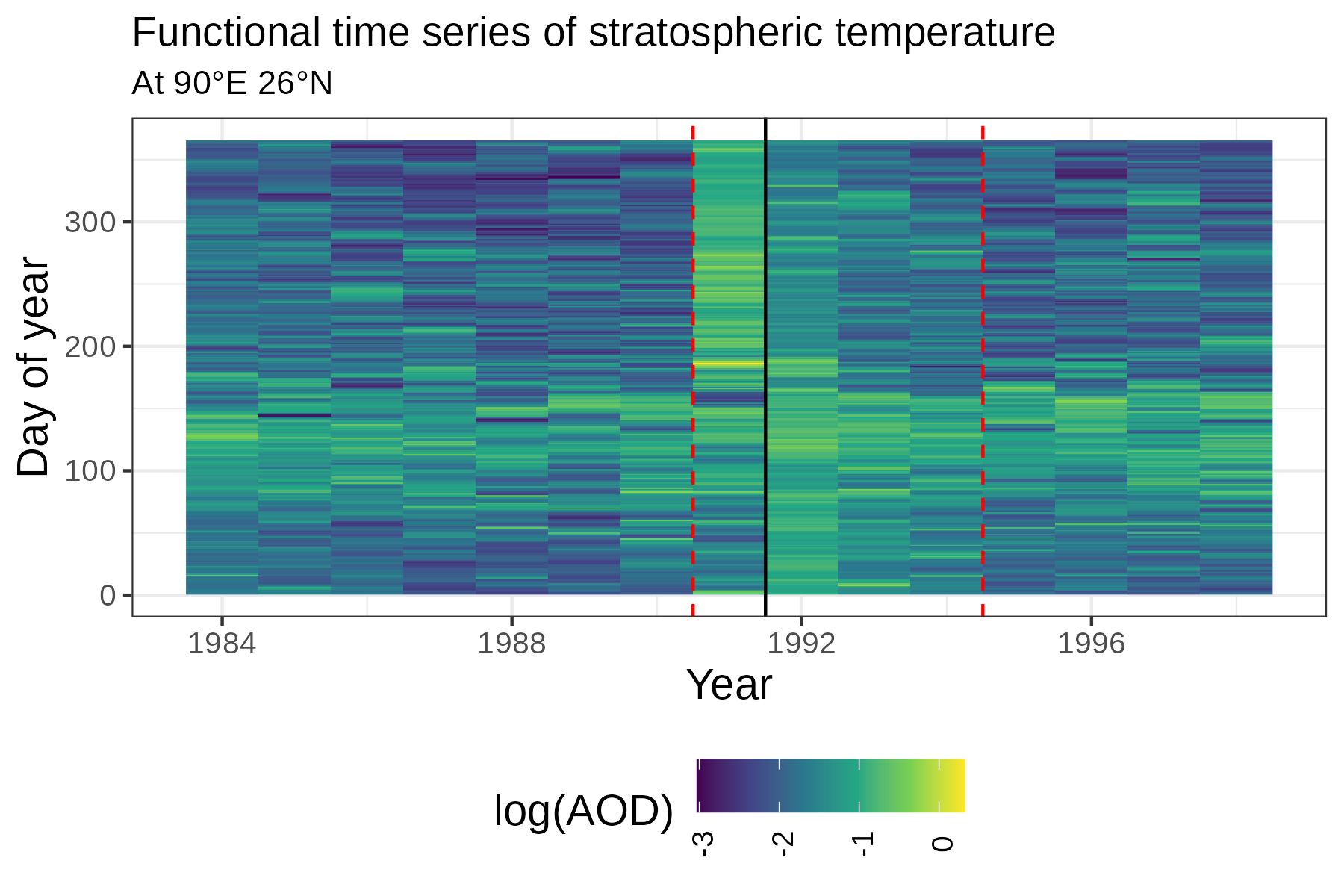}
\includegraphics[width = .48\textwidth]{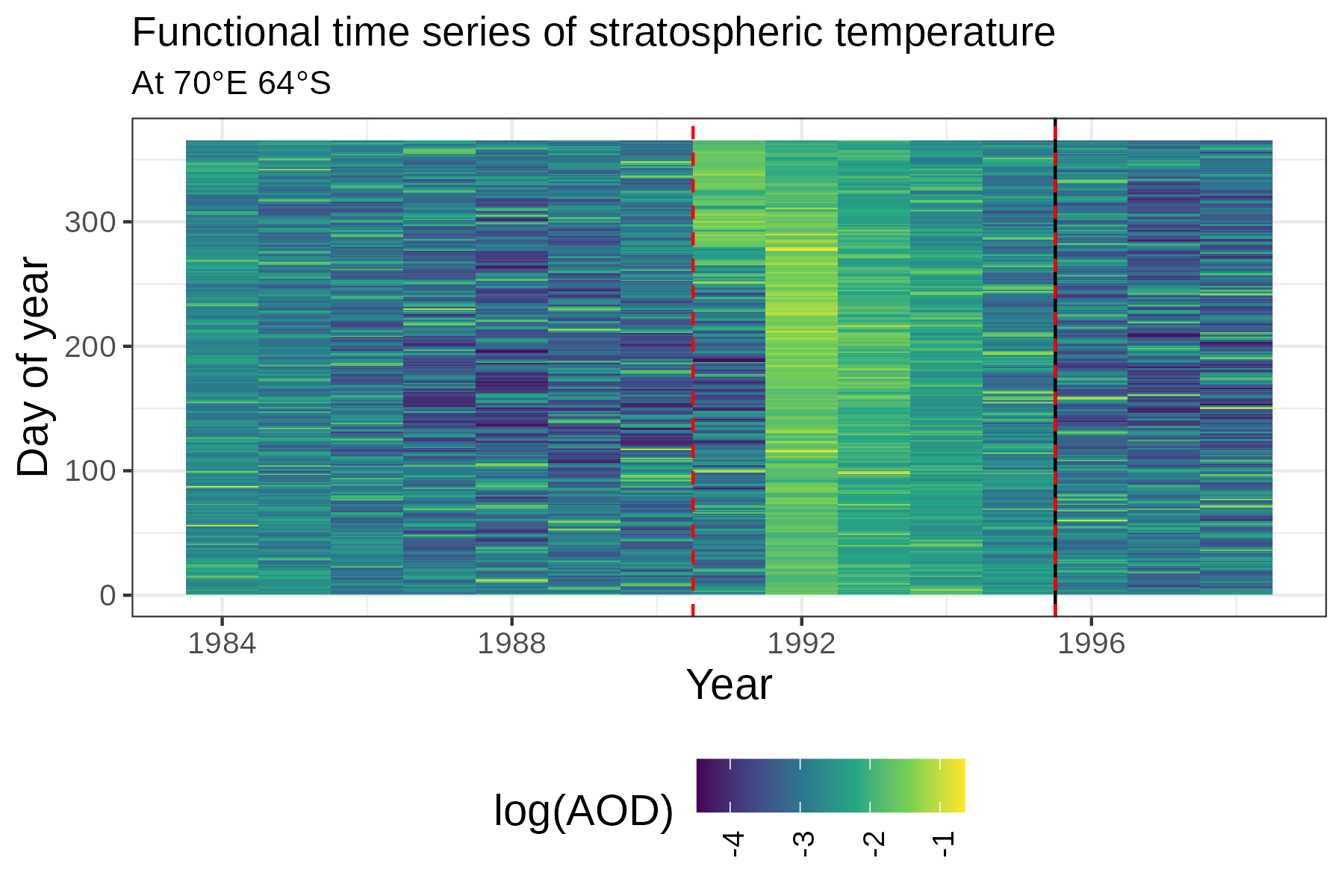}

\includegraphics[width = .48\textwidth]{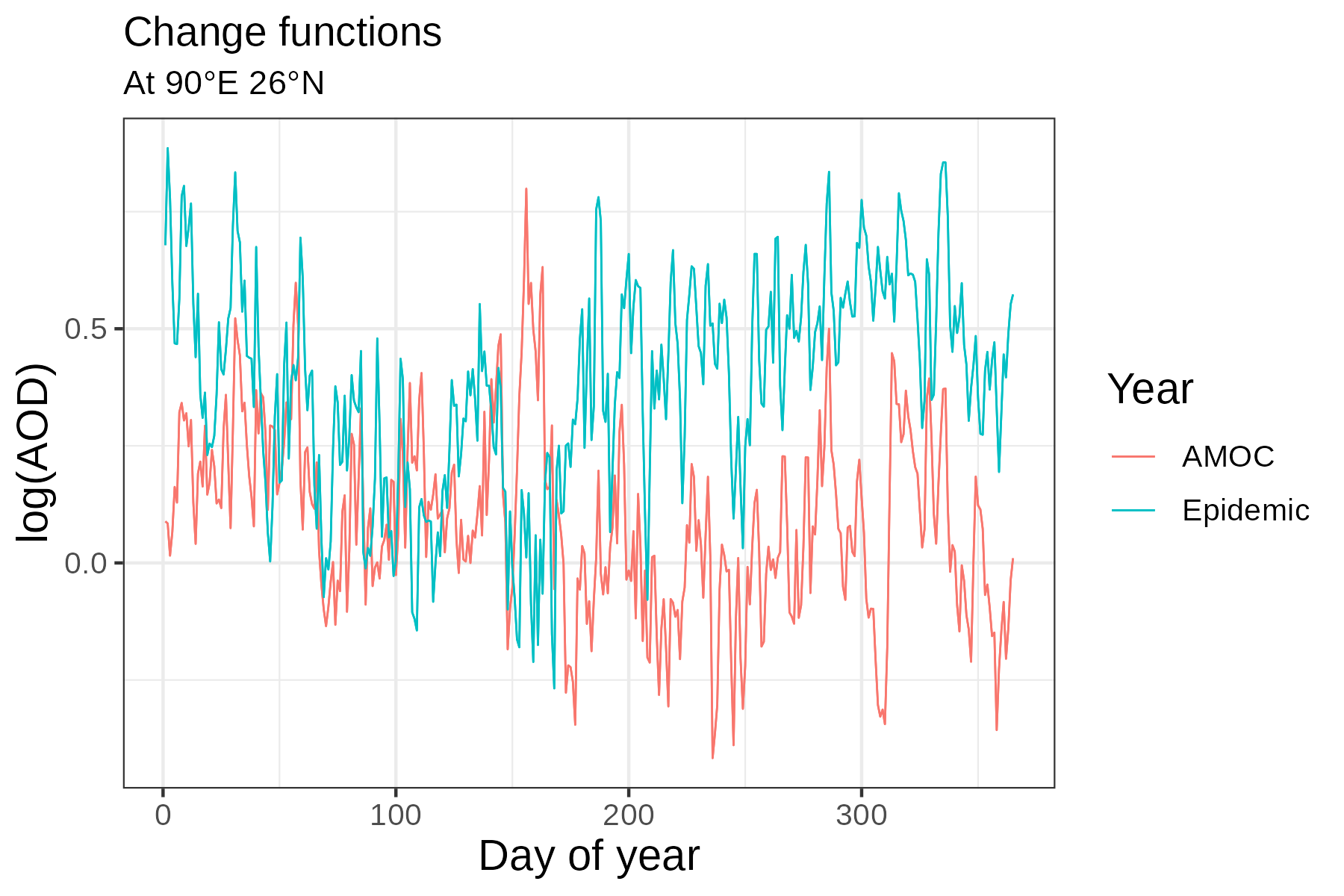}
\includegraphics[width = .48\textwidth]{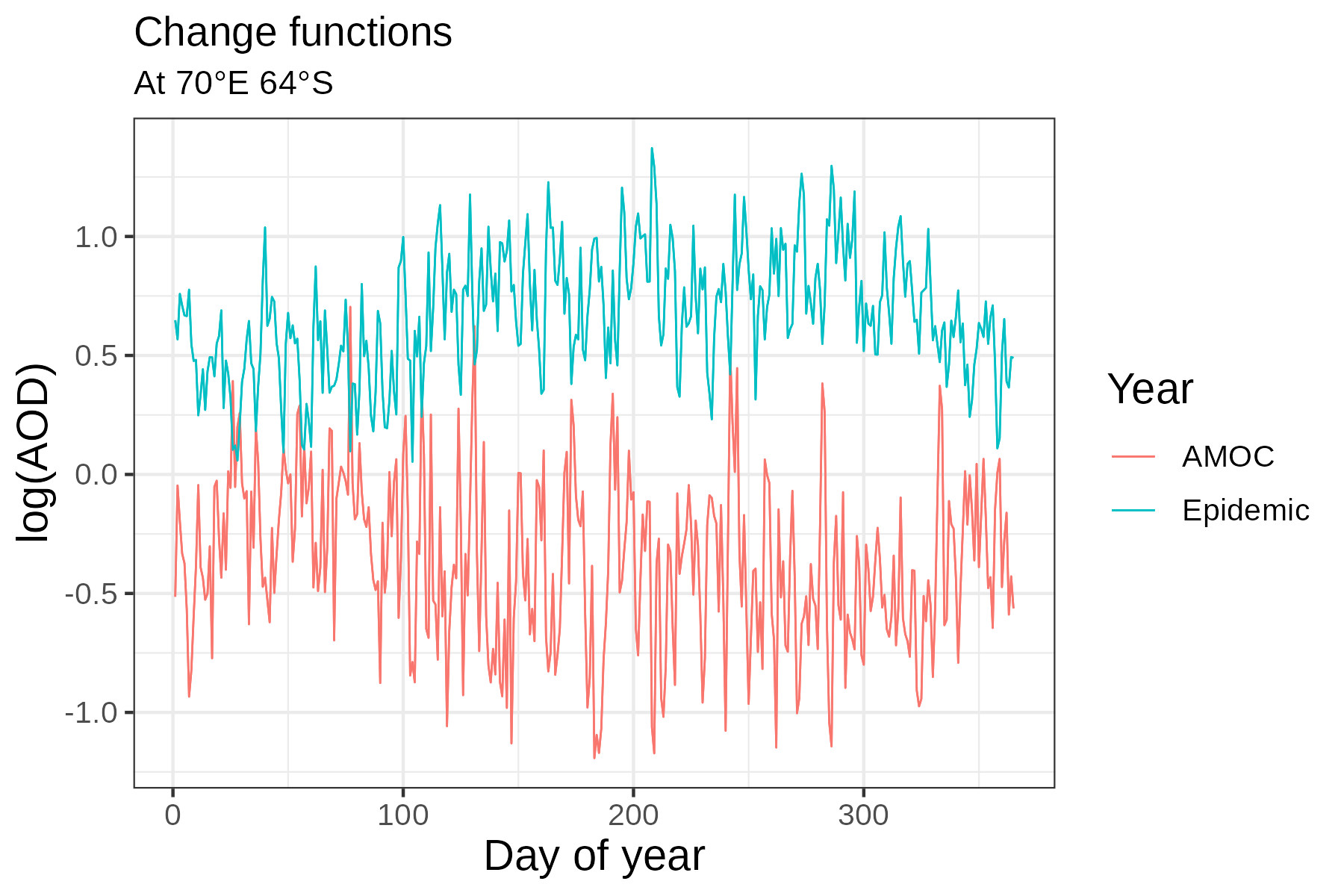}
\caption{Two examples of AOD and estimated change points. The solid black line represents the change detected using the AMOC model, with a p-value of $0.836$ (Left) and $ 0.281$ (Right). 
The two dashed red lines represent the change period detected using the epidemic model, with a p-value of $0.663$ (Left) and $0.512$ (Right). (Bottom) The estimated change functions for these two locations based on empirical averages without smoothing. }\label{fig:MERRA2_example_AOD}

\end{figure}

Data and results are presented for aerosol optical depth in Figures \ref{fig:MERRA2_data_AOD}, \ref{fig:merra2_amoc_AOD}, \ref{fig:merra2_epidemic_AOD}, \ref{fig:MERRA2_boxplot_AOD}, and \ref{fig:MERRA2_example_AOD}. 
Since AOD is a nonnegative quantity and potentially skewed, we first apply a log transformation to the data before applying our methodology. 


\section{Selection of hyperparameters}

Various approaches have been proposed to estimate the number of principal components $Q$ and the bandwidth for local regression $h$; see \cite{li2013selecting} and \cite{fan1996local}, respectively. 
Often these require retraining the model with varying $Q$ and $h$, a computationally expensive process. 
We choose $Q=10$ ensures we capture the top modes of variability without increasing computation too much.
For the simulation study, we set $h =  0.08$, while for the data analyses, we set $h=400$ kilometers. 
This gives the flexibility of estimating spatially-varying variances, but results in stable estimates. 
The number of basis functions was chosen to provide as much flexibility as computationally possible due to the complexity of the data.

\bibliographystyle{apalike2}

\bibliography{main.bib}